\documentclass[11pt]{article} 

\usepackage[top=1in,left=1in,left=1in,right=1in]{geometry}
\usepackage{indentfirst}
\usepackage{setspace}
\usepackage{pdflscape}

\usepackage{amsmath}
\usepackage{graphicx}

\usepackage{soul}
\usepackage[disable]{todonotes}

\usepackage{placeins}
\usepackage[aboveskip=1pt,labelfont=bf,labelsep=period,justification=raggedright,singlelinecheck=off]{caption}
\usepackage{subcaption}
\usepackage{booktabs} 
\usepackage{floatrow}

\usepackage{natbib}
 \bibpunct[, ]{(}{)}{,}{a}{}{,}%

\newcommand{\parencite}{\citep}
\newcommand{\textcite}{\citet}
\usepackage{hyperref}

\title{Strategic Choices of Migrants and Smugglers in the Central Mediterranean Sea}
\author{Katherine Hoffmann Pham\thanks{Corresponding author: khof@nyu.edu. The authors thank Panos Ipeirotis for his involvement in shaping this paper.} \and Junpei Komiyama }

\begin{document}
\maketitle

\begin{abstract}

The sea crossing from Libya to Italy is one of the world's most dangerous and politically contentious migration routes, and yet over half a million people have attempted the crossing since 2014. Leveraging data on aggregate migration flows and individual migration incidents, we estimate how migrants and smugglers have reacted to changes in border enforcement, namely the rise in interceptions by the Libyan Coast Guard starting in 2017 and the corresponding decrease in the probability of rescue at sea. We find support for a deterrence effect in which attempted crossings along the Central Mediterranean route declined, and a diversion effect in which some migrants substituted to the Western Mediterranean route. At the same time, smugglers adapted their tactics. Using a strategic model of the smuggler's choice of boat size, we estimate how smugglers trade off between the short-run payoffs to launching overcrowded boats and the long-run costs of making less successful crossing attempts under different levels of enforcement. Taken together, these analyses shed light on how the integration of incident- and flow-level datasets can inform ongoing migration policy debates and identify potential consequences of changing enforcement regimes. 
\end{abstract}

\doublespacing

\section{Introduction}

There are approximately {272 million} international migrants around the world \parencite{iom_world_2019}, 
and estimates suggested that a {quarter} of the international migrant stock were irregular migrants as of 2009 \parencite{undp_overcoming_2009}.
\footnote{We avoid value judgment and use the term ``irregular'' interchangeably with ``illegal,'' ``undocumented,'' ``unauthorized,'' etc. to refer to crossings that are not part of a country's formal channels for accepting migrants. In the discussions below, the term ``migrant'' may include refugees and asylum-seekers as well. Appendix~\ref{app:context} describes the legal context for sea migrants.} Smuggling networks often enable such flows, moving an estimated {2.5 million people in 2016 for an annual profit of \$5.5 - 7 billion} \parencite{un_office_on_drugs_and_crime_global_2018}. 
Limiting human smuggling may be desirable from a security perspective (e.g., in order to enforce the sovereignty of borders) or from a humanitarian perspective (e.g.,  to reduce exploitation and trafficking of undocumented people and discourage migrants from risking their lives on dangerous crossings). However, past research has found that efforts to limit human smuggling by increasing border enforcement may lead to unintended consequences. For example, researchers have identified a ``deterrence-diversion tradeoff'' in which some migrants forego the journey altogether, but others adapt by switching to alternative routes which may place them at higher risk \parencite{sorensen_effects_2007}.  As a result, it is difficult to estimate the effects of migration policies, and there is a need for system-level models to estimate how migrants and smugglers react to changes in the crossing environment.

\begin{figure}
        \centering
        \includegraphics[width=6in]{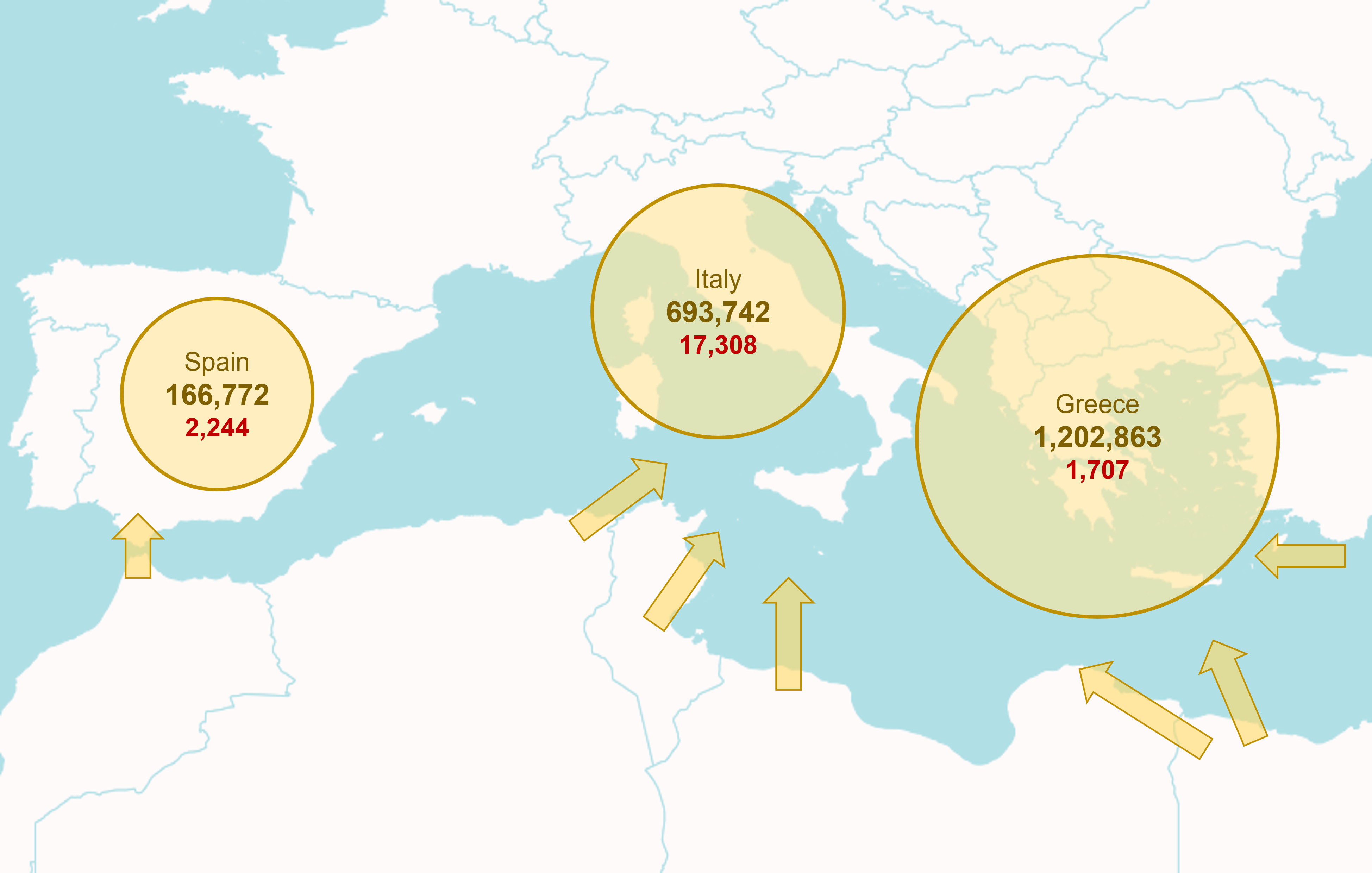}
        \caption{Flows along the Western, Central, and Eastern Mediterranean routes from 2014 - 2020}
        \floatfoot{The gold number shows the total number of arrivals to the primary destination country for each route \parencite{unhcr_mediterranean_2020-1}. The red number below shows the number of dead or missing migrants along the route \parencite{iom_missing_2021}. Graphic adapted from \cite{unhcr_mediterranean_2020-1}. }
        \label{fig:unhcr_flow}
\end{figure}

In this paper, we aim to build such models and apply them to study sea crossings on the Central Mediterranean route from Libya to Italy.  {Since 2014, over two million people have arrived in Europe by sea and over 20,000 people have died or gone missing}~\parencite{unhcr_mediterranean_2020-1}.  There are three primary routes through the Mediterranean (illustrated in Figure~\ref{fig:unhcr_flow}): the Eastern route from Turkey to Greece (approximately {58\%} of crossings from 2014 to 2020), the Central route from Libya to Italy and Malta (approximately {34\%}), and the Western route from Morocco to Spain (approximately  {8\%})~\parencite{unhcr_mediterranean_2020-1}.  
We focus on the Central Mediterranean route, which is the longest of the three and represents approximately {81\%} of all casualties observed in this period \parencite{iom_missing_2021}. 
In recent years, this route has been at the center of a contentious policy debate about the role of non-governmental organization-led (NGO-led) rescues and coast guard interceptions in encouraging or deterring risky migration attempts by sea.

While a number of researchers have attempted to model adaptive responses to migration conditions in the Central Mediterranean \parencite{steinhilper_contested_2018, deiana_migration_2019, cusumano_sea_2019, naiditch_matching_2020, camarena_political_2020}, much of this work focuses on analyzing the rise in NGO rescues and the surge in crossings from 2015 - 2017, when arrivals to Italy by sea peaked at over {27,000} people in a given month~\parencite{unhcr_mediterranean_2020}.
There has been less systematic analysis of the region after mid-2017, when the Libyan Coast Guard (LCG) began to intercept a growing share of migrants with Italian and European Union (EU) support, and information on the activity of migrants and smugglers became more elusive. In the face of the drastic change of environment caused by the increased role of the LCG, we seek to answer the following policy questions. 
    \begin{enumerate} 
        \item \textit{To what extent do Libyan interceptions deter the crossing of migrants?} Do increased interceptions on the Central route divert migrants towards the other two migration routes? Is there any causal relationship that we can identify from the data?
        \item \textit{How do smugglers, who help the migrants to cross the sea, elude interception?} Is there any evidence around how they adapt their strategy on the incident level, and what are the outcomes of these strategy adaptations?
        \item \textit{Can we generalize the ideas behind the analyses above to other irregular migration contexts, in order to estimate the consequences of policy changes?} What kind of models are able to support such counterfactual inferences? 
    \end{enumerate}

Answering these questions is significantly more challenging than in previous periods due to the scarcity and bias of incident datasets after the increase in Libyan interceptions in mid-2017. In particular, compared with the reports produced by NGOs which are motivated to disclose their rescue incidents, details on interceptions by the LCG (as well as casualties that occur as a consequence of their activities) are unpublished and the number of interceptions is reported only in aggregate. The most comprehensive dataset from the EU border control agency (Frontex) does not generally cover boats intercepted by the LCG at all. Therefore, we have a combination of biased but high-quality incident data, and comprehensive but low-quality flow data.  While this type of data is common in many mixed migration settings (where interceptions are often performed by non-transparent enforcement authorities), it is difficult to analyze because many of the conventional causal analysis methods do not apply to the situation. Indeed, few researchers\footnote{\textcite{deiana_migration_2019} and \textcite{camarena_political_2020} use incident data, but aggregate it at the daily level for their analysis.} have analyzed incident-level data on crossings, limiting their ability to study the evolution of smuggler strategy at the boat level. 

Our main contribution is therefore the causal analysis of Central Mediterranean crossing behavior both before and after the rise in Libyan interventions, with careful robustness checks. Unlike existing papers that focus on the causal impact of NGO presence~\citep{cusumano_sea_2019, deiana_migration_2019} or civil disorder and discontinuous policy change~\citep{camarena_political_2020}, we focus directly on the rate of successful crossing, which determines the benefit that migrants expect to obtain by their crossing attempts. We integrate the flow and the incident data by obtaining a reasonable estimate of the overall rate of interceptions across time, which we then associate with each recorded crossing attempt. We confirm that the rate of successful crossing affects multiple aspects of the Central Mediterranean crossing, from the volume of crossing attempts to crossing characteristics (namely, the boat size). 

We conduct two primary analyses. First, we build a time series model to estimate how the flow of crossings on the Central Mediterranean route responds to the growing rate of LCG interceptions, which increases the proportion of people returned to Libya and reduces migrants' probability of successful rescue to Europe. We employ an Error Correction Model (ECM) to avoid finding false causal relationships due to spurious correlations. We find a long-run positive relationship between rescue probability and attempted crossings. This is consistent with the somewhat surprising resurgence of sea migrants observed in the third quarter of 2019, as probability of interception declined.
The ECM's second-stage model of the number of crossings has an explanatory power of $R^2= 0.20$ (adjusted $R^2= 0.16$), which implies that the probability of successful rescue has a meaningful effect on the flow size. We estimate that a decline in rescue probability from {90\% to 50\% corresponds to over 10,000 fewer attempted monthly crossings on average, from over 14,600 expected crossings to approximately 3,400 crossings.}

Second, we analyze incident datasets to document the strategic response of smugglers to the increased probability of interception. As Libyan interceptions rise, anecdotal and observational evidence suggests that smugglers and migrants begin to prepare for longer voyages towards Europe. They increase the use of wooden boats relative to rubber rafts, and reduce the average number of people on board. To systematically explore crossing strategy, we construct a theoretical model of smuggler utility as a function of boat size. We build a discrete choice model of utility because it connects the rate of interception (flow-level data) and the incident-level data (boat size, type of boat) and enables counterfactual estimation with a limited amount of incident records. We estimate the strategic tradeoff between the short-run incentive to crowd more passengers onto wooden boats, and the long-run incentive to avoid Libyan interception by using smaller boats. {For rubber boats, we estimate that smaller boats ($\leq 50$ people) begin to dominate larger boats ($>100$ people) as the preferred alternative once the rate of Libyan interception approaches 60\%. } These results are consistent with the rise in the use of smaller boats in early 2019, when the intensity of LCG activity peaked. This implies that our model can capture continuous changes in the enforcement situation, unlike before-after discontinuity models.

In summary, we build on the evidence base for evaluating policy changes in the Central Mediterranean. The policy implications of our results are as follows:
    \begin{itemize}
        \item Our analysis of the flow-level dataset is consistent with the claim that increased interceptions decrease the odds of crossing. We find that this adjustment occurs relatively quickly in responses to changes in the rate of interception, and in the limit, we estimate that the total number of crossings could fall as low as 300 per month. However, we note that the crossings which \textit{do} continue despite increased enforcement are more perilous because it is less likely that distressed boats will be met with a rescue response. 
        \item  Our analysis finds evidence of a constrained diversion effect in which some migrants switch to the Western Mediterranean route when the chance of successful crossing on the Central route is low. 
        Descriptive analysis suggests that the extent of substitution varies by nationality and likely depends on the ease with which migrants can reach coastal departure points from their respective countries. This suggests that rather than focusing on conducting interceptions at individual crossing points (which can simply push migrants towards other crossing points that are less well policed and potentially more treacherous), policymakers should take a more comprehensive view of smuggling routes more generally. This is consistent with broader calls for integrated regional approaches to addressing the underlying drivers of migration, and for the expansion of safe and legal crossing routes \parencite{moas_introduction_2020}.  
        \item Our analysis of incident data suggests that smugglers adapt to the changes in the interception rate, and migrants are still elusive in the current environment. Even at times when the LCG operated most actively to block migrants transiting the Libyan coastal zone, its effectiveness was limited by smugglers' strategic response. Boats with smaller numbers on board are estimated to have had an advantage in passing the coastal region and reaching European search and rescue zones, and our model makes it possible to estimate how smugglers weigh this advantage against the possibility of collecting more revenue by adding passengers. 
    	 Conditional on rescue to Europe, our analysis suggests that the risk of death for passengers has not changed from one period to the next, despite this shift in strategy.
    	 However, as smugglers switch to smaller boats which are rescued farther out to sea, it is possible that there is an increasing number of boats which sink without ever being detected for a rescue attempt, thus biasing recent casualty estimates downwards. 
    \end{itemize}

The rest of the paper is structured as follows. In Sections~\ref{sec:context} and~\ref{sec:literature}, we provide background on the Central Mediterranean policy environment and a review of the literature on migration strategy. In Section~\ref{sec:flow_analysis}, we analyze the overall flow of crossings along the Central Mediterranean route. In Section~\ref{sec:incident_analysis}, we analyze the incident dataset and present a utility model of smuggling. Section~\ref{sec:conclusions} concludes the paper.\FloatBarrier

\section{Background}\label{sec:context}
The difficulty of policing maritime borders has long made the Mediterranean an attractive route into the European Union. In particular, the Central Mediterranean route from Libya to Italy and Malta draws migrants and refugees who have lived and worked in Libya for years, as well as those who use Libya as a transit country.
All together, since 2014 UNHCR reports that over {690,000} people have attempted the Central Mediterranean sea crossing \parencite{unhcr_mediterranean_2020} 
 and over {56,000} have been returned to Libya \parencite{unhcr_libya_2020},  
 while the International Organization for Migration (IOM) estimates that over {17,000} people have gone dead or missing along this route \parencite{iom_missing_2021}. 
 
Given the risks involved in the crossing, European policymakers are divided between the humanitarian imperative to save lives in the Central Mediterranean through search and rescue, and the desire to stop irregular migration flows and discourage risky migration. On the one hand, {migrants have legitimate reasons for fleeing Libya, where they have faced discrimination, human trafficking, detention in inhumane conditions, and the risk of air strikes from the civil war}~\parencite{amnesty_international_libyas_2017, amnesty_international_libya_2020}.
Conditional on successfully departing from Libya, {they may be protected from a forcible return, since international law requires that rescued migrants, refugees, and asylum-seekers be transported to a ``place of safety''}~\parencite{unhcr_desperate_2018}. On the other hand, smugglers are aware of these protections and actively manipulate them by sending migrants to sea in under-equipped boats that will require rescue. As crossings surged in 2016 the Italian authorities were reporting as many as {30} rescue operations a day, leading to accusations that search and rescue operations were acting as a {``ferry service'' for migrants and creating a ``pull factor''} which encouraged people to place their lives at risk~\parencite{deutsche_welle_italy_2016, baczynska_ferry_2017}.

\subsection{Policy Context}

Below, we characterize the recent policy response in the Central Mediterranean according to three main phases:
the dominance of Italian and EU naval missions; the rise of the NGO rescue response; and the growing role of the Libyan Coast Guard. We describe the phases in terms of the activities of these key actors, and thus some parts of them are overlapping.

\subsubsection{Phase 1: The Dominance of Italian and European Naval Missions (up to mid-2015)}\label{sec:eu_ops}
Large-scale search and rescue operations off the coast of Libya began as early as {2013}, when the Italian Navy launched the one-year-long Operation Mare Nostrum which assisted over {150,000} individuals~\parencite{marina_militare_mare_2014}.  At the end of {2014}, Mare Nostrum was replaced by the joint EU Operation Triton. While Mare Nostrum was {explicitly a life-saving operation, Triton emphasized border control and initially restricted naval patrols to a smaller area of the sea} \parencite{deiana_migration_2019}. The cost of this policy change became clear in {April 2015 when two major shipwrecks claimed over 1,000 lives}~\parencite{heller_death_2016}.  {In response to these tragedies, the EU launched Operation Sophia in June 2015, which supplemented Triton with an emphasis on preventing smuggling and destroying migrant ships so they could not be re-used in the future}~\parencite{eunavfor_med_operation_sophia_about_2018}. 

\subsubsection{Phase 2: The Rise in NGO Rescues (from mid-2015 to mid-2017)}
{The first NGO to operate in the Mediterranean was the Migrant Offshore Aid Station (MOAS), which began conducting rescues in 2014}. In the {spring of 2015} it was joined by M\'edecins sans Fronti\`eres (MSF, also known as Doctors without Borders), and a number of other NGOs have since followed suit. When crossings peaked in 2016 - 2017 there were as many as {13 different boats operating in the region}~\parencite{zandonini_how_2017}. The NGO presence increased the coordination of rescues, as NGOs would deploy off the coast of Libya and react quickly to any boats that were identified in international waters. However, this led to accusations that NGOs were colluding with smugglers and facilitating irregular migration.  

In {July 2017}, Italy proposed an NGO Code of Conduct which would impose a number of constraints on rescue ships wishing to use its ports \parencite{spagnolo_salvataggi_2017,cusumano_straightjacketing_2019}. European countries also began other efforts to limit the operations of NGO ships, for example by {initiating legal proceedings against them, preventing them from leaving port, or denying permission to disembark rescued migrants}~\parencite{european_union_agency_for_fundamental_rights_2019_2019}.

\begin{figure}[htb]
    \centering
    \includegraphics[width=5in]{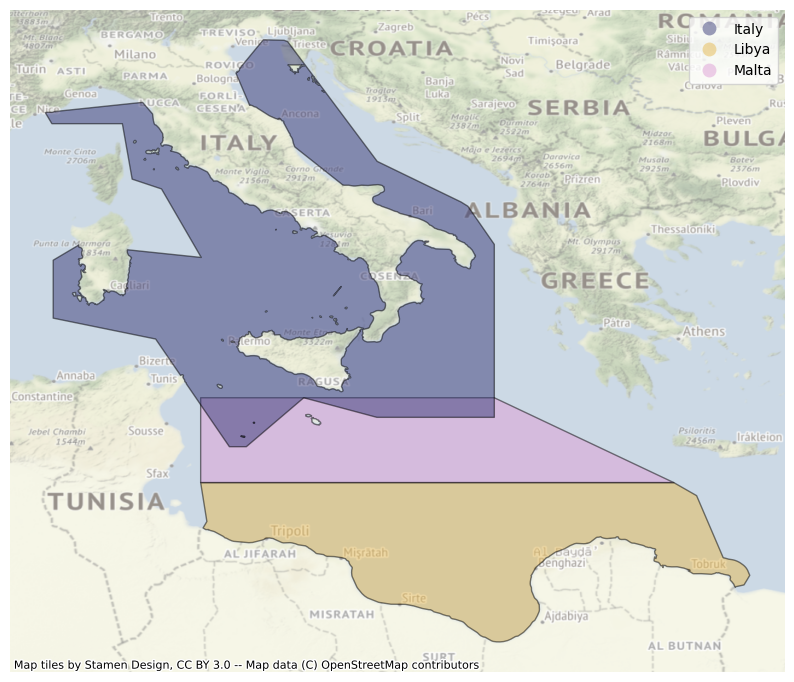}
    \caption{Map of Italian, Maltese, and Libyan Search and Rescue zones}
    \label{fig:sar_zones_map}
    \floatfoot{The spatial extent of each country's search and rescue zone was manually calculated by combining Natural Earth boundaries with the coordinates reported in the {International Maritime Organization's global search and rescue plan, updated in 2017 with Libya's rescue zone}~\parencite{natural_earth_110m_2020, international_maritime_organization_sar8circ4_2012, international_maritime_organization_ncsr_2017}.}
\end{figure}

\subsubsection{Phase 3: Interceptions by the Libyan Coast Guard (since mid-2017)}
Search and rescue (SAR) activity is typically managed by an {international network of coastal countries, which formally declare rescue zones and then supervise the response to incidents within their zones}.\footnote{{Formally, the coastal zone within 12 nautical miles (NM) from a country's shore is considered \textit{territorial waters}, and from a legal perspective is essentially treated like the land within the country's borders. Between 12 and 24 NM from shore are a country's contiguous waters, a zone which is not technically part of the country's territory but in which the country may still enforce some of its laws. Beyond this 24 NM boundary are the search and rescue (SAR) zones (Figure~\ref{fig:sar_zones_map}), in which the corresponding country's coast guard and/or naval forces are responsible for coordinating rescue operations}~\parencite{human_rights_watch_eu_2017}.} Key SAR zones in the Central Mediterranean are shown in Figure~\ref{fig:sar_zones_map}.
Prior to the formal recognition of the Libyan SAR zone in 2018, the Maritime Rescue Coordination Center  (MRCC) in Rome coordinated rescue efforts off the coast of Libya, and rescued migrants were typically taken to Europe. {However, as of October 2016 the EU began training the LCG, and as of May 2017 MRCC Rome began assigning rescues to LCG boats}~\parencite{human_rights_watch_eu_2017}. {This policy shift was formalized by the Malta Declaration in February 2017, in which the EU committed to support the training and capacity building of the LCG for the purpose of migration enforcement, and in June 2018, when Libya's request to formally establish a SAR zone was approved}~\parencite{human_rights_watch_eu_2017, human_rights_watch_euitalylibya_2018}. 

{While international ships are generally prevented from returning migrants to Libya because it is not considered a place of safety}~\parencite{unhcr_desperate_2018}, the LCG does not abide by these restrictions, and the increasing interventions of the LCG have clearly imperiled migrants. There are {reports} that the LCG is poorly trained, unprofessional, and ill-equipped to coordinate rescues~\parencite{eunavfor_med_monitoring_2018}; that it has ties to traffickers and has used its resources in the Libyan civil war~\parencite{scavo_trattativa_2019,scavo_migranti_2019, scavo_tripoli_2019, tondo_libya_2019}; and that it has perpetrated abuses against migrants in the course of rescues~\parencite{heller_its_2018, amnesty_international_libyas_2017}. 

\subsection{Smuggling Operations and Strategy}\label{sec:smuggling_ops}
The United Nations Office on Drugs and Crime (UNODC) estimates that {almost all} migrants who cross the Central Mediterranean rely on the help of smugglers
\parencite{un_office_on_drugs_and_crime_global_2018}. 
Migrants pay smugglers for passage, with payment depending on the type of boat, the location of the migrant on the boat, the nationality of the migrant, and the month and year of departure.\footnote{For example, in 2015 a report found that on the same vessel, Syrians in preferred locations might pay {\$2,500} while sub-Saharan Africans in the hold might pay just {\$800}~\parencite{aziz_changing_2015}. By 2017, the price of crossing in a rubber boat had fallen to {\$90} or below~\parencite{micallef_anti-human_2017}.}  

Migrants typically {depart at night}, most commonly in a wooden fishing boat or a rubber raft. Since the EU-led anti-smuggling Operation Sophia began destroying vessels in 2015, the {incentive to purchase cheaper disposable rafts has increased}~\parencite{uk_house_of_lords_european_union_committee_operation_2016}.
Wooden boats can hold up to {800} people \parencite{grunau_tragedy_2016},\footnote{ Sometimes as a result of {modifications which remove important structural components of the boat,} which can make the passage more dangerous~\parencite{micallef_human_2017}.} whereas rafts have a maximum capacity of approximately {150-200} people. When boats are overcrowded, the risk of sinking and injury to passengers on board is expected to rise.\footnote{For example, due to burns caused by leaked fuel mixing with salt water on the floor of the boat, or falling off the boat.}  Therefore, smugglers face a trade-off between collecting additional revenue per passenger and the risks incurred by overloading the boats.

Migrant boats are optionally equipped with life vests and a satellite phone, and one of the migrants may be chosen to act as navigator.  At the peak of the rescue response, boats were often given a {limited amount of fuel,} with the goal of reaching international waters \parencite{baker_rescue_2016}; once they passed the 24-nautical-mile boundary from the Libyan shore, they could use their satellite phone (if available) to request a rescue from the MRCC in Rome. As the LCG has grown more active in intercepting ships, anecdotal reports suggest that smugglers are {equipping boats with more fuel in order to help them evade the coast guard} and get further out to sea before requesting assistance \parencite{unhcr_desperate_2018}; the space taken by the fuel may in turn reduce the passenger capacity of the boats.

While the LCG is formally an adversary of the smuggling operations (since the coast guard is charged with intercepting migrant boats), there is evidence that coast guard members {coordinate with, profit from, and/or are involved in} smuggling operations \parencite{tondo_libya_2019,michael_making_2019, office_to_monitor_and_combat_trafficking_in_persons_2020_2020}.\FloatBarrier

\section{Related Work}\label{sec:literature}
Our study of Central Mediterranean crossings fits into the larger literature on human migration, which models movement patterns as a function of costs (such as travel expenses) and benefits (such as employment opportunities).
Informal migration is differentiated by the presence of a third factor: {internal and external border enforcement} \parencite{orrenius_undocumented_2015}. While there not been much focus on the \textit{process} of migration in the management science and operations research literature, related work has studied how to best allocate border patrol and coast guard efforts \parencite{papadaki_patrolling_2016,uzun_determining_2016}; strategies for humanitarian logistics \parencite{celik_humanitarian_2012,besiou_humanitarian_2020}; and the optimal placement and integration of migrants and refugees \parencite{ahani_placement_2021,haliassos_incompatible_2017,abujarour_understanding_2017}.

\subsection{Strategic Models of Migrants and Smugglers}

Smuggling and trafficking involve {sophisticated, diversified organizations}: ``the business is remarkably responsive to change and seems always to remain one or several steps ahead of those seeking to control it''~\parencite{salt_migration_1997}. Adaptation has been a key theme in existing research on the US-Mexican border, particularly with respect to 
the geographic intensity of US border patrol activities. For example,  \textcite{sorensen_effects_2007} posit that enforcement has {two primary effects:} (1) a \textit{deterrence effect} in which the policy discourages migrant crossings; and (2) a \textit{diversion effect} in which migrants shift their crossings to other parts of the border. As a result of the diversion effect, the overall volume of crossings can be relatively {inelastic} with respect to border enforcement.  Additional empirical research has found a {low impact of enforcement on overall crossing volumes, but substitution to other border sectors with higher crossing times and crossing risk and an increase the relative proportion of deaths from environmental factors such as dehydration}~\parencite{gathmann_effects_2008, cornelius_death_2001}. 

\subsection{Analyses of Central Mediterranean Crossings Before Phase 3}

In the Central Mediterranean, the deterrence-diversion debate has been shaped by two competing narratives: {a security/border control logic, and a humanitarian/crisis discourse}~\parencite{steinhilper_contested_2018}. A core research question is whether limiting rescue activity and increasing interceptions will discourage attempted crossings, or simply cause migrants to undertake increasingly risky crossings which are unassisted or even unobserved by state and humanitarian actors. To date, some studies have suggested that rescue presence {does not increase migrant departures or deaths}~\parencite{steinhilper_contested_2018,cusumano_sea_2019}.

\textcite{deiana_migration_2019} construct a theoretical model of how migrants and smugglers react to crossing risk in order to analyze behavioral responses to policy changes. {In their model,} migrants make a strategic decision about whether to cross in a safe boat, an unsafe boat, or not at all. They predict that SAR efforts lower crossing risks conditional on boat type, and therefore: encourage more migrants to undertake the journey; lead a larger fraction of crossings to use unsafe boats; and consequently, make departures more sensitive to crossing conditions. 

\textcite{naiditch_matching_2020} approach the problem from a different angle, building an {inter-temporal matching model} between migrants and smugglers. They predict that greater NGO presence in the Mediterranean will increase the number of migrants and smugglers, lower the costs of smuggling, and increase the likelihood of successful crossing.  Smugglers will benefit but the effect on migrant welfare and crossing prices is ambiguous.

\subsection{Analyses of Central Mediterranean Crossings Including Phase 3}

Policy analysis of the Central Mediterranean crossings given the recent rise in LCG intervention faces two key challenges. First, because of the changing political and economic environment, it can be difficult to isolate the causal effect of policy changes on crossing behavior. Prior works have studied the growing role of the LCG as a discontinuous policy change \citep{camarena_political_2020}, and studied how crossings correlate with NGO presence or capacity on a daily basis \citep{cusumano_sea_2019}. However, there have been no studies that assess how continuous changes in the overall border enforcement regime affect crossings. We address this gap with the use of an error correction model which allows us to estimate how crossing decisions on the Central and Western Mediterranean routes respond to rescue probabilities in the short and long term. With the help of this model we are able to analyze recent crossing activity through the end of 2019, when we observe a recovering flow of migrants.

Second, the lack of data on events involving the Libyan Coast Guard makes it difficult to connect smuggler choices (i.e. departure ports, boat type, or boat capacity) to outcomes at the incident level, since there is almost no data on boats that are intercepted. This hinders efforts to identify smuggler strategy because it hides the smuggler's reward function. To address this limitation, we borrow from discrete choice models and their connection to inverse reinforcement learning~\citep{abbeel_apprenticeship_2004, ziebart_maximum_2008, ermon_learning_2015}. Using data on the characteristics of a given incident and the choices made by the smugglers, we attempt to infer the parameters of the smuggler's utility function. Specifically, we study the question of boat crowding, and estimate the value that smugglers place on the revenue collected from adding more passengers, relative to the reduced chance of success when using larger boats (which in turn depends on the overall level of Libyan enforcement). We estimate this choice model with newly released Frontex data which, to our knowledge, has not been analyzed in its entirety to study this context.

\subsection{Limitations of Our Research}

Our approach has two key limitations. First, we assume that migrants are free to leave (though they may soon be intercepted at sea), and therefore we do not account for efforts to stop migrants from departing in the first place. \citet{camarena_political_2020} note that reductions in migrant flows during and after 2017 may be a function of either increased coast guard interceptions, or agreements with militias to reduce the availability of smuggling services and prevent departures in the first place; our analysis addresses the former. 

Second, while we believe that Frontex collects the most comprehensive incident data in the region, this dataset is biased because it generally does not include Libyan interceptions. This may lead us to overestimate the extent to which smugglers are strategic (because non-strategic actions are likely to be filtered out of the dataset by coast guard interceptions). However, LCG activities are erratic, with gaps in activity on certain days \citep{eunavfor_med_monitoring_2018}, which should help to ensure that a more diverse sample of incidents ultimately enters the Frontex dataset.\FloatBarrier

\section{Analysis of the Aggregate Flow Dataset}\label{sec:flow_analysis}

A common argument in favor of stricter border enforcement is that a decreased chance of successful crossing will deter crossing attempts. We assess whether or not this hypothesis is consistent with empirical estimates by investigating how crossings relate to the likelihood of rescue at sea and successful arrival in Italy or Malta. 

Since the crossing trend is highly non-stationary, a na\"ive regression can misidentify the model due to spurious regressions \citep{granger_spurious_1974}. To address this concern, we adopt a time-series error correction model to analyze the long- and short-term effects of rescue probability on the log odds of crossing \citep{box-steffensmeier_time_2014}. In Section \ref{sec:ecm_main}, we show that a reduced chance of successful crossing results in a smaller number of attempted crossings. Section~\ref{sec:spillover} similarly analyzes spillovers to the Western Mediterranean route, and finds significant but limited substitution to this route. Finally, Section \ref{subsec:nationality} presents descriptive analysis showing that spillover effects vary widely among different nationalities of migrants. 

\subsection{Data and Setup}\label{sec:log_odds}

We collect data on overall migration flows from the International Organization for Migration (IOM), which provides data on {aggregate sea arrivals\footnote{We assume that the number of sea arrivals is essentially equivalent to the number of people rescued because very few boats reach Europe independently.} to Italy and Malta; interceptions by the Libyan and Tunisian Coast Guards}; and dead or missing migrants along the
Central Mediterranean route~\parencite{iom_data_2019}. This data can be used to calculate the total number of crossing attempts and the likelihood of successful crossing. 

For a given month $t$, we define the number of people crossing on a given route as the sum of people who were rescued, intercepted, or reported dead or missing:
     $$N_{t,crossing} = N_{t,rescue} + N_{t,intercept} + N_{t,death}. $$
Note that when estimating the ECM model below, we take the unit of $N_{t,crossing}$ to be in thousands in order to yield coefficients that are more comparable in magnitude.

The probability of rescue is therefore:
    $$P_{t,rescue} \quad=\quad \frac{N_{t,rescue}}{N_{t,crossing}}.$$
    
\begin{figure}
        \centering
        \includegraphics[width=6in]{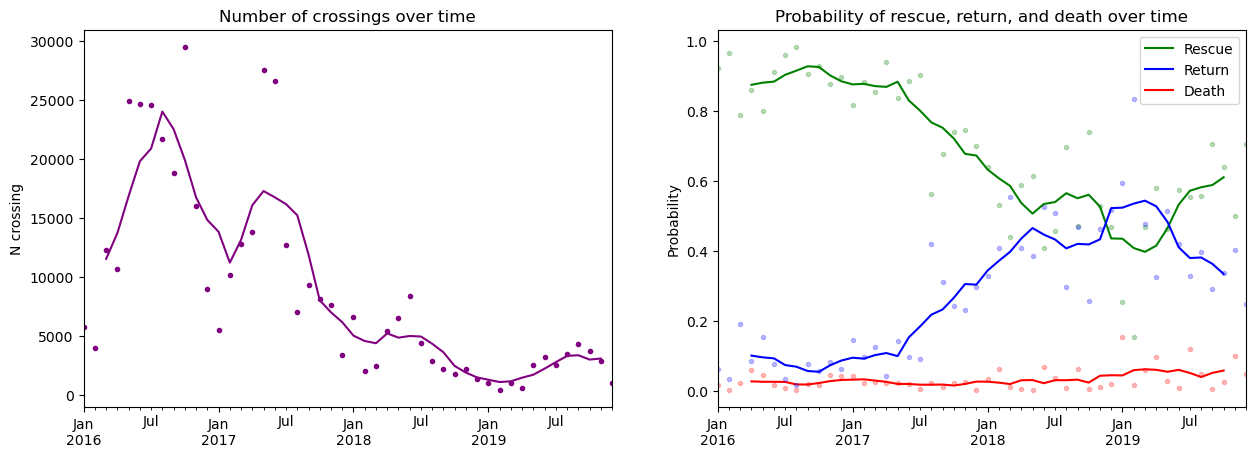}
        \caption{Crossing decisions and crossing outcomes on the Central Mediterranean route over time}
        \label{fig:n}
        	\floatfoot{Each dot represents the monthly number of crossings or the monthly probability of the respective outcome, whereas the lines represent centered six-month moving average trends. Data source: \textcite{iom_data_2019}.}
    \end{figure}

The number of people crossing on the Central Mediterranean route, as well as the probability of each outcome (rescue, return, and sinking), are shown in Figure~\ref{fig:n}. From inspecting the figure, it is clear that the number of people crossing and the probability of rescue have fallen over time, which coincides with the growing intervention by the LCG in Phase 3. However, it is unknown whether both series simply follow a common downward trend, or whether one series reflects changes in the other over the short or long term. In the short term, crossings might respond to increased rescue probability because migrants already in Libya could depart when it appears that the chances of success are high. In the long run, crossings might respond to increased rescue probability because additional migrants could travel to Libya in order to cross, and because smuggling operations could reconfigure to increase the overall volume of migrants they are able to launch.

\subsection{Model} \label{sec:ecm_overview}

In this section we provide a brief overview of the ECM, following the exposition in \citet[Section 6]{box-steffensmeier_time_2014}. The development of the ECM was motivated by the observation that when running Ordinary Least Squares (OLS) regressions using non-stationary dependent and independent variables (in our case, $N_{t,cross}$ and $P_{t, rescue}$, respectively), there is an elevated risk of finding a significant relationship between the two even when none exists (i.e., a spurious regression \parencite{granger_spurious_1974}). One solution in this case is to take first differences in order to obtain stationary variables, and then to fit a regression on the first differences. However, this provides insights only about short-run relationships between the variables, and does not account for the fact that these variables may react to each other in the longer term.  In fact, it is possible that two non-stationary series have a cointegrating relationship, in which a linear combination of the series is stationary (i.e. they have a stable long-run relationship). For example, it may be the case that in equilibrium:
    \begin{eqnarray}N_{t,cross} ~~=~~ \beta_0 ~~+~~ \beta_1~ P_{t,rescue}, \label{eq:equilib}\end{eqnarray}
where $N_{t,cross}$ is the monthly number of crossings (in thousands) and $P_{t,rescue}$ is defined as above.  The ECM essentially allows for both of these short- and long-run dynamics. Specifically, we estimate an ECM of the form:
    \begin{eqnarray}\Delta  N_{t,cross} ~~=~~ \alpha_0  ~~+~~ \alpha_1~e_{t-1} ~~+~~ \alpha_2~\Delta P_{t-1, rescue}~~+~~ \epsilon_t, \label{eq:ecm}\end{eqnarray}
where 
$\Delta N_{t,cross} = N_{t,cross} - N_{t-1,cross}$; 
$\alpha_0$ is a constant term; 
$e_{t-1} =  N_{t-1,cross}~ -~ \beta_0 ~-~ \beta_1 ~P_{t-1, rescue}$, the observed deviations from equilibrium as defined by Equation~\ref{eq:equilib};  
$\Delta P_{t-1, rescue}$ is defined analogously to $\Delta N_{t,cross}$; and 
$\epsilon_t$ is a random error term. 
In this case, $\alpha_1$ reflects the long run adjustment behavior that results from divergence between $P_{t, rescue}$ and $N_{t,cross}$, whereas $\alpha_2$ reflects the short run adjustment in $N_{t,cross}$ that results from a change in $P_{t-1, rescue}$.

We conduct our estimation using the Engle-Granger method \citep[Section 6.3.1]{box-steffensmeier_time_2014} with the help of the \verb|egranger| package in Stata SE 12.0; this is a two-step procedure in which Equation~\ref{eq:equilib} is first estimated from the data and the lagged residuals are then used as an estimate for ${e}_{t-1}$ in Equation~\ref{eq:ecm}.

\subsection{Evidence that Migrant Crossings on the Central Route Respond to Changes in the Probability of Rescue} \label{sec:ecm_main}
\FloatBarrier

We begin by estimating the relationship between crossing behavior and the probability of rescue on the Central Mediterranean route. The maximum number of crossings ever observed on the route was {29,478 in October 2016}. The results of estimating Equation~\ref{eq:ecm} using the monthly differenced number of crossings (in thousands) as the dependent variable are shown in Table~\ref{tbl:ecm_central}.

 We find significant evidence of long-run adjustment behavior, suggesting that the number of people crossing increases in the probability of rescue. 
 When crossings and rescue probability diverge from their equilibrium relationship, adjustment occurs fairly quickly: our results suggest that the log number crossing falls by approximately 40\% of the deviation from equilibrium {in each period after the divergence}. In other words, within four months over 85\% of the adjustment needed to restore equilibrium has occurred. Interestingly, we find no significant evidence of a short-run relationship between crossings and rescue probability. 
 
{The equilibrium relationship is estimated to be:}
    \begin{eqnarray*}
    N_{t,crossing}^{central} &\quad=\quad& -10.58 ~~+~~ 28.01  ~ P_{t,rescue}^{central}. \label{eq:coint2}
    \end{eqnarray*}
To place this equation in context, during our period of observation, we have seen the probability of rescue fall from approximately 90\% to 50\%.  In the long run, this is expected to correspond to a reduction of about {11,200} people per month according to our model.\footnote{That is, $(28.01\times 0.9) - (28.01\times 0.5)$.}

\begin{table}
    \centerline{\footnotesize{
\def\sym#1{\ifmmode^{#1}\else\(^{#1}\)\fi}
\begin{tabular}{l*{1}{c}}
\toprule
                                                       &\multicolumn{1}{c}{(1)}\\
                                                       &\multicolumn{1}{c}{$\Delta N^{central}_{t, cross}$}\\
\midrule
$\hat{e}_{t-1} $                                            &   -0.402\sym{***}\\
                                                       &  (0.123)         \\
\addlinespace
$\Delta ~P~^{central}_{t-1, rescue}$                   &   -3.249         \\
                                                       &  (5.477)         \\
\addlinespace
Constant                                                &    0.060         \\
                                                       &  (0.698)         \\
\midrule
R$^2$                                                     &    0.200         \\
R$^2$ - adjusted                                          &    0.163         \\
N Obs.                                                 &       46         \\
Mean Dep. Var.                                         &    -0.07         \\
\bottomrule
\multicolumn{2}{l}{\footnotesize Standard errors in parentheses}\\
\multicolumn{2}{l}{\footnotesize \sym{*} \(p<0.10\), \sym{**} \(p<0.05\), \sym{***} \(p<0.01\)}\\
\end{tabular}
}
}
    \caption{{Results of estimating the error correction model: Crossings on the Central route (in thousands) vs. rescues on the Central route}}\label{tbl:ecm_central}
\end{table}

\textbf{Robustness checks:}
We conducted Dickey-Fuller tests and Engle-Granger cointegration tests to verify the appropriateness of the error correction model, which we discuss in Appendix~\ref{subsec:ecm_check}. In general, our results are robust to different choices of the dependent variable and to several different specifications of the model, which we present in Appendix~\ref{sec:ecm_robustness}.

\subsection{Evidence That Migrants Substitute Strategically from the Central to the Western Route}
\label{sec:spillover}

Next, we test whether the probability of rescue on the Central Mediterranean route affects crossings on the Western Mediterranean route through Spain. As above, we gather flow data from IOM, which reports the monthly number of sea arrivals in Spain as well as the estimated number of deaths along the Western Mediterranean route.\footnote{Note that interception data for coast guards along the Western Mediterranean route is not included in the IOM dataset (presumably because it is not available to IOM) so the number crossing includes only those who are rescued and those who are reported dead or missing.} The maximum number of monthly crossings for the Western Mediterranean route was 10,598 in October 2018. 

In Table~\ref{tbl:ecm_western}, we re-estimate the error correction model from Section~\ref{sec:ecm_main} using the differenced number of crossings (in thousands) on the Western route as the dependent variable. As above, our estimates suggest significant long-run adjustments in response to deviation from the equilibrium relationship between Western Mediterranean crossings and the Central Mediterranean probability of rescue. However, the speed of adjustment is considerably slower (20\% vs. 40\%). 
As before, we find no significant short-term effect of the probability of rescue on crossings.

\begin{table}
    \centerline{\footnotesize{
\def\sym#1{\ifmmode^{#1}\else\(^{#1}\)\fi}
\begin{tabular}{l*{1}{c}}
\toprule
                                                       &\multicolumn{1}{c}{(1)}\\
                                                       &\multicolumn{1}{c}{$\Delta N~^{western}_{t,cross}$}\\
\midrule
$\hat{e}_{t-1} $                                             &   -0.204\sym{**} \\
                                                       &  (0.092)         \\
\addlinespace
$\Delta ~P~^{central}_{t-1,rescue}$                     &    0.421         \\
                                                       &  (1.525)         \\
\addlinespace
\addlinespace
Constant                                               &    0.040         \\
                                                       &  (0.199)         \\
\midrule
R$^2$                                                     &    0.102         \\
R$^2$ - adjusted                                          &    0.060         \\
N Obs.                                                 &       46         \\
Mean Dep. Var.                                         &     0.03         \\
\bottomrule
\multicolumn{2}{l}{\footnotesize Standard errors in parentheses}\\
\multicolumn{2}{l}{\footnotesize \sym{*} \(p<0.10\), \sym{**} \(p<0.05\), \sym{***} \(p<0.01\)}\\
\end{tabular}
}
}
    \caption{{Results of estimating the error correction model: Crossings on the Western route (in thousands) vs. rescues on the Central route}}\label{tbl:ecm_western}
\end{table}

{The equilibrium relationship is estimated to be:}
    \begin{eqnarray*}
        N_{t,cross}^{western} &\quad=\quad& 5.32 ~~-~~ 4.31~ P_{t,rescue}^{central}\label{eq:coint_spain}  
    \end{eqnarray*}
This suggests a negative relationship between the two: as the probability of rescue along the Central Mediterranean route decreases, Western Mediterranean crossings increase. In particular, if the probability of rescue falls from approximately 90\% to 50\%, this model predicts about {1,700} additional crossings per month according to our model.\footnote{That is,  $(4.31\times 0.9) - (4.31\times 0.5)$.} Therefore, in addition to the slower rate of adjustment, we find that the log odds of crossing on the Western Mediterranean route exhibit a smaller absolute response to changes in the probability of rescue on the Central route. Taken together, our models estimate that when the probability of rescue on the Central Mediterranean route falls from 90\% to 50\% (roughly corresponding to the shift between January 2017 and January 2019), approximately 15\% of the 10,300 people who are deterred from crossing on the Central Mediterranean route will shift to the Western Mediterranean route. 

\subsection{Evidence that Substitution Behavior Varies by Nationality}\label{subsec:nationality}

\begin{figure}
    \centerline{\includegraphics[width=5.5in]{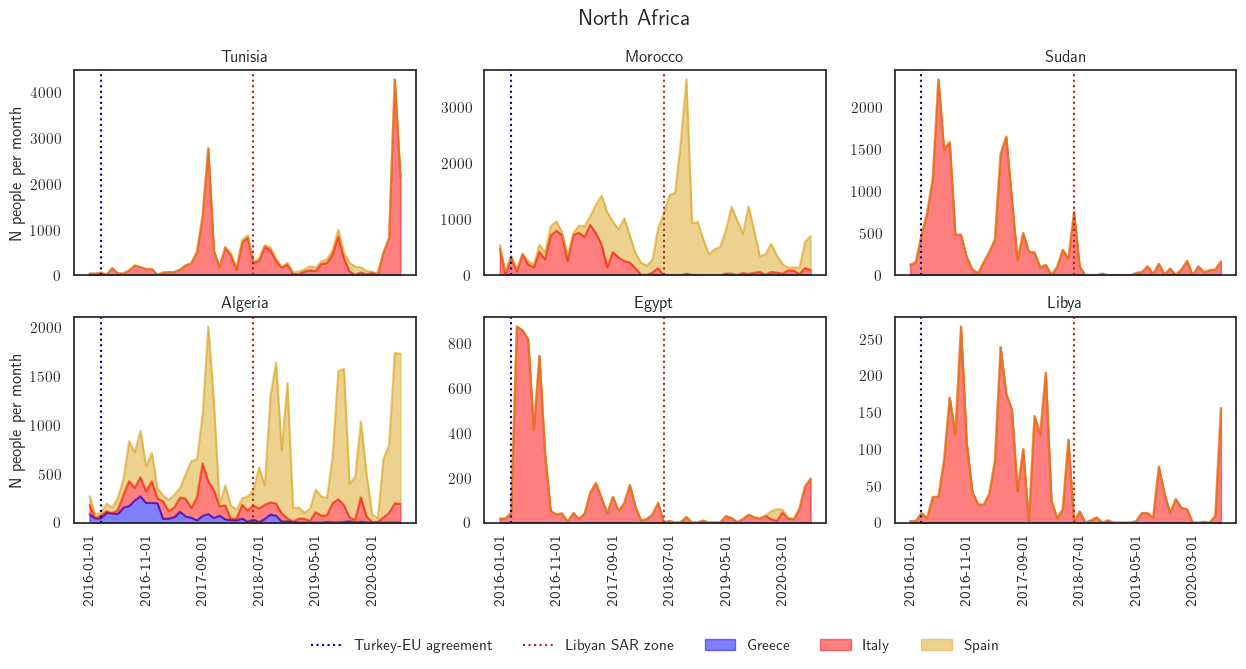}}
    \centerline{\includegraphics[width=5.5in]{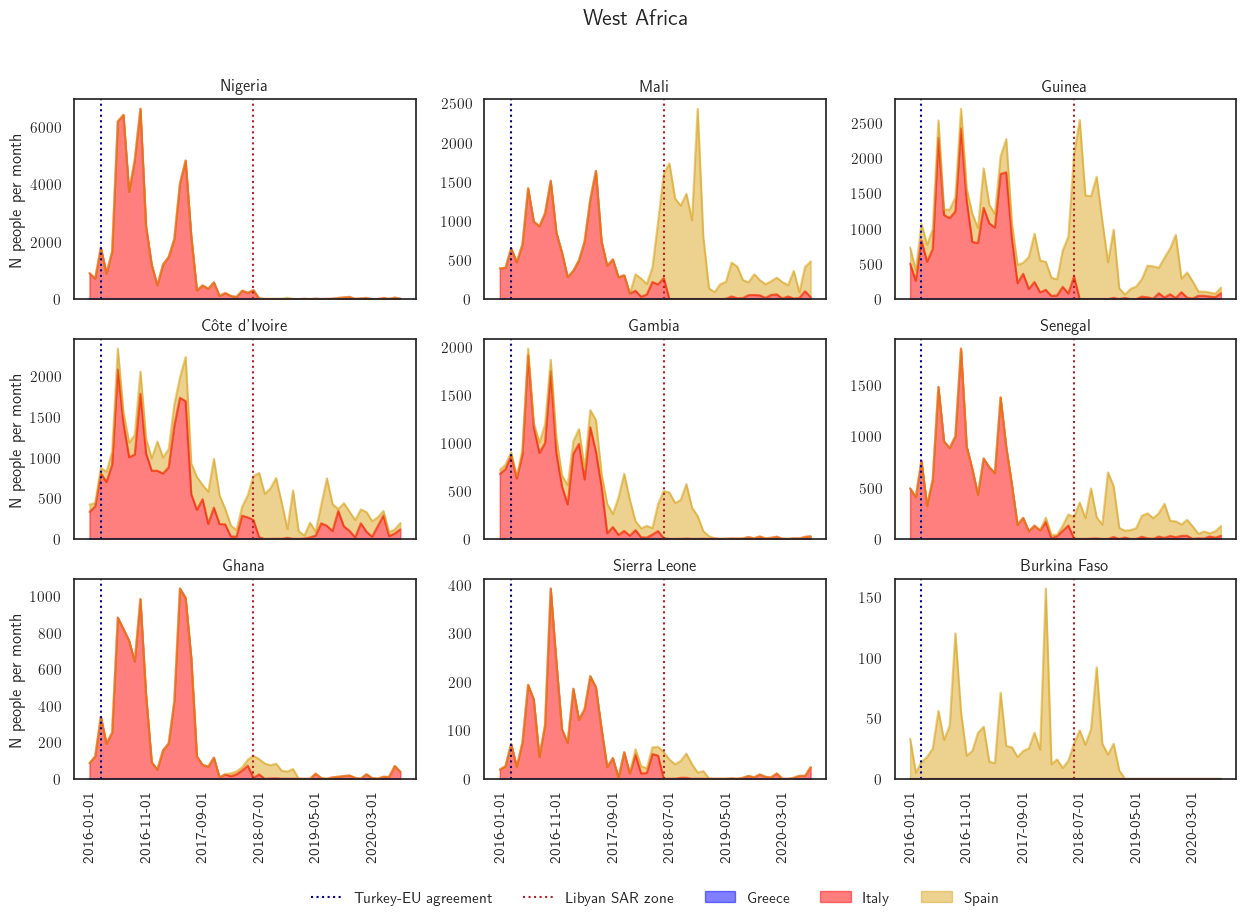}}
    \floatfoot{The following West African countries had less than 150 monthly arrivals in all time periods (all via the Spanish route) and were omitted from the plots:
    Mauritania, Guinea-Bissau, Togo, and Liberia. Data source: \textcite{unhcr_europe_2019}.}
    \caption{The number of migrants crossing by nationality and route: North and West Africa}\label{fig:gravity_a}
\end{figure}
\begin{figure}
    \centerline{\includegraphics[width=5.5in]{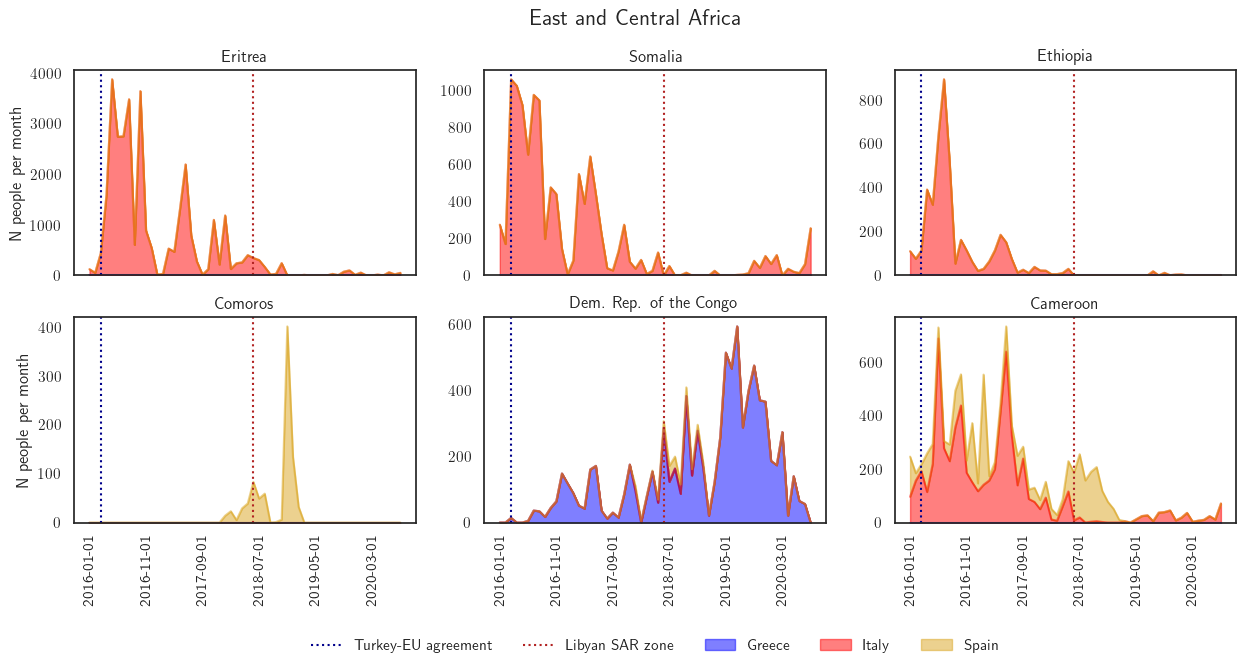}}
    \floatfoot{The following Central African countries had less than 50 monthly arrivals in all time periods (all via the Spanish route) and were omitted: the Republic of the Congo, the Central African Republic, and Chad. Data source: \textcite{unhcr_europe_2019}.}
    \caption{The number of migrants crossing by nationality and route: East and Central Africa}\label{fig:gravity_b}
\end{figure}    

We also investigate route choice by nationality, and find that the choice of route is highly correlated with country of origin. We gather additional information on flows from UNHCR, which reports data on {monthly sea arrivals to Italy, Greece, Spain and Cyprus} broken down by country of origin~\parencite{unhcr_europe_2019, unhcr_mediterranean_2020-1}. Figures~\ref{fig:gravity_a} and~\ref{fig:gravity_b} illustrate the share of African migrants crossing on each route by nationality over time. From Figure~\ref{fig:gravity_a}, we can see that 
North Africans primarily take the Central Mediterranean route, with the exception of Algerians (who seem to take advantage of all three routes) and Moroccans (who tend to take the Western Mediterranean route since Morocco is a key departure point for Spain, but who surprisingly preferred the Central Mediterranean route when crossing was easy on this route).

West Africans have generally substituted from the Central Mediterranean route to the Western Mediterranean route over time. This is likely due to the fact that from West Africa, multiple overland routes exist to either Western or Central Mediterranean departure points. In contrast, East Africans from the horn of Africa favor the Central Mediterranean route with very little substitution over routes, most likely because there are well-established smuggling routes from the horn to Libya.  Interestingly, migrants from two nationalities which are farther from common overland smuggling routes -- Comoros and the Democratic Republic of the Congo -- do not always choose the most geographically proximate points; we also see little substitution by these nationalities over time. 

Taken together, we can see that many nationalities' preference of migration route varies over time. Furthermore, substitution between routes seems to coincide with major policy changes, such as the growing role of the Libyan Coast Guard in intercepting migrants {as formalized by the establishment of the SAR zone in June 2018.} However, substitution appears constrained by geographic proximity to departure points and by the availability of overland smuggling routes.\footnote{Two other relevant factors are entry requirements for any borders that must be crossed in order to reach the target departure points (to the extent that border crossings are formally monitored), and the likelihood that members of a given nationality will be approved for an asylum claim (which may give some nationalities an incentive to select routes with a lower probability of detection).} Therefore, both the routes chosen and the sensitivity of this choice to crossing conditions vary substantially by region of origin. 

 \FloatBarrier
\section{Analysis of the Individual Incident Dataset}\label{sec:incident_analysis}

Thus far, we have analyzed the data on migration flows over the Central Mediterranean routes and found that migrants' decision to cross on the Central Mediterranean route exhibits a long-term response to changes in the probability of rescue. 

While migrants make a strategic decision of whether and where to cross depending on crossing conditions, smugglers may also respond to these conditions by varying their strategy.\footnote{We assume that smugglers have control over the details of the passage, such as boat type, the number of people on the boat, and other details of the crossings \parencite{salt_migration_1997,tamura_migrant_2010}.} As the LCG claimed increasing control over the coastal zone, migrant boats located in the Libyan SAR zone were more likely to be intercepted and less likely to be rescued to Europe. Consequently, we observe two high-level shifts in the strategic actions of smugglers along the Central Mediterranean route. 

First, the use of wooden boats increased, and the average size of boats departing from Libya (as measured by the number of people on board) grew smaller in this phase. The more aggressive the LCG's interception activities are, the greater the distance migrant boats need to cross to secure a rescue by NGOs or European authorities. As a result, anecdotal reports suggest that smugglers have been using space on boats to load {more fuel} \citep{unhcr_desperate_2018}, rather than collecting more revenue (and potentially, slowing boats down) by adding additional passengers.

Second, the share of boats departing from Tunisia relative to Libya, {which was very small at the beginning of Phase 3}, increased. In this section, we focus on boats departing from Libya because it is not clear whether local smuggling networks in Libya can choose to launch boats from Tunisia, and because the Libyan route has been historically more popular and has more total incidents than the Tunisian route. 

We first describe the incident data in Section~\ref{subsec:incidentdata}.
In Sections~\ref{sec:incident_model} through~\ref{sec:utility_results}, we build and estimate a choice model of Libyan smugglers' strategic shift to smaller boats in response to LCG interceptions. In Section~\ref{subsec:incident_counterfactual}, we conduct counterfactual estimation of boat sizes under different levels of Libyan interceptions, which explains the snowballing shift towards small boats observed in 2018 - 2019. Appendix~\ref{sec:incident_desc} provides further descriptive analysis of the incident datasets for context.

\FloatBarrier\FloatBarrier
\subsection{Description of the Incident Dataset}
\label{subsec:incidentdata}

We collect incident data from Frontex, the European border control agency which supervises the deployment of {aerial and naval assets} to patrol the EU's maritime borders. The agency records border-related incidents in its {Joint Operations Reporting Application (JORA)} database. In response to our request for public access to documents, Frontex has released records of incidents which occurred under {Operations Hermes, Triton, and Themis from 2014 - 2019}, including: {information on the date of detection; the departure country of the migrants; the number of people involved; the number of deaths; the boat type; and whether the boat was detected inside or outside of Frontex's operational area}~\parencite{asktheeu_jora_2020}. In total, the datasets we received contained 
{4,365} incidents originating in Libya from 2014 - 2019, including rescues involving actors outside of Frontex, such as NGOs and merchant ships.\footnote{We excluded incidents where the transportation type was land-based (i.e. ``bus'', ``camper van'', ``on foot'', and ``car''). The dataset we used to estimate the choice model was further reduced to {1,851} incidents, because it was limited to: incidents occurring from 2016 onwards (since Libyan interception data is only available from 2016); incidents involving rubber boats; and incidents where the number of people on board and the number of vessels involved were reported (i.e., not missing).} Each incident corresponds to a boat or (occasionally) a set of multiple boats that is acknowledged by Frontex.

\subsubsection{Other Datasets and Analyses}

We have also gathered four other incident datasets (from M\'edecins sans Fronti\`eres, Watch The Med, Broadcast Warnings, and the IOM), which contain incidents recorded by NGOs, monitoring efforts, and emergency calls. After comparing these different datasets, we found that Frontex is the most comprehensive dataset in terms of volume. Therefore, we solely use the Frontex dataset for our main analyses and use the other datasets for supplementary analyses. 

In Appendix \ref{sec:data}, we conduct a brief comparative analysis of these different datasets which suggests that after 2017Q3, there was (1) a decrease in boat size and (2) an increase in the share of wooden boats compared with rubber boats. We also observe (3) an increase in incident distance from the SAR border, which implies that the smugglers are traveling further before detection and that the mobility of the boats is increasingly important. Finally, we (4) confirm the relation between the mobility and boat size, showing that the probability of boats being found in the Frontex operational area (i.e., farther from the coast of Libya) increases when the boat size is small. 

\subsection{Strategy Model for the Shift in Boat Size}
\label{sec:incident_model}
A key challenge with this incident dataset is that it is not a representative sample of all movements in the region. In particular, the Frontex data does not cover LCG interceptions.\footnote{There are several other datasets that may include individual LCG interception incidents, such as IOM's Missing Migrants dataset or the narrative data produced by Watch The Med/Alarm Phone. However, these datasets are not very comprehensive. As a robustness check, we conduct a descriptive analysis that utilizes these other datasets in Appendix~\ref{sec:incident_desc}.} As a result, we do not have the data to estimate how a given set of inputs (e.g. boat type or boat size) translates to outcomes (i.e. rescue, interception, or sinking) at the incident level. Instead, we proceed in two stages. First, we identify the overall probability of interception in a given quarter. Then, we examine how the behavior of smugglers changes as a function of the quarterly interception probability. 

Our model of the smuggler's utility depends on two primary factors: the number of people on board a given boat, and the estimated probability that a boat will be intercepted in the Libyan SAR zone. 
We aggregate the data by quarter because this secures a sufficient number of incidents sampled per quarter, and because we estimate that three to four months is approximately the amount of time it takes for crossing behavior to react to changes in the probability of rescue (see Section~\ref{sec:flow_analysis} for further discussion on the rate of adjustment), which suggests that this may also be a sensible time scale at which to analyze smuggler responses.

\subsubsection{Estimating the Probability of Interception in the Libyan SAR zone}\label{subsec:assumptions}

We begin by estimating the quarterly probability that a boat that departs from Libya is intercepted, $p_{interception}^{qL}$, where $q$ denotes the quarter and $L$ denotes the fact that the boat originated in Libya. This part utilizes the flow data as well as the incident data.
While the IOM flow data includes the number of interceptions off the coast of Libya $(N_{interception}^{qL})$, it includes only the \textit{total} number of arrivals ($N_{arrivals}^{q})$ and deaths $(N_{death}^{q})$ for the Central Mediterranean route, which might include incidents originating in nearby countries such as Egypt or Tunisia. To proceed with the estimation of $p_{interception}^{qL}$, we therefore rely on the incident-level data from Frontex and make the following two assumptions:

\begin{enumerate}
\item  \textit{Frontex's recorded incidents do not cover interceptions. However, Frontex data is a uniform and nearly comprehensive sample of rescue incidents in the region.} This assumption is justified by the fact that the total population in the Frontex dataset generally matches the total number of arrivals reported by IOM (see Figure \ref{fig:representativeness_of_frontex_data}). 
\item \textit{The smuggler's decision focuses on the probability of interception and does not independently weigh the probability of sinking, which is small. Therefore, excluding this outcome will not substantially affect the model results.} This assumption is justified by the fact that the casualty rate is low relative to the number of interceptions and crossings (see Figure \ref{fig:n}), and that migrants are typically very distressed by the prospect of LCG interception.
\end{enumerate}
Additional discussion of these assumptions is provided in Appendix~\ref{sec:justify_assumptions}.

\begin{figure}[htb]
    \centerline{\includegraphics[width=\linewidth]{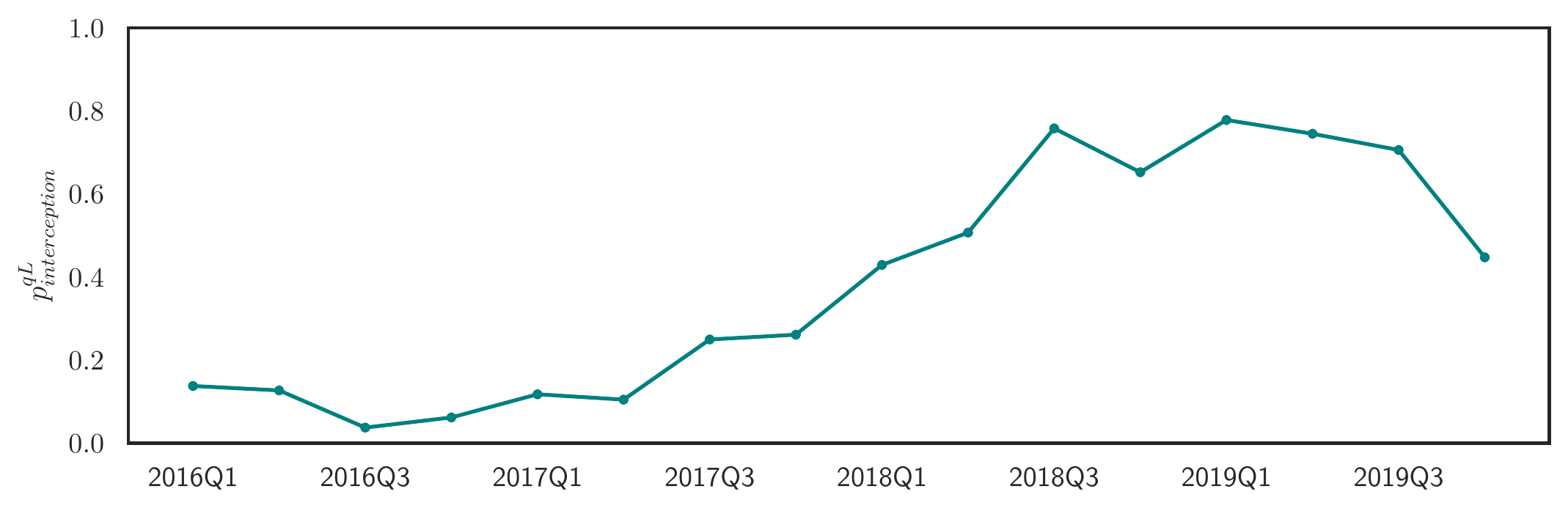}}
    \caption{The estimated probability of interception for boats departing Libya by quarter.}
    \label{fig:pinterception}
\end{figure}

Using Assumption (1), the estimated number of arrivals from Libya can be estimated as:
$$N^{qL}_{rescue} \quad=\quad s^{qL}_{rescue}~*~ N^{q}_{rescue},$$
where $s^{qL}_{rescue}$ is the quarterly share of migrants rescued who originate from Libya as opposed to other origin countries, which we estimate from the Frontex data.
Using assumption (2), the smuggler's expected probability of interception given departure from Libya is:
$$p^{qL}_{interception} \quad=\quad \frac{N^{qL}_{interception}}{N^{qL}_{rescue}~+~N^{qL}_{interception}}.$$
The calculated probability of interception over time is illustrated in Figure~\ref{fig:pinterception}. For all remaining sections of the paper, we simplify the notation by removing the $L$ superscript and use $p^{q}_{interception}$ to refer to the estimated rate of interception for boats departing Libya only.

\subsubsection{The Smuggler's Utility Function}

Having calculated the quarterly interception probability, we proceed to analyze incident-level decision-making.
 We next assume that the smuggler makes a  choice of boat size. We discretize boat size into bins of $n \in \{1-50,~ 50-100,~100+\}$ migrants.
The smuggler then selects $n$ to maximize his payoff according to his utility function, which takes the form:
    \begin{equation}\label{eq:utility}
    U_{in} \quad=\quad \alpha_n  \quad+\quad \beta_n~p^{q}_{interception} \quad+\quad \varepsilon_{in}. 
    \end{equation}
Here, $i$ represents the choice setting (i.e., the incident); $n$ represents the choice of boat size; $\alpha_n$ represents a boat size fixed effect; $p^{q}_{interception}$ is the estimated quarterly probability that a boat departing Libya will be intercepted in the Libyan SAR zone, for the quarter that corresponds to incident $i$; $\beta_n$ is a boat size-specific coefficient on the quarterly probability of interception; and $\varepsilon_{in}$ is a choice-specific idiosyncratic error term.

In other words, the smuggler's utility is a function of two deterministic factors: 
    \begin{enumerate} 
        \item \textbf{The short-run marginal payoff} to the number of migrants chosen ($\alpha_n$), i.e. the total rent collected from all $n$ migrants in exchange for the crossing. We generally expect this payoff to increase in $n$, since each passenger is charged a fare for the journey.
        \item \textbf{The expected long-run reputational payoff} of crossing ($\beta_n~p^{q}_{interception}$), which depends on the smugglers' risk of interception.
        \footnote{While there may be no immediate consequences of interception for smugglers (since migrants have typically already paid for passage), one can imagine that smuggling businesses will ultimately suffer if they are unable to secure rescues to Europe for their passengers.} Since the probability of interception should vary by boat type, but the precise extent of variation is unknown, we allow for boat-size-specific coefficients on this probability. These coefficients are effectively an estimate of how changes in the interception probability affect utility for different choices of boat size. If small boats provide an advantage when interceptions are high, we expect $\beta_{0-50}~>~\beta_{50-100} ~>~\beta_{100+}$, whereas if large boats provide an advantage we expect the opposite to be true. 
    \end{enumerate}

The addition of an idiosyncratic error term $\varepsilon_{in}$ allows for random variation in the behavior of smugglers due to unobservables, such that not all smugglers will select the same boat size choice even under the same conditions.\footnote{For example, this term could include unobserved variation in smuggler costs, resources, or preferences. This was one of the motivations for using a choice model, since it generates variation in the smugglers' optimal boat size choices even under the same observable conditions.} Assuming that $\varepsilon_{in}$ takes a {Type-I extreme value distribution, the probability of choosing boat size $n$ takes the form of the standard logit probabilities} \citep[Section 3.1]{train_discrete_2009}. Let $\mathcal{N} =\{1-50,~ 50-100,~100+\}$ be the set of possible boat size choices and let $V_{in} = U_{in} - \varepsilon_{in}$; that is, let $V_{in}$ represent the deterministic part of the utility function. Then,
    $$P_{in} \quad=\quad \frac{e^{V_{in}}}{\sum_{n'\in \mathcal{N}} ~e^{V_{in'}}}.$$

This model is also sometimes referred as the Maximum Entropy Inverse Reinforcement Learning (IRL) model \parencite{abbeel_apprenticeship_2004,ziebart_maximum_2008,ermon_learning_2015}, in which the smuggler's choice of boat size is drawn proportional to his exponentiated expected reward:
    \begin{equation*}\label{eq:maxent_irl}
        P_{in}(\alpha_n, \beta_n) \quad\propto\quad e^{~V_{in}(\alpha_n,\beta_n)}.
    \end{equation*}

\subsection{Estimation of the Utility Function from Incident-level Data}\label{sec:mnl_incident}

Using the model in Section~\ref{sec:incident_model}, we empirically estimate the parameters $\alpha_n$ and $\beta_n$ in Equation~\ref{eq:utility}. Recall that $\alpha_n$ describes the payoff to a given boat size which is independent of $p_{interception}^q$, whereas $\beta_n$ describes the long-term payoff associated with the probability of interception vs. rescue, conditional on boat size. We attempt to recover $\alpha_n$ and $\beta_n$ using incident-level data from Frontex.

We restrict our analysis to incidents involving rubber boats. This choice was made for two primary reasons. First, rubber boats tend to have a more uniform physical size, which means that the number of people is a reasonable proxy for crowding; this is not the case for wooden boats, which may vary dramatically in their capacity. Second, while rubber boats can be imported cheaply, the supply of large wooden boats in the region has become increasingly scarce over time as these boats have sunk or been destroyed by anti-smuggling operations (see Section~\ref{sec:smuggling_ops}); therefore, the choice of how many people to place on board a wooden boat may be exogenously affected by the scarcity of large ships.  Restricting our analysis to rubber boats preserves the majority of incidents in the Frontex dataset: {74\% of incidents originating in Libya involve rubber boats, relative to just 14\% of incidents which involve wooden boats.} Further details on the distribution of boat sizes and the relationship between boat size and quarterly interception probabilities are discussed in Appendix~\ref{sec:frontex_descriptives}.

In our model, the {log likelihood of the data is:}
    \begin{eqnarray}
  \log L(\alpha, \beta) \quad=\quad \sum_{i \in \mathcal{I}}~\sum_{n \in \mathcal{N}} ~~d_{in}~~
  \log(~P_{in}(\alpha_n, \beta_n)~),
    \end{eqnarray}
where $\mathcal{I}$ is the set of incidents in the dataset and $d_{in}$ is equal to one if $n$ is the boat size actually chosen in incident $i$, and zero otherwise. Estimation was performed in Stata SE 12.0 using the \verb|clogit| {command,} {which optimizes the log likelihood function using Newton's method.} 

\subsection{Results}\label{sec:utility_results}
\FloatBarrier

Results from the estimation of the model are shown in Table~\ref{tbl:utility_params}.\footnote{Note that there are a limited number of observations in 2018 and 2019 compared with 2016 and 2017 (see Table~\ref{tbl:mnl_support} of Appendix~\ref{sec:frontex_descriptives} for detailed information). For this reason, we re-weight all incidents such that each quarter in the dataset is given equal weight; for further discussion of the role of the weights and a comparison with unweighted estimates, see Appendix~\ref{sec:robustness_mnl_weights}. } Because choice probabilities are relative, we must fix the utility of one choice category in order to identify the others. We normalize the utility of the $0-50$ boat size category by setting $\alpha_{~0-50}$ and $\beta_{~0-50}$ to zero. The remaining coefficients are then interpreted in relation to this base category.

We can see that when there is no chance of interception, large boats are generally preferred ($\alpha_{~0-50} ~<~ \alpha_{~50-100} ~<~ \alpha_{~100+}$). This is consistent with our expectations, since smugglers can likely extract a higher rent by launching more crowded rubber boats. However, our results suggest that larger boat sizes face a disadvantage when the probability of interception is nonzero ($\beta_{~0-50} ~>~ \beta_{~50-100} ~>~ \beta_{~100+}$). This is consistent with the empirical evidence that large boats are chosen less frequently when interceptions are high, as they may struggle to evade the LCG and reach the EU SAR zones. Taken together, these results support the hypothesis that there is a tradeoff between the short-run payoff to launching large boats and the long-run cost to interception as boats grow more crowded. 
    \begin{table}[htb]
        \centerline{{
\def\sym#1{\ifmmode^{#1}\else\(^{#1}\)\fi}
\begin{tabular}{l*{2}{c}}
\toprule
                    &&\multicolumn{1}{c}{$p_{in}$}\\
\midrule
$\alpha_{~50-100}$       &&    1.786\sym{***}\\
                    & &  (0.413)         \\
\addlinespace
$\alpha_{~100+}$          &&    3.849\sym{***}\\
                    &&  (0.604)         \\
\addlinespace
$\beta_{~50-100}$     &&   -3.587\sym{***}\\
                    &&  (0.955)         \\
\addlinespace
$\beta_{~100+}$     &&   -6.511\sym{***}\\
                    &&  (1.998)         \\
\midrule
pseudo $R^2$           &&    0.268         \\
N choice alternatives              & &    5,553         \\
N choices              & &    1,851         \\
\bottomrule
\multicolumn{3}{l}{\footnotesize Standard errors in parentheses}\\
\multicolumn{3}{l}{\footnotesize \sym{*} \(p<0.10\), \sym{**} \(p<0.05\), \sym{***} \(p<0.01\)}\\
\end{tabular}
}
}
        \caption{Parameter estimates for the smuggler's utility model}
        \label{tbl:utility_params}
    \end{table}

\begin{figure}[htb]
        \centering
        \includegraphics[width=4in]{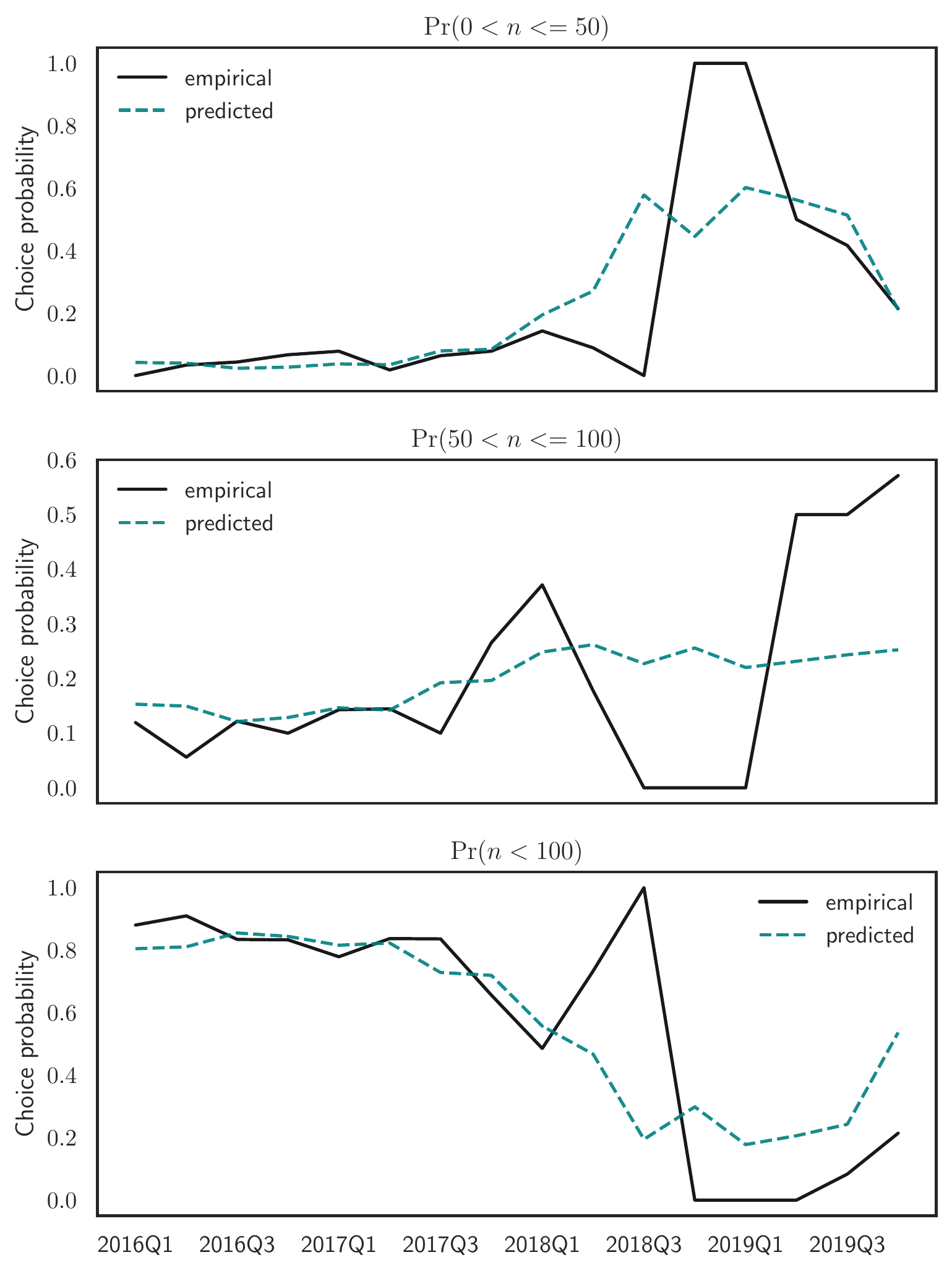}
        \caption{Predicted distribution of boat sizes by quarter}
        \label{fig:predicted_choices_by_quarter}
    \end{figure}

In Figure~\ref{fig:predicted_choices_by_quarter} we compare our model-based estimates for the distribution of boat size to the empirical distribution of boat sizes by quarter. We can see that generally, the predictions of the model follow the overall trends in the empirical choice of boat distribution. Although we do not fully match the fluctuations in average boat size by quarter that occur from 2018 onward, we note that these fluctuations occur in part because of the small number of incidents in the last two years of our dataset.

\textbf{Robustness Checks:} We find that the drop in boat size between Phase 2 and Phase 3 is statistically significant, which we detail in Appendix~\ref{sec:ttests}. We also consider different weighting schemes in Appendix~\ref{sec:robustness_mnl_weights}.

\subsection{Counterfactual Estimation}
\label{subsec:incident_counterfactual}

\begin{figure}
   \centerline{\includegraphics[width=4in]{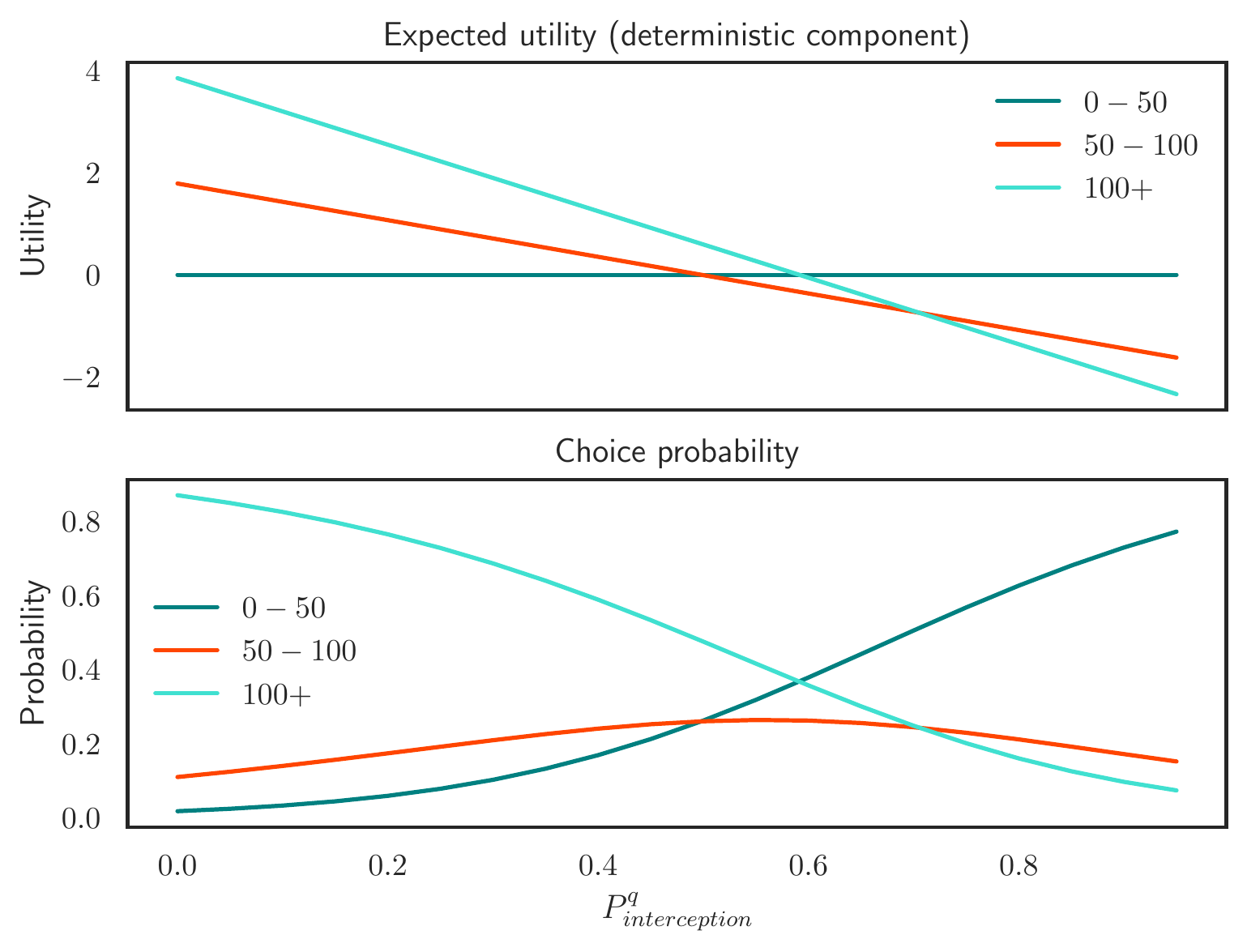}}
    \caption{Predicted utility and distribution of boat sizes, by probability of interception}
    \label{fig:predicted_choices_by_pnonrescue}
\end{figure}

Equipped with these model estimates, we can conduct simulations to illustrate how smugglers are expected to react to a change in interception rates. Using the incident-level parameter estimates for $\alpha_{n}$ and $\beta_{n}$, Figure~\ref{fig:predicted_choices_by_pnonrescue} shows how the expected utility (i.e., the deterministic component of the utility) and choice probability for each boat size varies with the quarterly rate of interceptions. We estimate that large boat sizes are preferred until the interception rate approaches 60\%, which occurs starting in the third quarter of 2018 (see Figure~\ref{fig:pinterception}). After this point, small boats are preferred. Interestingly, midsize boats are never the dominant choice in expectation.

\begin{figure}[htb]
    \centerline{\includegraphics[width=4in]{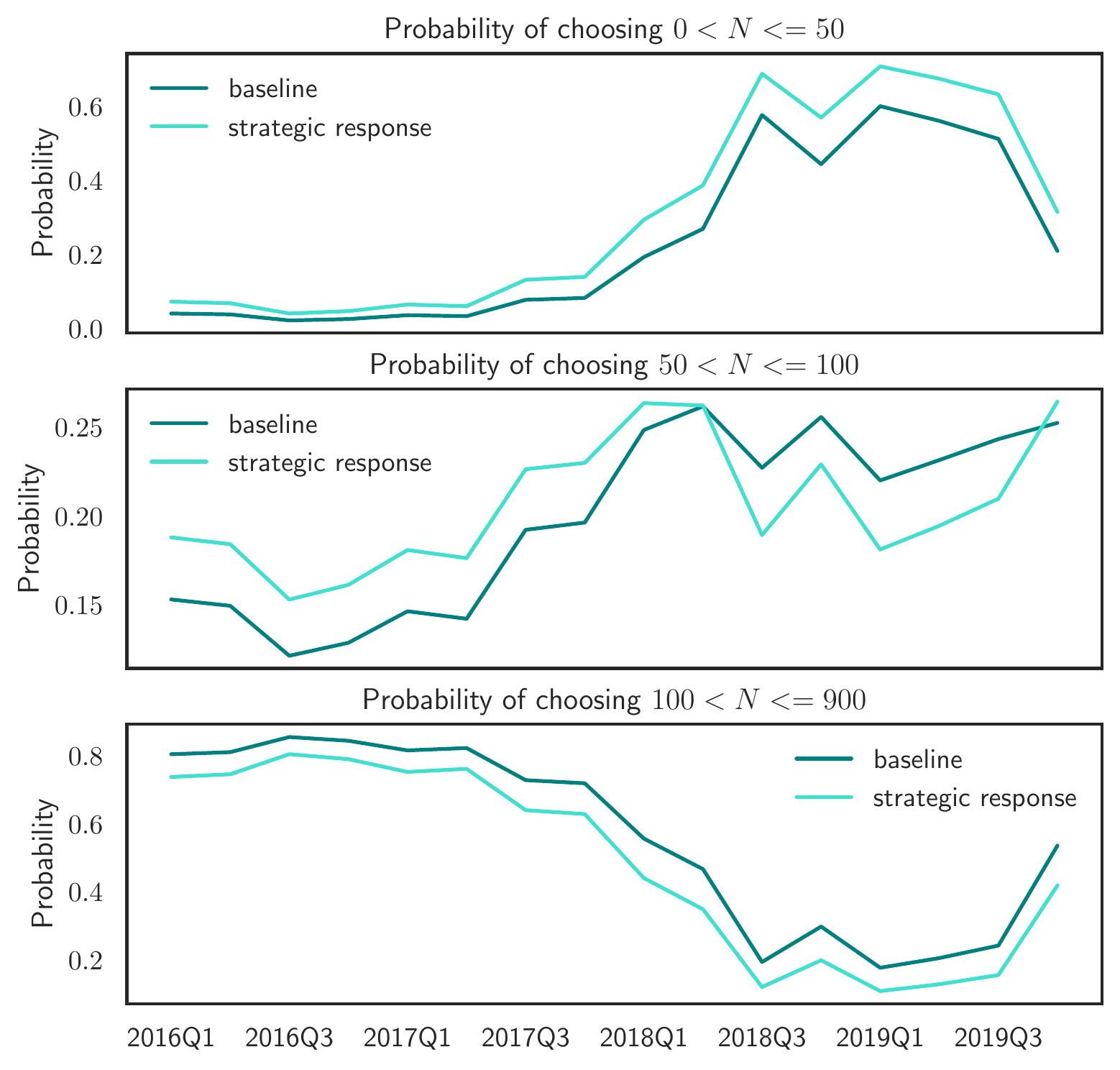}}
    \floatfoot{``Baseline'' refers to the ``true'' scenario in which the interception rate matches reality.  ``Strategic response'' refers to the scenario in which smugglers adapt their choice probabilities to reflect the higher probability of interception.}
    \caption{Predicted distribution of boat sizes, given a 10 percentage point increase in the probability of interception}
    \label{fig:predicted_choices_increased_pnonrescue}
\end{figure}

Figure~\ref{fig:predicted_choices_increased_pnonrescue} illustrates a hypothetical scenario for rubber boats in which the number of incidents per quarter remains the same, but the probability of interception is increased by 10 percentage points (relative to the baseline interception rate) across all quarters. Using the model and these two different interception levels, we estimate the choice probabilities for each scenario. Across all quarters in our dataset, we predict that the smugglers will respond to the changing environment by increasing the use of small boats and decreasing the use of large boats; the effect on the use of mid-size boats varies by period. 

\FloatBarrier
\FloatBarrier
\section{Discussion and Conclusions}\label{sec:conclusions}
Prior work on migration flows in the Central Mediterranean has focused on assessing the claim that NGO rescues endanger migrants by encouraging crossings and incentivizing riskier trips. In response to these claims, Italy and the EU more generally have acted to increase the capacity of the LCG and encourage a more aggressive regime of interceptions and returns. However, to date there has been little theoretical analysis of how migrants have responded to these changes in recent years.

In this paper, we analyze the flow of Central Mediterranean migrants over time, including their response to the rise in Libyan interceptions that {started in mid-2017.} From the flow data we find that the number of crossings on the Central route increase with the success probability of crossing, i.e., that the overall flow of migrants does respond to the growing rate of interceptions. Our analysis indicates spillovers from the Central Mediterranean route to the Western Mediterranean route, although we estimate that only a {fraction} of migrants substitute across routes and that the propensity to substitute appears to vary by geography of origin. These findings correspond to the deterrence and diversion effects previously identified in the literature \parencite{sorensen_effects_2007}. 

Our analysis supports the claim that the growing rate of Libyan interceptions has discouraged migrant crossings on the Central Mediterranean route. However, crossings have continued despite a very high rate of interceptions {(approaching almost 80\% for boats departing Libya)}. 
Therefore, we also analyze how smugglers have adapted to the changing interception environment. When comparing incident-level datasets (Appendix~\ref{sec:incident_desc}), we observed evidence of strategic responses along several dimensions. Namely, there appears to be a decline in the number of people per boat and a shift towards the use of wooden boats. These strategy shifts coincide with an overall tendency for incidents to occur closer to the EU SAR zones, and we find a general correlation between boat type/boat size choice and incident locations in these later periods. 

To formally analyze these changes, we build a utility model of smuggling. Using a discrete choice model, we estimate that there is a positive payoff to launching larger/more crowded boats, but that this is counterbalanced by a penalty on larger boats that rises with the overall rate of interceptions. Therefore smugglers trade off between collecting higher rents by continuing to crowd migrants on to boats, and the reputational costs of launching crowded boats that have a lower probability of success.

The trend towards the use of smaller boats may have several implications for migrants crossing on the Central Mediterranean route. On the one hand, boats with fewer passengers may be less likely to sink, since overcrowding can make boats less seaworthy.\footnote{For example, overcrowded boats sit low in the water and may be more likely to take on waves. In rubber rafts, overcrowding risks the collapse of the inflatable tubes that support the boat, whereas wooden boats can tip if they are over capacity and passengers suddenly move to one side (e.g., in the course of a rescue).} On the other hand, boats that are physically smaller in size may be less likely to be detected at sea and may be more susceptible to weather conditions on the open water. We do see a {rise in the rate of deaths reported by the IOM along the Central Mediterranean route after 2017} (see Figure~\ref{fig:n}), suggesting that deaths may be occurring in the course of LCG interceptions and/or that ships are sinking without a rescue or interception ever being initiated. 

Consistent with other findings in the literature, our work suggests that increased border enforcement induces strategic responses on a number of different dimensions, which can lead to unpredictable or unforeseen consequences for the crossing experience. When debating an increase in border surveillance, it is important to consider implications for both migrant and smuggler strategy, as well as the possible long-run impact of these strategy shifts on fatalities. In general, analyzing these implications is difficult because data on informal migration is typically incomplete and/or biased (if it is available at all), and because causal inference is challenging in such a complex environment. We address the latter challenge through the use of an error correction model to reduce the risk of spurious regressions in the flow data, and address the former by combining biased incident data with more representative flow data in our multinomial choice model.

Future work could expand along several different dimensions. First, there is a clear need for careful empirical work which collects and combines different sources of data in order to produce comprehensive migration incident datasets at the regional level. For example, we have experimented with the application of dataset matching methods from other fields, namely capture-recapture models~\parencite{manrique-vallier_capture-recapture_2019}, for this purpose. Second, existing strategy models could be expanded to more sophisticated  settings, for example by building nested choice models in which smugglers choose a boat type before selecting a boat size, or select between departure ports before making boat-specific decisions.

While the Mediterranean is currently the world’s deadliest border, new maritime crossing routes have continued to emerge globally as vulnerable people flee economic and political hardship in Venezuela, escape ethnic persecution in Myanmar, and seek economic opportunity in the United States~\parencite{casey_hungry_2016,pearlman_who_2015,chivers_risking_2021}. By integrating different methods and data sources we aim to contribute to a growing body of literature on sea crossings, providing evidence on how these crossings are similar to or different from land-based informal migration. We hope that a better understanding of crossing behavior and strategy can ultimately contribute to safer and more humane global border regimes. 
\FloatBarrier


\newpage
\singlespacing
\bibliographystyle{informs2014}
\bibliography{Migrants}

\begin{thebibliography}{78}
\providecommand{\natexlab}[1]{#1}
\providecommand{\url}[1]{\texttt{#1}}
\providecommand{\urlprefix}{URL }

\bibitem[{Abbeel \protect\BIBand{} Ng(2004)}]{abbeel_apprenticeship_2004}
Abbeel P, Ng AY (2004) Apprenticeship learning via inverse reinforcement
  learning. \emph{Proceedings of the twenty-first international conference on
  {Machine} learning},
  \urlprefix\url{http://dx.doi.org/10.1145/1015330.1015430}.

\bibitem[{AbuJarour \protect\BIBand{}
  Krasnova(2017)}]{abujarour_understanding_2017}
AbuJarour S, Krasnova H (2017) Understanding the {Role} of {ICTs} in
  {Promoting} {Social} {Inclusion}: {The} {Case} of {Syrian} {Refugees} in
  {Germany}. \emph{Proceedings of the 25th {European} {Conference} on
  {Information} {Systems} ({ECIS})}, 1792--1806 (Guimaraes, Portugal),
  \urlprefix\url{https://aisel.aisnet.org/ecis2017_rp/115}.

\bibitem[{Ahani et~al.(2021)Ahani, Andersson, Martinello, Teytelboym,
  \protect\BIBand{} Trapp}]{ahani_placement_2021}
Ahani N, Andersson T, Martinello A, Teytelboym A, Trapp AC (2021) Placement
  {Optimization} in {Refugee} {Resettlement}. \emph{Operations Research}
  69(5):1468--1486, ISSN 0030-364X,
  \urlprefix\url{http://dx.doi.org/10.1287/opre.2020.2093}, publisher: INFORMS.

\bibitem[{{Amnesty International}(2017)}]{amnesty_international_libyas_2017}
{Amnesty International} (2017) Libya's {Dark} {Web} of {Collusion}: {Abuses}
  {Against} {Europe}-bound {Refugees} and {Migrants}. Technical report,
  \urlprefix\url{https://www.amnesty.org/en/documents/mde19/7561/2017/en/},
  library Catalog: www.amnesty.org.

\bibitem[{{Amnesty International}(2020)}]{amnesty_international_libya_2020}
{Amnesty International} (2020) Libya: {Renewal} of migration deal confirms
  {Italy}’s complicity in torture of migrants and refugees.
  \urlprefix\url{https://www.amnesty.org/en/press-releases/2020/01/libya-renewal-of-migration-deal-confirms-italys-complicity-in-torture-of-migrants-and-refugees/},
  library Catalog: www.amnesty.org.

\bibitem[{{asktheEU}(2020)}]{asktheeu_jora_2020}
{asktheEU} (2020) {JORA} variables for {Boat} {Interceptions} under
  {Operations} {Triton}, {Themis}, {Sophia}, and {Irini} (2014-2020) - a
  {Freedom} of {Information} request to {European} {Border} and {Coast} {Guard}
  {Agency}.
  \urlprefix\url{https://www.asktheeu.org/en/request/jora_variables_for_boat_intercep}.

\bibitem[{Aziz et~al.(2015)Aziz, Monzini, \protect\BIBand{}
  Pastore}]{aziz_changing_2015}
Aziz NA, Monzini P, Pastore F (2015) The {Changing} {Dynamics} of
  {Cross}-border {Human} {Smuggling} and {Trafficking} in the {Mediterranean}.
  Technical report, Instittuto Affari Internazionali, Rome, Italy.

\bibitem[{Baczynska(2017)}]{baczynska_ferry_2017}
Baczynska G (2017) Ferry service or humanitarian rescue boat: {EU}'s
  {Mediterranean} dilemma. \emph{Reuters}
  \urlprefix\url{https://www.reuters.com/article/us-europe-migrants-eu-libya-idUSKBN1881CK}.

\bibitem[{Baker(2016)}]{baker_rescue_2016}
Baker A (2016) Rescue at {Sea} - {A} week on board a refugee recovery ship.
  \emph{TIME} \urlprefix\url{https://time.com/refugee-rescue/}, library
  Catalog: time.com.

\bibitem[{Besiou \protect\BIBand{}
  Van~Wassenhove(2020)}]{besiou_humanitarian_2020}
Besiou M, Van~Wassenhove LN (2020) Humanitarian {Operations}: {A} {World} of
  {Opportunity} for {Relevant} and {Impactful} {Research}. \emph{Manufacturing
  \& Service Operations Management} 22(1):135--145, ISSN 1523-4614,
  \urlprefix\url{http://dx.doi.org/10.1287/msom.2019.0799}, publisher: INFORMS.

\bibitem[{Box-Steffensmeier et~al.(2014)Box-Steffensmeier, Freeman, Hitt,
  \protect\BIBand{} Pevehouse}]{box-steffensmeier_time_2014}
Box-Steffensmeier JM, Freeman JR, Hitt MP, Pevehouse JCW (2014) \emph{Time
  {Series} {Analysis} for the {Social} {Sciences}}. Analytical {Methods} for
  {Social} {Research} (Cambridge: Cambridge University Press), ISBN
  978-0-521-87116-7,
  \urlprefix\url{http://dx.doi.org/10.1017/CBO9781139025287}.

\bibitem[{Camarena et~al.(2020)Camarena, Claudy, Wang, \protect\BIBand{}
  Wright}]{camarena_political_2020}
Camarena KR, Claudy S, Wang J, Wright AL (2020) Political and environmental
  risks influence migration and human smuggling across the {Mediterranean}
  {Sea}. \emph{PLoS ONE} 15(7):e0236646, publisher: Public Library of Science
  San Francisco, CA USA.

\bibitem[{Casey(2016)}]{casey_hungry_2016}
Casey N (2016) Hungry {Venezuelans} flee in boats to escape economic collapse.
  \emph{The New York Times}
  \urlprefix\url{https://www.nytimes.com/2016/11/25/world/americas/hungry-venezuelans-flee-in-boats-to-escape-economic-collapse.html}.

\bibitem[{Celik et~al.(2012)Celik, Ergun, Johnson, Keskinocak, Lorca, Pekgun,
  \protect\BIBand{} Swann}]{celik_humanitarian_2012}
Celik M, Ergun O, Johnson B, Keskinocak P, Lorca A, Pekgun P, Swann J (2012)
  Humanitarian logistics. \emph{New directions in informatics, optimization,
  logistics, and production}, 18--49 (INFORMS).

\bibitem[{Chivers(2021)}]{chivers_risking_2021}
Chivers C (2021) Risking {Everything} to {Come} to {America} on the {Open}
  {Ocean}. \emph{The New York Times}
  \urlprefix\url{https://www.nytimes.com/2021/02/03/magazine/customs-border-protection-migrants-pacific-ocean.html}.

\bibitem[{Cornelius(2001)}]{cornelius_death_2001}
Cornelius WA (2001) Death at the border: {Efficacy} and unintended consequences
  of {US} immigration control policy. \emph{Population and development review}
  27(4):661--685, \urlprefix\url{https://www.jstor.org/stable/2695182},
  publisher: Wiley Online Library.

\bibitem[{Cusumano(2019)}]{cusumano_straightjacketing_2019}
Cusumano E (2019) Straightjacketing migrant rescuers? {The} code of conduct on
  maritime {NGOs}. \emph{Mediterranean Politics} 24(1):106--114, ISSN
  1362-9395, \urlprefix\url{http://dx.doi.org/10.1080/13629395.2017.1381400},
  publisher: Routledge \_eprint: https://doi.org/10.1080/13629395.2017.1381400.

\bibitem[{Cusumano \protect\BIBand{} Villa(2019)}]{cusumano_sea_2019}
Cusumano E, Villa M (2019) Sea rescue {NGOs}: {A} pull factor of irregular
  migration? Technical Report~22, Migration Policy Centre, publisher: European
  University Institute.

\bibitem[{Deiana et~al.(2019)Deiana, Maheshri, \protect\BIBand{}
  Mastrobuoni}]{deiana_migration_2019}
Deiana C, Maheshri V, Mastrobuoni G (2019) Migration at {Sea}: {Unintended}
  {Consequences} of {Search} and {Rescue} {Operations}. Technical Report
  3454537, SSRN, Rochester, NY.

\bibitem[{{Deutsche Welle}(2016)}]{deutsche_welle_italy_2016}
{Deutsche Welle} (2016) Italy rescues 4,500 migrants in one day. \emph{Deutsche
  Welle}
  \urlprefix\url{https://www.dw.com/en/italy-rescues-4500-migrants-in-one-day/a-19380550}.

\bibitem[{Ermon et~al.(2015)Ermon, Xue, Toth, Dilkina, Bernstein, Damoulas,
  Clark, DeGloria, Mude, \protect\BIBand{} Barrett}]{ermon_learning_2015}
Ermon S, Xue Y, Toth R, Dilkina B, Bernstein R, Damoulas T, Clark P, DeGloria
  S, Mude A, Barrett C (2015) Learning large-scale dynamic discrete choice
  models of spatio-temporal preferences with application to migratory
  pastoralism in {East} {Africa}. \emph{Proceedings of the {AAAI} {Conference}
  on {Artificial} {Intelligence}}, volume~29, issue: 1.

\bibitem[{{EUNAVFOR Med}(2018)}]{eunavfor_med_monitoring_2018}
{EUNAVFOR Med} (2018) Monitoring {Report}: {October} 2017 - {January} 2018.
  Technical report, EUNAVFOR Med Operation Sophia, Rome, Italy,
  \urlprefix\url{https://g8fip1kplyr33r3krz5b97d1-wpengine.netdna-ssl.com/wp-content/uploads/2019/02/ENFM-Monitoring-of-Libyan-Coast-Guard-and-Navy-Report-October-2017-January-2018.pdf}.

\bibitem[{{EUNAVFOR Med Operation
  Sophia}(2018)}]{eunavfor_med_operation_sophia_about_2018}
{EUNAVFOR Med Operation Sophia} (2018) About us.
  \urlprefix\url{https://www.operationsophia.eu/about-us/}, library Catalog:
  www.operationsophia.eu.

\bibitem[{{European Union Agency for Fundamental
  Rights}(2019)}]{european_union_agency_for_fundamental_rights_2019_2019}
{European Union Agency for Fundamental Rights} (2019) 2019 update - {NGO} ships
  involved in search and rescue in the {Mediterranean} and criminal
  investigations.
  \urlprefix\url{https://fra.europa.eu/en/publication/2019/2019-update-ngo-ships-involved-search-and-rescue-mediterranean-and-criminal}.

\bibitem[{Gathmann(2008)}]{gathmann_effects_2008}
Gathmann C (2008) Effects of enforcement on illegal markets: {Evidence} from
  migrant smuggling along the southwestern border. \emph{Journal of Public
  Economics} 92(10-11):1926--1941,
  \urlprefix\url{http://dx.doi.org/10.1016/j.jpubeco.2008.04.006}, publisher:
  Elsevier.

\bibitem[{Granger \protect\BIBand{} Newbold(1974)}]{granger_spurious_1974}
Granger CW, Newbold P (1974) Spurious regressions in econometrics.
  \emph{Journal of Econometrics} 2(2):111--120, publisher: Wiley Online
  Library.

\bibitem[{Grunau(2016)}]{grunau_tragedy_2016}
Grunau A (2016) The tragedy of the 2015 {Catania} migrant shipwreck.
  \urlprefix\url{https://www.dw.com/en/the-tragedy-of-the-2015-catania-migrant-shipwreck/a-36729526}.

\bibitem[{Haliassos et~al.(2017)Haliassos, Jansson, \protect\BIBand{}
  Karabulut}]{haliassos_incompatible_2017}
Haliassos M, Jansson T, Karabulut Y (2017) Incompatible {European} {Partners}?
  {Cultural} {Predispositions} and {Household} {Financial} {Behavior}.
  \emph{Management Science} 63(11):3780--3808, ISSN 0025-1909,
  \urlprefix\url{http://dx.doi.org/10.1287/mnsc.2016.2538}, publisher: INFORMS.

\bibitem[{Heller \protect\BIBand{} Pezzani(2016)}]{heller_death_2016}
Heller C, Pezzani L (2016) Death by rescue: {The} lethal effects of the {EU}'s
  policies of non-assistance at sea. \urlprefix\url{https://deathbyrescue.org}.

\bibitem[{Heller et~al.(2018)Heller, Pezzani, Mann, Moreno-Lax, Weizman, Adams,
  Bye, Ellick, Varjacques, \protect\BIBand{} Strasser}]{heller_its_2018}
Heller C, Pezzani L, Mann I, Moreno-Lax V, Weizman E, Adams T, Bye K, Ellick
  AB, Varjacques L, Strasser M (2018) ‘{It}’s an {Act} of {Murder}’:
  {How} {Europe} {Outsources} {Suffering} as {Migrants} {Drown}. \emph{The New
  York Times} ISSN 0362-4331,
  \urlprefix\url{https://www.nytimes.com/interactive/2018/12/26/opinion/europe-migrant-crisis-mediterranean-libya.html,
  https://www.nytimes.com/interactive/2018/12/26/opinion/europe-migrant-crisis-mediterranean-libya.html}.

\bibitem[{{Human Rights Watch}(2017)}]{human_rights_watch_eu_2017}
{Human Rights Watch} (2017) {EU}: {Shifting} {Rescue} to {Libya} {Risks}
  {Lives}.
  \urlprefix\url{https://www.hrw.org/news/2017/06/19/eu-shifting-rescue-libya-risks-lives}.

\bibitem[{{Human Rights Watch}(2018)}]{human_rights_watch_euitalylibya_2018}
{Human Rights Watch} (2018) {EU}/{Italy}/{Libya}: {Disputes} {Over} {Rescues}
  {Put} {Lives} at {Risk}.
  \urlprefix\url{https://www.hrw.org/news/2018/07/25/eu/italy/libya-disputes-over-rescues-put-lives-risk},
  library Catalog: www.hrw.org.

\bibitem[{{International Maritime
  Organization}(2012)}]{international_maritime_organization_sar8circ4_2012}
{International Maritime Organization} (2012) Sar.8/{Circ}.4 {Availability} of
  {Search} and {Rescue} {Services}.

\bibitem[{{International Maritime
  Organization}(2017)}]{international_maritime_organization_ncsr_2017}
{International Maritime Organization} (2017) {NCSR} 5/{INF}.17 {Further}
  {Development} of the {Provision} of {Global} {Maritime} {SAR} {Services} -
  {Libyan} {Maritime} {Rescue} {Coordination} {Centre} {Project}.

\bibitem[{{IOM}(2019{\natexlab{a}})}]{iom_data_2019}
{IOM} (2019{\natexlab{a}}) Data on {Attempted} {Crossings} of the
  {Mediterranean} {Sea} 2016-2019.
  \urlprefix\url{https://missingmigrants.iom.int/sites/default/files/Annex_Med%20arrivals%20interceptions%20deaths_0.xlsx}.

\bibitem[{{IOM}(2019{\natexlab{b}})}]{iom_world_2019}
{IOM} (2019{\natexlab{b}}) World {Migration} {Report} 2020. Technical report,
  International Organization for Migration, Geneva, Switzerland,
  \urlprefix\url{https://publications.iom.int/system/files/pdf/wmr_2020.pdf}.

\bibitem[{{IOM}(2021)}]{iom_missing_2021}
{IOM} (2021) Missing {Migrants} {Project}.
  \urlprefix\url{https://missingmigrants.iom.int/downloads}.

\bibitem[{Manrique-Vallier et~al.(2019)Manrique-Vallier, Ball,
  \protect\BIBand{} Sadinle}]{manrique-vallier_capture-recapture_2019}
Manrique-Vallier D, Ball P, Sadinle M (2019) \emph{Capture-{Recapture} for
  {Casualty} {Estimation} and {Beyond}: {Recent} {Advances} and {Research}
  {Directions}} (Submitted).

\bibitem[{{Marina Militare}(2014)}]{marina_militare_mare_2014}
{Marina Militare} (2014) Mare {Nostrum}: one year into the operation.
  \urlprefix\url{https://www.marina.difesa.it/EN/Conosciamoci/notizie/Pagine/20141024_omnoneyear.aspx},
  library Catalog: www.marina.difesa.it.

\bibitem[{Micallef(2017)}]{micallef_human_2017}
Micallef M (2017) The {Human} {Conveyor} {Belt}: trends in human trafficking
  and smuggling in post-revolution {Libya}. Technical report, The Global
  Initiative Against Transnational Organized Crime, Geneva, Switzerland,
  \urlprefix\url{https://globalinitiative.net/wp-content/uploads/2017/03/GI-Human-Conveyor-Belt-Human-Smuggling-Libya-2017-.pdf}.

\bibitem[{Micallef \protect\BIBand{} Reitano(2017)}]{micallef_anti-human_2017}
Micallef M, Reitano T (2017) The anti-human smuggling business and {Libya}’s
  political end game. Technical report, Global Initiative against Transnational
  Organized Crime,
  \urlprefix\url{https://globalinitiative.net/analysis/the-anti-human-smuggling-business-and-libyas-political-end-game/}.

\bibitem[{Michael et~al.(2019)Michael, Hinnant, \protect\BIBand{}
  Brito}]{michael_making_2019}
Michael M, Hinnant L, Brito R (2019) Making misery pay: {Libya} militias take
  {EU} funds for migrants. \emph{AP NEWS}
  \urlprefix\url{https://apnews.com/article/9d9e8d668ae4b73a336a636a86bdf27f},
  section: Outsourcing Migrants.

\bibitem[{{MOAS}(2020)}]{moas_introduction_2020}
{MOAS} (2020) An {Introduction} to {Safe} and {Legal} {Routes}.
  \urlprefix\url{https://www.moas.eu/blog-an-introduction-to-safe-and-legal-routes/}.

\bibitem[{{MSF}(2020)}]{msf_search_2020}
{MSF} (2020) Search and {Rescue} - {Interactive} {Map}.
  \urlprefix\url{http://searchandrescue.msf.org/map.html}.

\bibitem[{Naiditch \protect\BIBand{} Vranceanu(2020)}]{naiditch_matching_2020}
Naiditch C, Vranceanu R (2020) A {Matching} {Model} of the {Market} for
  {Migrant} {Smuggling} {Services}. Technical Report 3530001, SSRN, Rochester,
  NY,
  \urlprefix\url{https://papers.ssrn.com/sol3/papers.cfm?abstract_id=3530001}.

\bibitem[{{National Geospatial-Intelligence
  Agency}(2018)}]{national_geospatial-intelligence_agency_broadcast_2018}
{National Geospatial-Intelligence Agency} (2018) Broadcast {Warnings}.
  \urlprefix\url{https://msi.nga.mil/NavWarnings}.

\bibitem[{{Natural Earth}(2020)}]{natural_earth_110m_2020}
{Natural Earth} (2020) 1:10m {Cultural} {Vectors}.
  \urlprefix\url{https://www.naturalearthdata.com/downloads/10m-cultural-vectors/},
  library Catalog: www.naturalearthdata.com.

\bibitem[{{Office to Monitor and Combat Trafficking in
  Persons}(2020)}]{office_to_monitor_and_combat_trafficking_in_persons_2020_2020}
{Office to Monitor and Combat Trafficking in Persons} (2020) 2020 {Trafficking}
  in {Persons} {Report}: {Libya}. Technical report, US Department of State,
  \urlprefix\url{https://www.state.gov/reports/2020-trafficking-in-persons-report/libya/}.

\bibitem[{Orrenius \protect\BIBand{}
  Zavodny(2015)}]{orrenius_undocumented_2015}
Orrenius P, Zavodny M (2015) Undocumented immigration and human trafficking.
  \emph{Handbook of the economics of international migration}, volume~1,
  659--716 (Elsevier),
  \urlprefix\url{https://doi.org/10.1016/B978-0-444-53764-5.00013-X}.

\bibitem[{Papadaki et~al.(2016)Papadaki, Alpern, Lidbetter, \protect\BIBand{}
  Morton}]{papadaki_patrolling_2016}
Papadaki K, Alpern S, Lidbetter T, Morton A (2016) Patrolling a {Border}.
  \emph{Operations Research} 64(6):1256--1269, ISSN 0030-364X,
  \urlprefix\url{http://dx.doi.org/10.1287/opre.2016.1511}, publisher: INFORMS.

\bibitem[{Pearlman(2015)}]{pearlman_who_2015}
Pearlman J (2015) Who are the {Rohingya} boat people? \emph{The Telegraph} ISSN
  0307-1235,
  \urlprefix\url{https://www.telegraph.co.uk/news/worldnews/asia/burmamyanmar/11620933/Who-are-the-Rohingya-boat-people.html}.

\bibitem[{Salt \protect\BIBand{} Stein(1997)}]{salt_migration_1997}
Salt J, Stein J (1997) Migration as a {Business}: {The} {Case} of
  {Trafficking}. \emph{International Migration} 35(4):467--494, ISSN 1468-2435,
  \urlprefix\url{http://dx.doi.org/10.1111/1468-2435.00023}, \_eprint:
  https://onlinelibrary.wiley.com/doi/pdf/10.1111/1468-2435.00023.

\bibitem[{Scavo(2019{\natexlab{a}})}]{scavo_trattativa_2019}
Scavo N (2019{\natexlab{a}}) La trattativa nascosta. {Dalla} {Libia} a {Mineo},
  il negoziato tra l'{Italia} e il boss. \emph{L'Avvenire}
  \urlprefix\url{https://www.avvenire.it/attualita/pagine/dalla-libia-al-mineo-negoziato-boss-libico}.

\bibitem[{Scavo(2019{\natexlab{b}})}]{scavo_migranti_2019}
Scavo N (2019{\natexlab{b}}) Migranti. {Libia}, i guardacoste sono spariti.
  \emph{L'Avvenire}
  \urlprefix\url{https://www.avvenire.it/attualita/pagine/libia-i-guardacoste-sono-spariti}.

\bibitem[{Scavo(2019{\natexlab{c}})}]{scavo_tripoli_2019}
Scavo N (2019{\natexlab{c}}) Tripoli interrompe i soccorsi in mare e usa le
  navi italiane per la guerra. \emph{L'Avvenire}
  \urlprefix\url{https://www.avvenire.it/attualita/pagine/libia-nessuno-pattuglia-mare-sar}.

\bibitem[{Sorensen \protect\BIBand{}
  Carrion-Flores(2007)}]{sorensen_effects_2007}
Sorensen T, Carrion-Flores C (2007) The {Effects} of {Border} {Enforcement} on
  {Migrants}’ {Border} {Crossing} {Choices}: {Diversion} or {Deterrence}?,
  \urlprefix\url{https://pdfs.semanticscholar.org/966a/367f5105d942399e2c1cea1fd2103d0bf4c2.pdf}.

\bibitem[{Spagnolo(2017)}]{spagnolo_salvataggi_2017}
Spagnolo V (2017) Salvataggi. {Codice} di condotta {Ong}, spiragli verso
  un'intesa. \emph{L'Avvenire}
  \urlprefix\url{https://www.avvenire.it/attualita/pagine/codice-di-condotta-ong-lunedi-la-firma},
  section: attualita.

\bibitem[{Steinhilper \protect\BIBand{}
  Gruijters(2018)}]{steinhilper_contested_2018}
Steinhilper E, Gruijters RJ (2018) A contested crisis: {Policy} narratives and
  empirical evidence on border deaths in the {Mediterranean}. \emph{Sociology}
  52(3):515--533, \urlprefix\url{http://dx.doi.org/10.1177/0038038518759248},
  publisher: SAGE Publications Sage UK: London, England.

\bibitem[{Tamura(2010)}]{tamura_migrant_2010}
Tamura Y (2010) Migrant smuggling. \emph{Journal of public economics}
  94(7-8):540--548,
  \urlprefix\url{http://dx.doi.org/10.1016/j.jpubeco.2010.03.005}, tex.ids:
  tamura\_migrant\_2010-1 publisher: Elsevier.

\bibitem[{Tondo(2019)}]{tondo_libya_2019}
Tondo L (2019) Libya ordered arrest of alleged trafficker who attended {Italy}
  migration talks. \emph{The Guardian} ISSN 0261-3077,
  \urlprefix\url{https://www.theguardian.com/world/2019/oct/28/libya-orders-arrest-of-trafficker-who-attended-italy-migration-talks}.

\bibitem[{Train(2009)}]{train_discrete_2009}
Train KE (2009) \emph{Discrete {Choice} {Methods} with {Simulation}} (Cambridge
  University Press),
  \urlprefix\url{https://eml.berkeley.edu/books/choice2.html}.

\bibitem[{{UK House of Lords, European Union
  Committee}(2016)}]{uk_house_of_lords_european_union_committee_operation_2016}
{UK House of Lords, European Union Committee} (2016) Operation {Sophia}, the
  {EU}’s naval mission in the {Mediterranean}: an impossible challenge.
  Technical Report 14th Report of Session 2015–16,
  \urlprefix\url{https://publications.parliament.uk/pa/ld201516/ldselect/ldeucom/144/14402.htm}.

\bibitem[{{UN Office on Drugs and
  Crime}(2018)}]{un_office_on_drugs_and_crime_global_2018}
{UN Office on Drugs and Crime} (2018) Global {Study} on {Smuggling} of
  {Migrants}. Technical report, Vienna, Austria,
  \urlprefix\url{https://www.unodc.org/documents/data-and-analysis/glosom/GLOSOM_2018_web_small.pdf}.

\bibitem[{{UN Treaty Collection}(1951)}]{un_treaty_collection_convention_1951}
{UN Treaty Collection} (1951) Convention {Relating} to the {Status} of
  {Refugees}.
  \urlprefix\url{https://treaties.un.org/doc/Treaties/1954/04/19540422%2000-23%20AM/Ch_V_2p.pdf}.

\bibitem[{{UN Treaty
  Collection}(1974)}]{un_treaty_collection_international_1974}
{UN Treaty Collection} (1974) International {Convention} for the {Safety} of
  {Life} at {Sea}.
  \urlprefix\url{https://treaties.un.org/doc/Publication/UNTS/Volume%201184/volume-1184-I-18961-English.pdf}.

\bibitem[{{UN Treaty
  Collection}(1979)}]{un_treaty_collection_international_1979}
{UN Treaty Collection} (1979) International {Convention} on {Maritime} {Search}
  and {Rescue}.
  \urlprefix\url{https://treaties.un.org/doc/Publication/UNTS/Volume%201405/volume-1405-I-23489-English.pdf}.

\bibitem[{{UNDP}(2009)}]{undp_overcoming_2009}
{UNDP} (2009) \emph{Overcoming barriers: {Human} mobility and development}.
  Number 2009 in Human development report (Houndmills: Palgrave Macmillan),
  ISBN 978-0-230-23904-3, oCLC: 553541644.

\bibitem[{{UNHCR}(2011)}]{unhcr_rescue_2011}
{UNHCR} (2011) Rescue at {Sea}, {Stowaways} and {Maritime} {Interception}, 2nd
  {Ed}. \urlprefix\url{https://www.unhcr.org/4ee1d32b9.pdf}.

\bibitem[{{UNHCR}(2018)}]{unhcr_desperate_2018}
{UNHCR} (2018) Desperate {Journeys} - {Refugees} and migrants arriving in
  {Europe} and at {Europe}'s borders.
  \urlprefix\url{https://www.unhcr.org/desperatejourneys/}.

\bibitem[{{UNHCR}(2019)}]{unhcr_europe_2019}
{UNHCR} (2019) Europe - {Refugee} and {Migrant} arrivals and dead and missing
  data. \urlprefix\url{https://data2.unhcr.org/en/documents/details/58460}.

\bibitem[{{UNHCR}(2020{\natexlab{a}})}]{unhcr_libya_2020}
{UNHCR} (2020{\natexlab{a}}) Libya: {Activities} at {Disembarkation} -
  {Monthly} {Update}.
  \urlprefix\url{https://data2.unhcr.org/en/search?country=&text=%22Libya:+Activities+at+Disembarkation+-+Monthly+update%22&type[0]=link&type[1]=news&type[2]=highlight&type[3]=document&type[4]=needs_assessment&type[5]=dataviz&partner=&sector=&date_from=&date_to=&country_json={%220%22:%22%22}&sector_json={%220%22:%22%22}&apply=&page=1}.

\bibitem[{{UNHCR}(2020{\natexlab{b}})}]{unhcr_mediterranean_2020-1}
{UNHCR} (2020{\natexlab{b}}) Mediterranean {Situation}.
  \urlprefix\url{https://data2.unhcr.org/en/situations/mediterranean}.

\bibitem[{{UNHCR}(2020{\natexlab{c}})}]{unhcr_mediterranean_2020}
{UNHCR} (2020{\natexlab{c}}) Mediterranean {Situation} - {Italy}.
  \urlprefix\url{https://data2.unhcr.org/en/situations/mediterranean/location/5205}.

\bibitem[{Uzun et~al.(2016)Uzun, Dağdeviren, \protect\BIBand{}
  Kabak}]{uzun_determining_2016}
Uzun G, Dağdeviren M, Kabak M (2016) Determining the {Distribution} of {Coast}
  {Guard} {Vessels}. \emph{Interfaces} 46(4):297--314, ISSN 0092-2102,
  1526-551X, \urlprefix\url{http://dx.doi.org/10.1287/inte.2016.0852}.

\bibitem[{{Watch the Med}(2020{\natexlab{a}})}]{watch_the_med_reports_2020}
{Watch the Med} (2020{\natexlab{a}}) Reports.
  \urlprefix\url{http://watchthemed.net/index.php/reports}.

\bibitem[{{Watch the Med}(2020{\natexlab{b}})}]{watch_the_med_search_2020}
{Watch the Med} (2020{\natexlab{b}}) Search and {Rescue} {Zone}.
  \urlprefix\url{https://watchthemed.net/json/layer/9}.

\bibitem[{Zandonini(2017)}]{zandonini_how_2017}
Zandonini G (2017) How the humanitarian {NGOs} work at sea.
  \urlprefix\url{https://openmigration.org/en/analyses/how-the-humanitarian-ngos-operate-at-sea/}.

\bibitem[{Ziebart et~al.(2008)Ziebart, Maas, Bagnell, \protect\BIBand{}
  Dey}]{ziebart_maximum_2008}
Ziebart BD, Maas AL, Bagnell JA, Dey AK (2008) Maximum entropy inverse
  reinforcement learning. \emph{{AAAI}}, volume~8, 1433--1438 (Chicago, IL,
  USA), \urlprefix\url{https://www.aaai.org/Papers/AAAI/2008/AAAI08-227.pdf}.

\end{thebibliography}
\doublespacing

\FloatBarrier
\newpage

\appendix
\renewcommand\thefigure{\thesection.\arabic{figure}} 
\renewcommand\thetable{\thesection.\arabic{table}} 

\setcounter{figure}{0}  
\setcounter{table}{0}

\FloatBarrier
\section{Legal Protections for Migrants Recovered at Sea} \label{app:context}

\begin{table}[htb]
\centerline{
\begin{tabular}{|p{2.5in}p{.25in}p{3.5in}|}
\hline
Agreement &Year & Provisions \\
\hline
The Convention Relating to the Status of Refugees, Ch. V, Art. 33  \cite{un_treaty_collection_convention_1951}
		& 1951 
		& 	``No Contracting State shall expel or return (`refouler') a refugee in any manner whatsoever to the frontiers of territories where his life or freedom would be threatened on account of his race, religion, nationality, membership of a particular social group or political opinion.'' \\
\hline
International Convention for the Safety of Life at Sea (SOLAS), Ch. V, Reg. 10* \cite{un_treaty_collection_international_1974}
		&1974
		& ``The master of a ship at sea, on receiving a signal from any source that a ship or air craft or survival craft thereof is in distress, is bound to proceed with all speed to the assistance of the persons in distress informing them if possible that he is doing so.''  \\
		\hline
SOLAS Amendment \cite{unhcr_rescue_2011} 
		& 2004 
		& ``This obligation to provide assistance applies regardless of the nationality or status of such persons or the circumstances in which they are found.'' \\
		\hline
International Convention on Maritime Search and Rescue, Ch. 2 \cite{un_treaty_collection_international_1979} 
		&  1979 
		& ``Parties shall ensure that assistance be provided to any person in distress at sea. They shall do so regardless of the nationality or status of such a person or the circumstances in which that person is found.'' \\
		\hline
SAR Convention Amendment \cite{unhcr_rescue_2011} 
		& 2004 
		& ``The Party responsible for the search and rescue region in which such assistance is rendered shall exercise primary responsibility for ensuring such coordination and cooperation occurs, so that survivors assisted are disembarked from the assisting ship and delivered to a place of safety, taking into account the particular circumstances of the case \ldots'' \\
		\hline
\end{tabular}}
\floatfoot{* In other versions, this is Regulation {33}.}
\caption{International agreements governing the rescue of migrants and refugees at sea}\label{tab:legal}
\end{table}

\FloatBarrier 
\newpage
\FloatBarrier
\section{Supplementary Details for the Flow Analysis}

Below, we provide additional details on the estimation of the error correction models in Sections~\ref{sec:ecm_main} and \ref{sec:spillover}.

\subsection{ECM of Crossings on the Central Mediterranean Route}
\subsubsection{Checks on the Model}
\label{subsec:ecm_check}

Dickey-Fuller tests are consistent with the hypothesis that  the total crossings, the log total crossings, the log odds of crossing, and the probability of rescue {are non-stationary but that their first differences are stationary,} suggesting that the ECM is appropriate in this setting. Similarly, the Engle-Granger test supports the hypothesis of cointegrating relationships between (1) the total crossings and the probability of rescue, (2) the log total crossings and the probability of rescue, and (3) the log odds of crossing and the probability of rescue.

\subsubsection{Alternative Specifications for the ECM of Crossings on the Central Mediterranean Route}
\label{sec:ecm_robustness}

Below, we discuss possible alternative specifications for the model presented in Section~\ref{sec:ecm_main}. 
In Table~\ref{tbl:ecm_robustness}, we test three different measures of crossing behavior as the dependent variable: the differenced total number of crossings in thousands, which is the dependent variable used in the main paper (columns 1-3); the differenced log number of crossings in thousands (columns 4-6); and the differenced log odds of crossing (columns 7-9).\footnote{In order to calculate the odds of crossing relative to staying, it is necessary to know how many people could potentially cross in a given period. We assume that the maximum potential number of people crossing is equal to the largest number of crossings observed in any period (29,478 for the Central Mediterranean route), multiplied by $\frac{10}{9}$ to ensure that there is no period in which all potential crossings occur (because this would result in division by zero when calculating the odds).}  We also experiment with different specifications of the ECM; the model described in Equation~\ref{eq:ecm} is shown in columns 2, 5, and 8,  but we also present the results when excluding the short-run adjustment term (columns 1, 4, and 7) and including lagged crossing behavior (columns 3, 6, and 9). 

We estimate a similar speed of adjustment across all dependent variables: in each period after a divergence from the equilibrium relationship between crossing behavior and the probability of rescue, we estimate that the number of crossings falls by approximately {40\%} of the deviation from equilibrium, whereas the log number of crossings falls by approximately {43\%} of the deviation from equilibrium and the log odds of crossing falls by approximately  {46\%} of the deviation from equilibrium. The estimated speed of adjustment is slightly slower when using a model that omits the differenced probability of rescue from the previous period (the short-run effect), and slightly faster when we also include the differenced dependent variable from the previous period. Our coefficients on the long-run speed of adjustment parameter remain significant across all specifications, and we never estimate a significant short-run adjustment.

The estimated equilibrium relationships for the log number of crossings and the log odds of crossing {are}:
\begin{eqnarray}
\log \left( N_{t,cross}\right) &=& -1.17 ~~+~~ 4.13~  ~ P_{t,rescue} \label{eq:coint1} \\ 
\log \left( \frac{P_{t,cross}}{P_{t,stay}}\right) &\quad=\quad& -5.33 ~~+~~ 5.70~ P_{t,rescue}.\label{eq:coint} 
\end{eqnarray}
These equations suggest that when the probability of rescue falls from approximately 90\% to 50\%,
 the number of monthly crossings will decline by approximately {10,300 - 12,200} people. 
 
	\begin{table}
    	\centerline{\footnotesize{
\def\sym#1{\ifmmode^{#1}\else\(^{#1}\)\fi}
\begin{tabular}{|l|ccc|ccc|ccc|}
\hline
                                                      
                                                       &\multicolumn{3}{c|}{$\Delta \text{N~}^{central}_{t,cross}$}
                                                       &\multicolumn{3}{c|}{$\Delta \log(\text{N~}^{central}_{t,cross})$}
                                                       &\multicolumn{3}{c|}{$\Delta \text{log\_odds}^{central}_{t}$}\\
                                                        &\multicolumn{1}{c}{(1)}&\multicolumn{1}{c}{(2)}&\multicolumn{1}{c|}{(3)}&\multicolumn{1}{c}{(4)}&\multicolumn{1}{c}{(5)}&\multicolumn{1}{c|}{(6)}&\multicolumn{1}{c}{(7)}&\multicolumn{1}{c}{(8)}&\multicolumn{1}{c|}{(9)}\\
\hline
$\hat{e}_{t-1} $                                             &   -0.351\sym{***}&   -0.402\sym{***}&   -0.489\sym{***}&   -0.365\sym{***}&   -0.426\sym{***}&   -0.432\sym{***}&   -0.394\sym{***}&   -0.462\sym{***}&   -0.513\sym{***}\\
                                                       &  (0.115)         &  (0.123)         &  (0.137)         &  (0.120)         &  (0.133)         &  (0.150)         &  (0.122)         &  (0.133)         &  (0.152)         \\
$\Delta \text{P~}_{t-1,rescue}^{central}$                                        &                  &   -3.249         &   -4.206         &                  &   -0.340         &   -0.356         &                  &   -0.679         &   -0.847         \\
                                                       &                  &  (5.477)         &  (5.464)         &                  &  (0.650)         &  (0.683)         &                  &  (0.982)         &  (1.015)         \\
$\Delta \text{N~}^{central}_{t-1, cross}$                                    &                  &                  &    0.210         &                  &                  &                  &                  &                  &                  \\
                                                       &                  &                  &  (0.152)         &                  &                  &                  &                  &                  &                  \\
$\Delta \log(\text{N~}^{central}_{t-1, cross})$                                &                  &                  &                  &                  &                  &    0.013         &                  &                  &                  \\
                                                       &                  &                  &                  &                  &                  &  (0.159)         &                  &                  &                  \\
$\Delta \text{log\_odds}^{central}_{t-1}$                                            &                  &                  &                  &                  &                  &                  &                  &                  &    0.111         \\
                                                       &                  &                  &                  &                  &                  &                  &                  &                  &  (0.156)         \\
Constant                                               &   -0.040         &    0.060         &    0.098         &   -0.024         &   -0.009         &   -0.009         &   -0.023         &   -0.002         &    0.002         \\
                                                       &  (0.685)         &  (0.698)         &  (0.691)         &  (0.078)         &  (0.079)         &  (0.080)         &  (0.120)         &  (0.122)         &  (0.123)         \\
\hline
R$^2$                                                     &    0.172         &    0.200         &    0.235         &    0.171         &    0.202         &    0.202         &    0.189         &    0.222         &    0.231         \\
R$^2$ - adjusted                                          &    0.153         &    0.163         &    0.180         &    0.153         &    0.164         &    0.145         &    0.170         &    0.186         &    0.177         \\
N Obs.                                                 &       47         &       46         &       46         &       47         &       46         &       46         &       47         &       46         &       46         \\
Mean Dep. Var.                                         &    -0.10         &    -0.07         &    -0.07         &    -0.04         &    -0.03         &    -0.03         &    -0.04         &    -0.03         &    -0.03         \\
\bottomrule
\multicolumn{10}{l}{\footnotesize Standard errors in parentheses}\\
\multicolumn{10}{l}{\footnotesize \sym{*} \(p<0.10\), \sym{**} \(p<0.05\), \sym{***} \(p<0.01\)}\\
\end{tabular}
}
}
    	\caption{Alternative specifications of the error correction model for crossings along the Central Mediterranean route}\label{tbl:ecm_robustness}
	\end{table}

\subsubsection{Stability of Coefficient Estimates and Backtesting}
In addition to testing alternative specifications of the model, we also tested the robustness of the model to being fit on different time windows, in order to gain insight into the stability of coefficient estimates. 

\begin{figure}
	\begin{subfigure}{\textwidth}
    	\centerline{\includegraphics[width=5in]{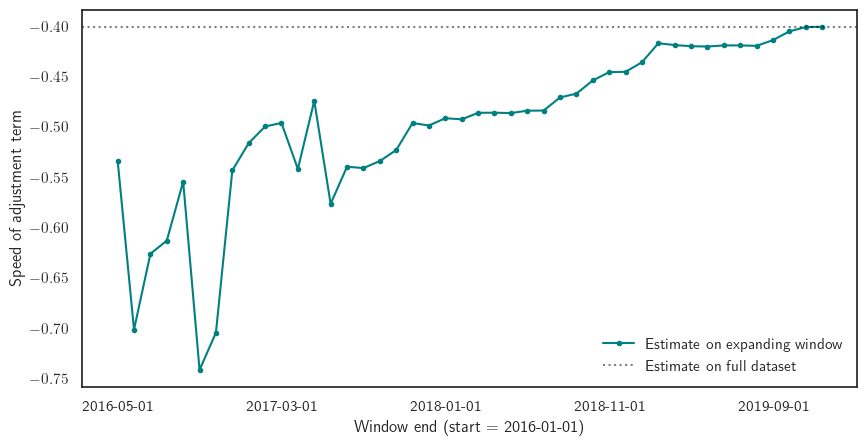}}
    	\centerline{\footnotesize }
    	\caption{Rate of adjustment, estimated over different expanding windows}\label{fig:ecm_expanding_window}
	\end{subfigure}
	\begin{subfigure}{\textwidth}
    	\centerline{\includegraphics[width=5in]{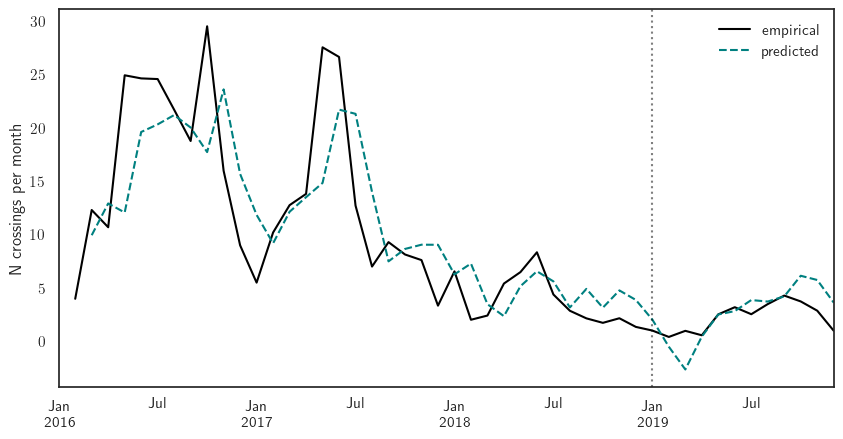}}
    	\caption{Observed vs. predicted absolute number of crossings}\label{fig:ecm_predict_N}
	\end{subfigure}
	\caption{Backtesting for the ECM model of crossings on the Central Mediterranean Route}\label{fig:backtesting}
\end{figure}

In Figure \ref{fig:ecm_expanding_window}, we re-estimate the speed of adjustment over different sliding windows, starting with the five-month window from January - May 2016 and ending with the full 48-month dataset spanning January 2016 - December 2019. This allows us to determine how the coefficient estimate changes as more (recent) data is added to the model. As we can see, the estimated speed of adjustment has decreased\footnote{Recall that the absolute magnitude of the rate of adjustment determines the speed whereas the sign determines the direction of the adjustment, so a less negative rate of adjustment is a ``slower'' rate of adjustment.} over time but appears to be stabilizing as more data points are used for estimation, suggesting that we are approaching a more consistent estimate.

The stability of coefficient estimates is particularly important for cases in which such estimates could be used to predict future behavior. Therefore, in Figure \ref{fig:ecm_predict_N}, we compare the observed and predicted period-to-period changes in the number of crossings along the Central Mediterranean route. We train the model on the years 2016-2018 and then predict the period-to-period changes in crossings in 2019 using the trained ECM model. For each data point, we then take the true observed arrivals from period $t-1$ and add $\Delta \hat{N}^{central}_{t, cross}$ as predicted by the model to determine the estimated number of arrivals. As shown in the figure, we are able to approximate the observed arrivals trend; the mean absolute error is 1,949 arrivals on the training dataset (relative to a monthly average of 11,226 arrivals) and 1,141 arrivals on the test dataset (relative to a monthly average of 2,199 arrivals).
While the model is fitting the training dataset, predictive power appears limited as the MAE from this ECM model is higher than that from the na\"ive approach of predicting no change since the previous period.

\subsection{Substitution from the Central to the Western Route}
\subsubsection{Checks on the Model} \label{subsec:ecm_check_west}

Dickey-Fuller tests suggest that the number of crossings (in thousands) in the Western Mediterranean is also non-stationary, and that its first differences are stationary. However, we do not find significant evidence of a cointegrating relationship according to the Engle-Granger test; given that we have only 48 observations for training the model, we may be under-powered to detect such a relationship. We also note that this model seems less stable; \textcite{box-steffensmeier_time_2014} suggest that when the independent and dependent variables in the equilibrium equation are reversed and the ECM is re-estimated, the long-run adjustment term should remain significant, which is not the case here. While we fit an ECM for comparability with the model of crossings on the Central Mediterranean route, the results of this model in the Western Mediterranean context should be interpreted cautiously.

\FloatBarrier  
\newpage
\FloatBarrier
\section{Analysis of Additional Incident Datasets}\label{sec:data}
\subsection{Supplementary Data Sources}

As described in Section~\ref{sec:incident_analysis}, our primary analysis relies on incident-level data from Frontex. In these appendices, we draw on additional datasets to conduct supplementary analysis. Specifically, we gather additional data on incidents from four primary data sources:

    \begin{itemize}
    
        \item\textbf{Watch The Med/Alarm Phone:}
        Watch The Med (WTM) is a platform that has monitored migration incidents in the Mediterranean since {2012} \parencite{watch_the_med_reports_2020}. Watch The Med data comes primarily from Alarm Phone, an emergency telephone hotline designed to help migrants and refugees at sea. Alarm Phone typically receives calls directly from migrant boats that have been equipped with a satellite phone, which also allows these boats to report their positions. Incident reports may contain information on the ship location and type, the number of people on board, the port of departure, and the details of the rescue.
        
        \item\textbf{Broadcast Warnings:} Broadcast Warnings are a form of general-purpose maritime communication, and are issued to alert ships of nearby operations, dangers, and emergencies. They contain announcements about ships in distress, often reporting the location, the number of people on board, and a description of the situation. We parse the text of all Broadcast Warnings issued for the Mediterranean since 2014, removing irrelevant incidents \parencite{national_geospatial-intelligence_agency_broadcast_2018}.\footnote{For example, we removed incidents that referred to {hazards or logistics (e.g., cable-laying operations or adrift fishing gear) and ship-related incidents that did not appear to be related to migration (e.g., disabled fishing vessels or ships that were referenced by name)}. Incidents were first filtered heuristically using keyword searches, and remaining incidents were then manually inspected for relevance.}
        
        \item\textbf{M\'edecins sans Fronti\`eres:}
        Data on rescues involving the NGO M\'edecins sans Fronti\`eres (MSF, or Doctors without Borders) is available through its online search and rescue portal, which tracks the activity of the organization's {eight} rescue ships since 2015~\parencite{msf_search_2020}. {This portal contains information on rescue missions and individual operations, including the location and time of the rescue; the number of people involved; and the boat type and weather conditions.}
        
        \item\textbf{Missing Migrants Project (IOM):}
        Finally, we incorporate data from the IOM's Missing Migrants Project, an initiative which has tracked migration-related deaths since {2014}~\parencite{iom_missing_2021}. The project collects data from {national authorities, NGOs, and media sources} to identify individual incidents. In particular, the missing migrants dataset contains information on {the location of the incident, the number of dead or missing, the number of survivors, and the estimated cause of death}. 
    
    \end{itemize}

Where geospatial information on incident locations is available, we restrict our dataset to incidents which occurred within the Italian, Maltese, or Libyan rescue zones.

\begin{landscape}
    {\footnotesize \begin{table}
	\caption{Overview of the incident-level datasets}\label{tbl:datasets}
~\\
	\centerline{\footnotesize
	\begin{tabular}{|p{.81in}|p{.75in}|p{0.65in}|l|p{2.5in}|p{3in}|}
	\hline
	Data source & Collected by & Time span & N& Description & Limitations \\
	\hline
	Frontex & Frontex 
			& 2014-01 -- 2019-12 & 4,365 
			& Boat detections and interceptions recorded by Frontex, including incidents coordinated by national authorities, NGOs, and commercial ships. 
			& This seems to be the most comprehensive incident-level dataset. However, many incidents are missing precise location data, and the dataset does not appear to include Libyan interceptions. Sometimes, multiple boats are reported as part of the same incident.
			\\
			\hline
	Watch The Med & Alarm Phone (primarily)
			&  2014-06 -- 2019-12 & 325 
			& Incident reports that are largely drawn from calls for help to the Alarm Phone NGO hotline. Alarm Phone has no rescue capacity but it advocates on behalf of migrants, tracks their progress, alerts rescue authorities, requests intervention, and tops up the credit of boats' satellite phones so they can continue making calls.
			& Calls typically come only from boats equipped with satellite phones, or from relatives on shore. Calls to Alarm Phone seem to be \textit{lower} in phases of high rescue activity, possibly because (1) many boats were independently detected by rescue NGOs in the region without a call; and (2) Migrants may have been more likely to call the MRCC in Rome directly since they have actual capacity to intervene. Once the MRCC started assigning rescues to the Libyan Coast Guard, we see more calls going directly to Alarm Phone. \\ 
			\hline
	Broadcast Warnings & National Geospatial-Intelligence Agency
			& 2014-01 -- 2020-04 & 1,043 
			& Maritime alerts about ships in distress, often with information about the number of people on board and whether a ship is sinking, disabled, etc. They also contain information on activities like naval operations, hazards, etc.
			& Only includes publicly issued requests for help; the MRCC seems to assign and coordinate some rescues behind the scenes without producing a Broadcast Warning. There is no formal information on whether an incident involves migrants or not; we estimate this based on the content of the report, keeping reports that we believe are migrant-related. Sometimes warnings reference multiple boats and positions; these have been divided into multiple distinct incidents when possible. \\
			\hline
	M\'edecins sans Fronti\`eres & M\'edecins sans Fronti\`eres 
			& 2015-05 -- 2019-12 & 444 
			& Records of rescues conducted by MSF ships, including information on weather conditions and {whether a rescue was initiated by the MRCC or the boat was detected by MSF or another vessel.} Records are grouped by {operations,} each of which may include several rescues.
			& Only covers the operations of MSF boats; rescues only occur when MSF boats have been deployed in the region. \\
			\hline
	Missing Migrants Project & IOM 
			& 2014-02 -- 2019-12 & 344 
			& Database of incidents in which migrants died or went missing, compiled from media sources, national authorities, etc.
			& Does not include incidents where everyone survived. 
			Sources are mixed in quality (though the dataset does provide a source quality estimate), and sometimes multiple incidents appear to be reported as one. \\
			\hline
	\end{tabular}}
\end{table}
}
\end{landscape}

\subsection{Comparison of Alternative Data Sources}\label{subsec:data_comparison}

\begin{figure}[t]
\centerline{\includegraphics[width=5in]{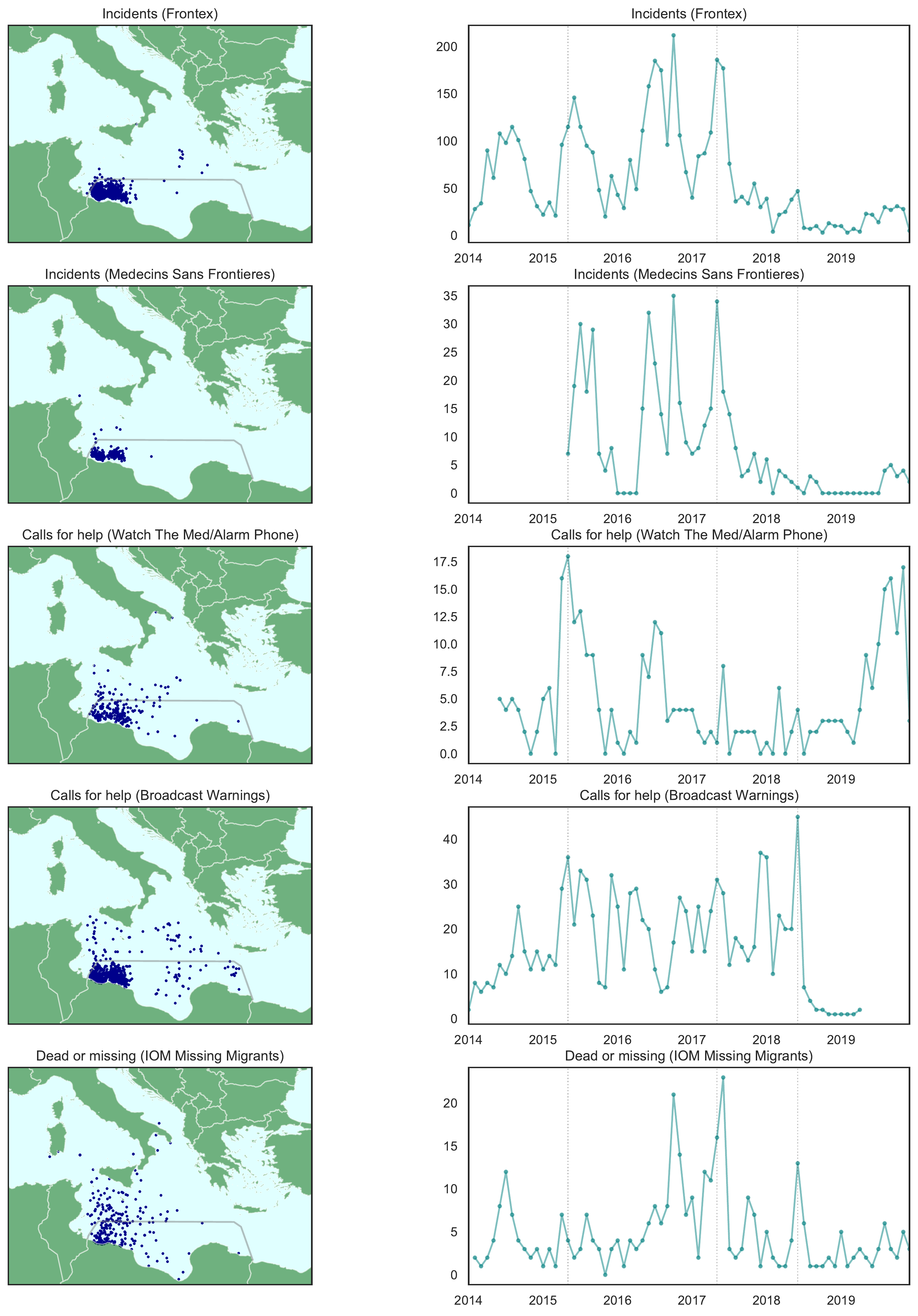}}
\caption{The location and frequency of incidents, by dataset}
\label{fig:incident_datasets}
\floatfoot{Country borders were obtained from Natural Earth~\parencite{natural_earth_110m_2020}. The linear boundary of Libya's search and rescue zone, shown in grey, was obtained from  Watch The Med~\parencite{watch_the_med_search_2020}. The vertical lines in the right panel represent the beginning of Phase 2; the beginning of Phase 3; and the formal recognition of the Libyan SAR zone, respectively.}
\end{figure}

The analysis below is one of the first to compare multiple incident-level data sources, which we summarize in more detail in Table~\ref{tbl:datasets}. We argue that comparing multiple data sources is important because each dataset captures different phenomena. This is illustrated in Figure~\ref{fig:incident_datasets}, which summarizes the distribution of observed incidents by dataset for the years 2014 - 2019. 

The left panel of the figure shows the spatial distribution of all geotagged incidents in each dataset (restricted to incidents which occurred in the Italian, Maltese, or Libyan SAR zones).
 In the Frontex dataset, location information is unavailable for all incidents {before November 2014 or after October 2017} and for incidents within Frontex's operational areas (which are primarily off the European coasts), 
so most geotagged incidents are clustered near Libya or in the sea between Egypt and Italy. We can see that the MSF and Broadcast Warnings data are also largely clustered off the coast of Libya, where the bulk of rescues occurred in 2016 - 2017. In contrast, the Watch The Med data and the IOM Missing Migrants data are more scattered, likely because these datasets capture incidents that may have ``fallen through the cracks'' of the standard rescue response, because they continue to capture recent incidents, and/or because their associated geocoordinates may be less accurate.\footnote{Watch The Med often associates a single set of geocoordinates with multiple different incidents. {IOM attempts to recover incident information from public data sources, and it seems they may use approximate coordinates when no precise data is available.}}

\begin{figure}[t]
\centerline{\includegraphics[width=6in]{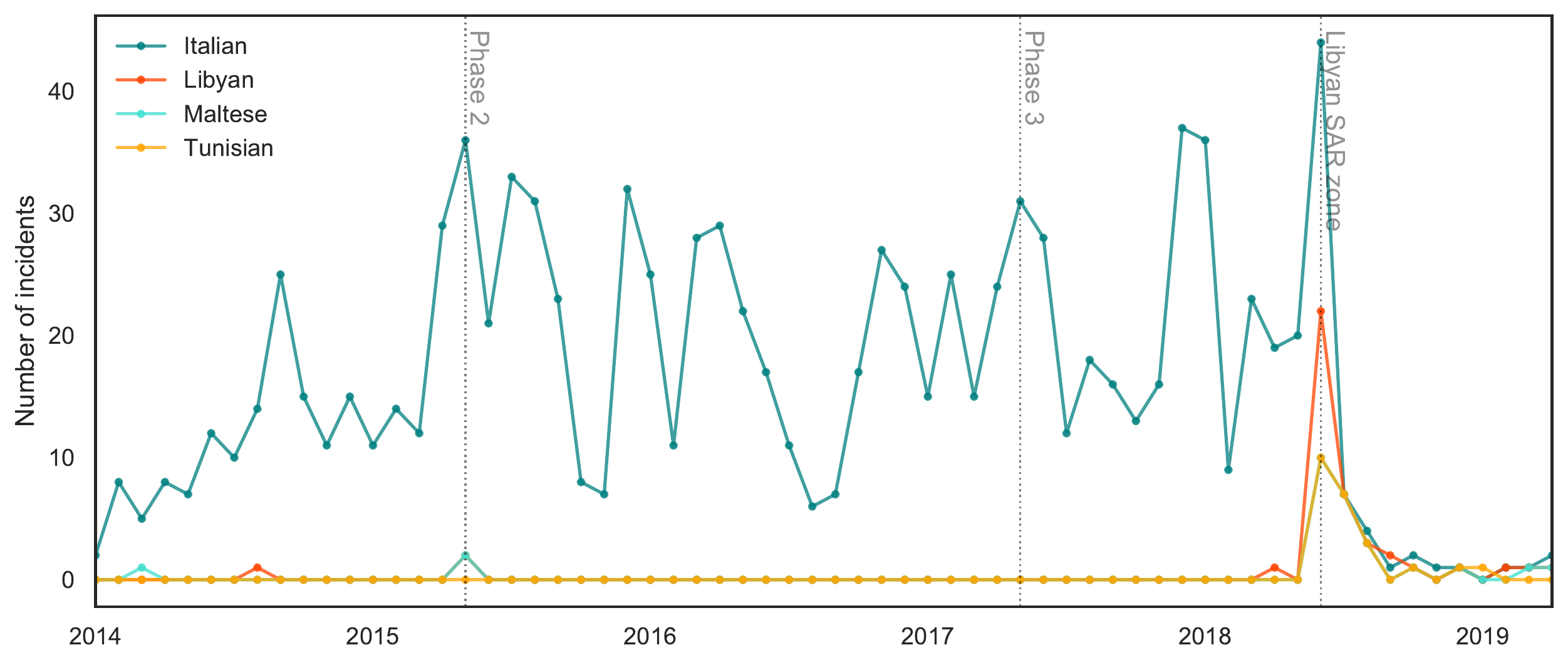}}
\floatfoot{Above, we show the monthly count of incidents referencing different rescue authorities, which we identified by searching the text of the Broadcast Warnings for the phrases {``MRCC Rome'', ``[J]RCC Libya'', ``Libyan Coast Guard'', ``RCC Malta'', and ``MRCC Tunis.''} Note that a single message might reference multiple rescue authorities, so the categories are not mutually exclusive.}
\caption{The number of incidents by month and rescue authority in the Broadcast Warnings dataset}\label{fig:bw_routing}
\end{figure}

From the right panel of Figure~\ref{fig:incident_datasets}, we can also see that the coverage of the datasets varies. Frontex reports {a peak of over 200 incidents a month, whereas the Watch The Med dataset never contains more than 20 incidents per month for this region.} It is also evident that with the growing restrictions on NGO activities, the volume of incidents handled by MSF has declined.

Furthermore, we can observe a shift in the distribution of incidents across datasets over time. For example, the Broadcast Warnings dataset recorded a large volume of incidents from 2015 - 2018. However, as the Italian authorities began turning over rescue responsibility for these calls to the Libyan authorities, it appears that some migrants have substituted to an alternative channel when requesting assistance: the volume of Broadcast Warnings has fallen even as calls to the independent NGO Alarm Phone have increased. In Figure~\ref{fig:bw_routing}, we analyze the text of the Broadcast Warnings to illustrate that the decline in calls directly coincided with the rise in referrals to non-Italian rescue authorities.

\FloatBarrier
\subsection{Descriptive Analysis of the Incident-level Data}\label{sec:incident_desc}
\subsubsection{Variation in Strategic Inputs (Boat Type and Size) Over Time}

\begin{figure}
	\begin{subfigure}{\textwidth}
    	\centerline{\includegraphics[width=5in]{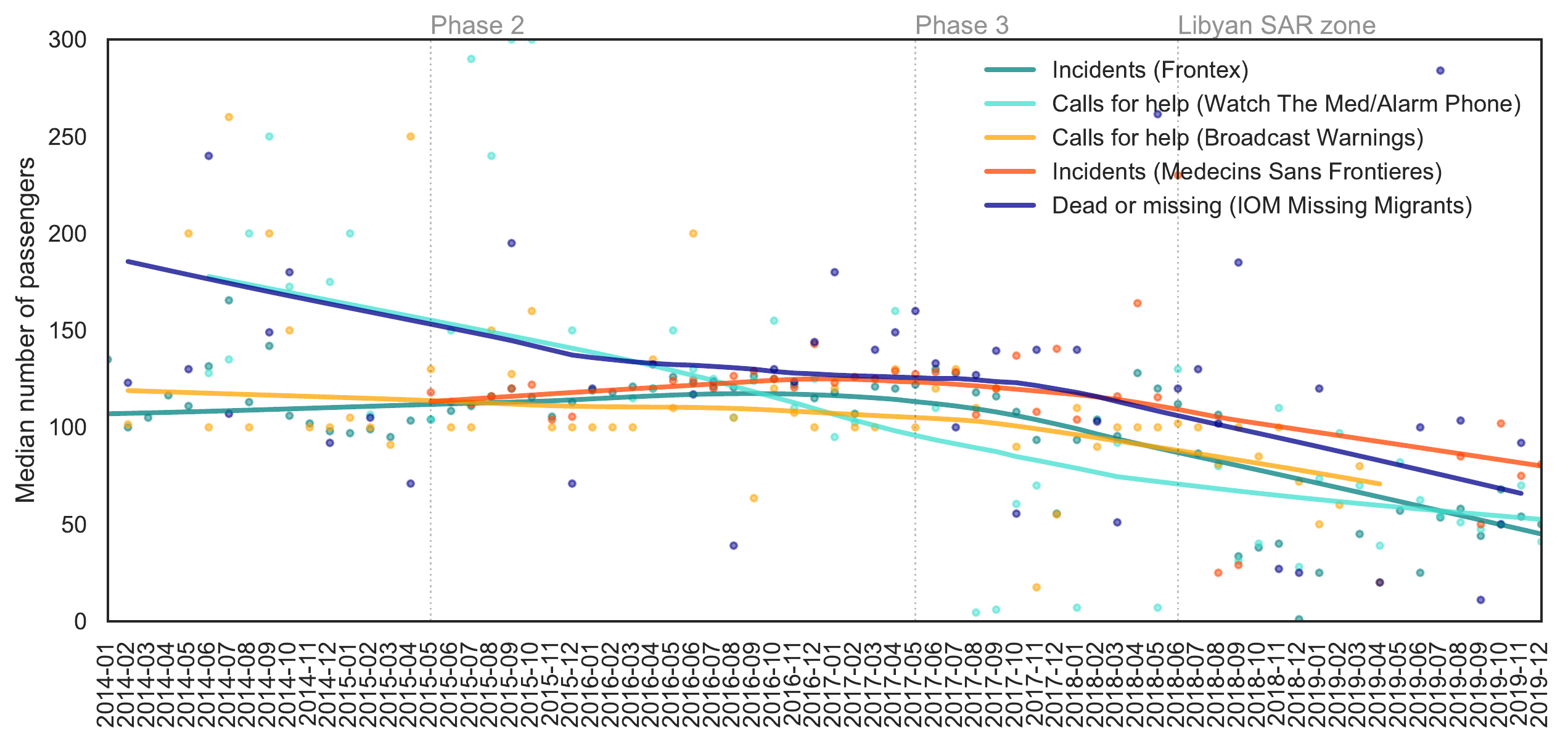}}
    	\caption{All boats}\label{fig:crowding_all}
	\end{subfigure}
	\begin{subfigure}{\textwidth}
    	\centerline{\includegraphics[width=5in]{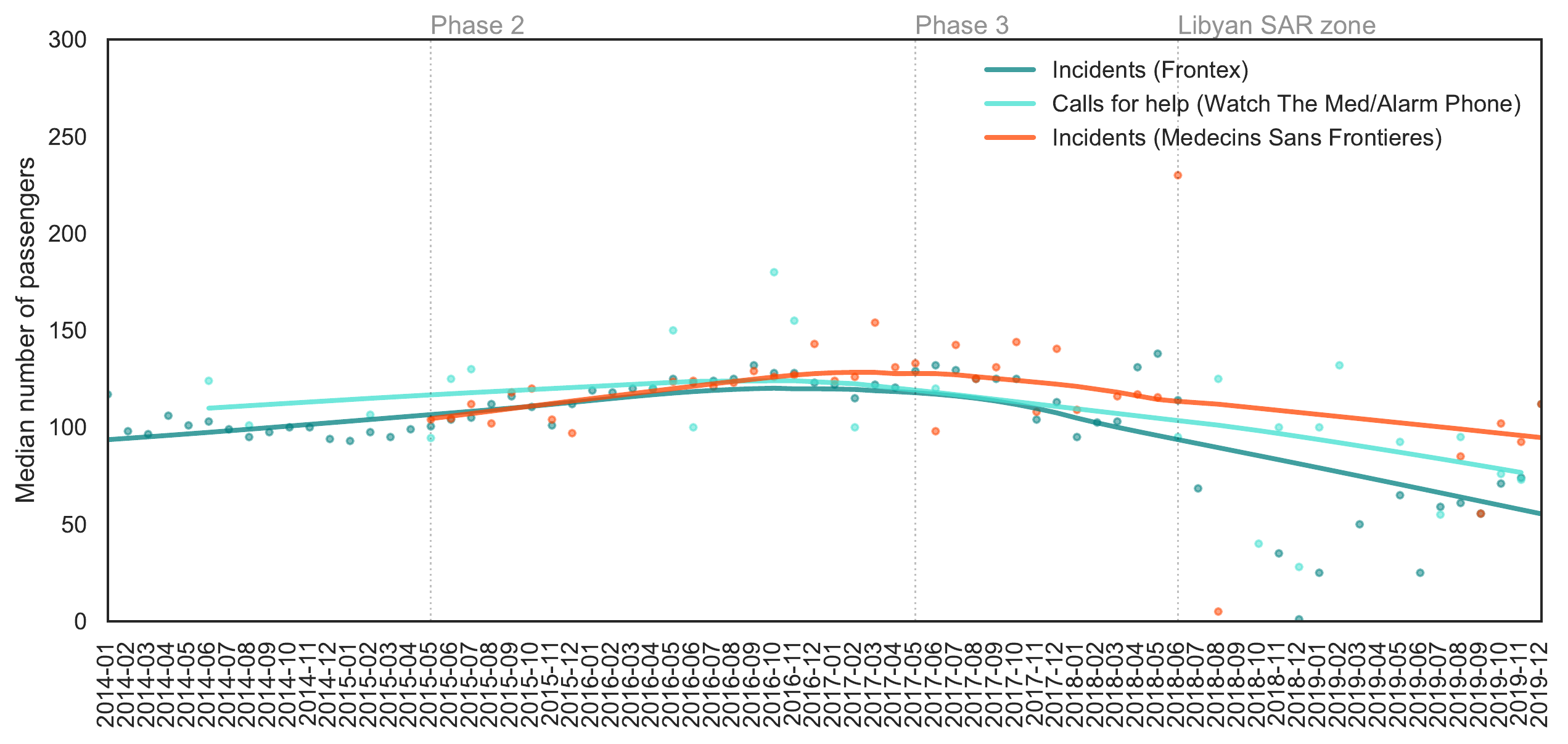}}
    	\caption{Rubber only}\label{fig:crowding_rubber}
	\end{subfigure}
	\begin{subfigure}{\textwidth}
    	\centerline{\includegraphics[width=5in]{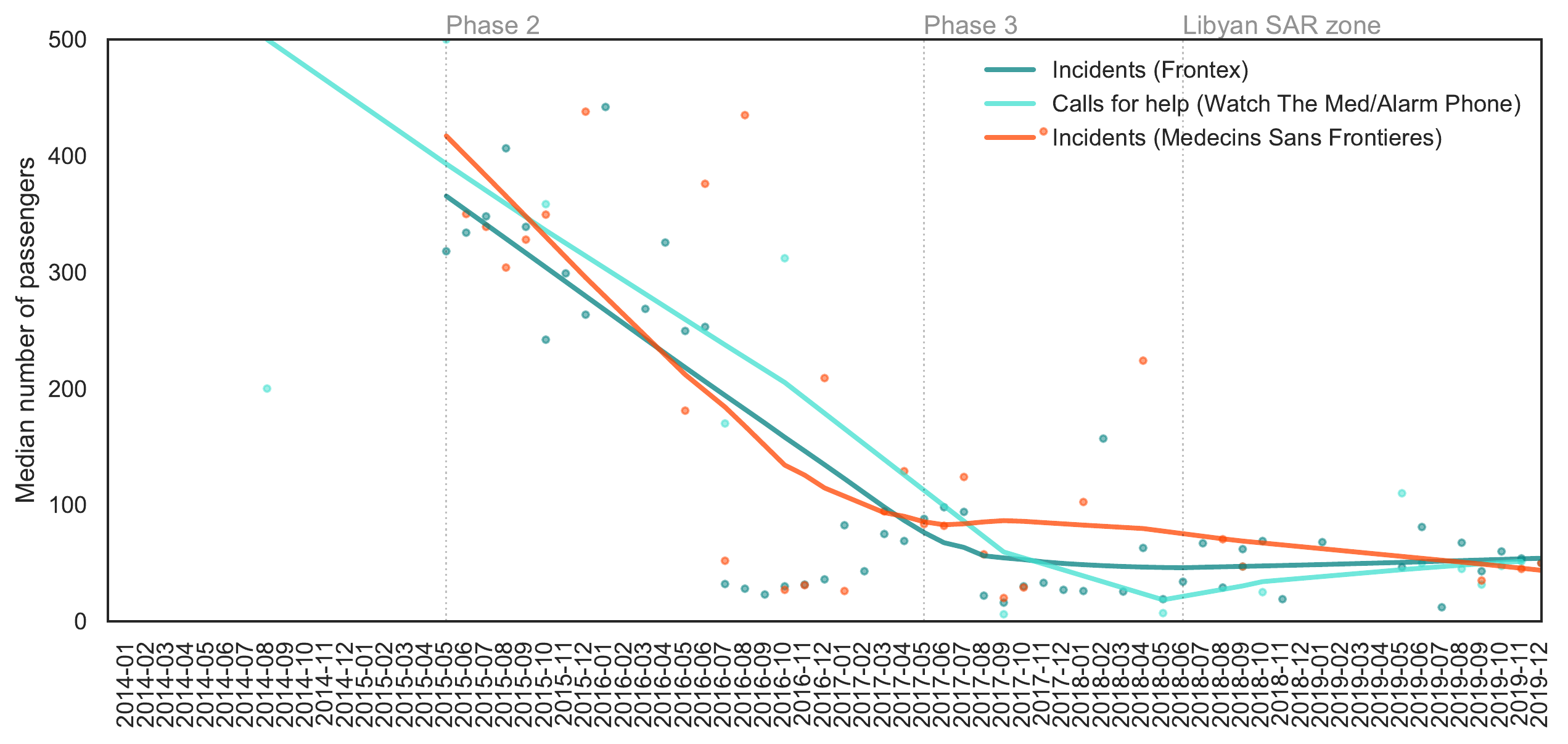}}
    	\caption{Wooden only}\label{fig:crowding_wooden}
    	\end{subfigure}
	\caption{The number of people per boat by month and dataset}\label{fig:crowding}
	\floatfoot{Each dot represents the median number of passengers per month for the respective dataset, whereas the lines represent locally weighted scatter plot smoothing (LOWESS) fits to the scatter plot trends.}
\end{figure}

Using a combination of incident-level data sources, we empirically analyze variations in two key strategic choices made by smugglers: how many migrants to place on each boat, and what type of boat to use. Figure~\ref{fig:crowding} illustrates the median number of passengers per boat by month for different datasets. As is evident from Figure~\ref{fig:crowding_all}, the median number of people has declined across almost all datasets during our period of study; this decline appears to have accelerated starting around the end of 2017. 

If we examine the number of people by boat type, a more nuanced picture emerges. From Figure~\ref{fig:crowding_wooden}, we can see that prior to mid-2017, the declining number of passengers per boat is driven primarily by the falling size of wooden boats. This may be due to the decreasing availability of large boats, in part because many wooden boats were destroyed by Operation Sophia starting in 2015 \parencite{uk_house_of_lords_european_union_committee_operation_2016}.
However, it may also result from the fact that rafts became an increasingly viable transport option during this period (since rafts are cheaper, and the growing intensity of rescue efforts made rescue more likely), thus decreasing the relative profitability of launching large wooden boats (which have very high fixed costs). If we examine rubber boats, we can find support for this hypothesis: Figure~\ref{fig:crowding_rubber} shows that crowding on rubber boats increased through early 2017 while rescue capacity was high, before beginning to decline towards 2018 - 2019.  Particularly interesting is the trend for the MSF rescues, which tend to {occur within the Libyan SAR zone}: we can see that crowding still remains higher for these NGO rescues than for other types of incidents. More generally, the fall in boat sizes is consistent with the notion that smugglers are currently using smaller, less crowded boats to move migrants further out to sea before detection, a hypothesis that we evaluate in more detail below. 

\begin{figure}
	\begin{subfigure}{\textwidth}
    	\centerline{\includegraphics[width=5in]{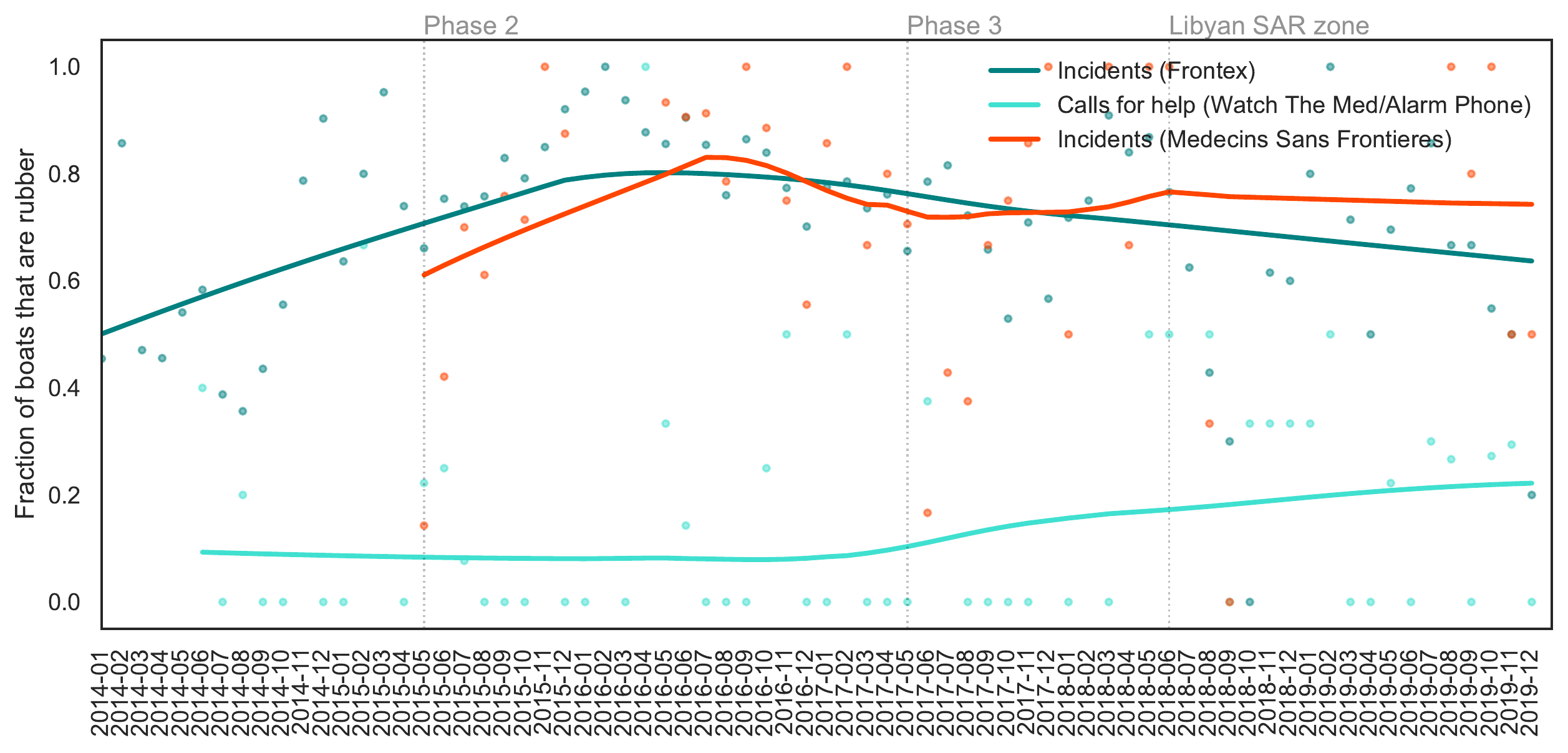}}
    	\caption{Monthly fraction of rubber boats}\label{fig:boat_type_rubber}
	\end{subfigure}
	\begin{subfigure}{\textwidth}
    	\centerline{\includegraphics[width=5in]{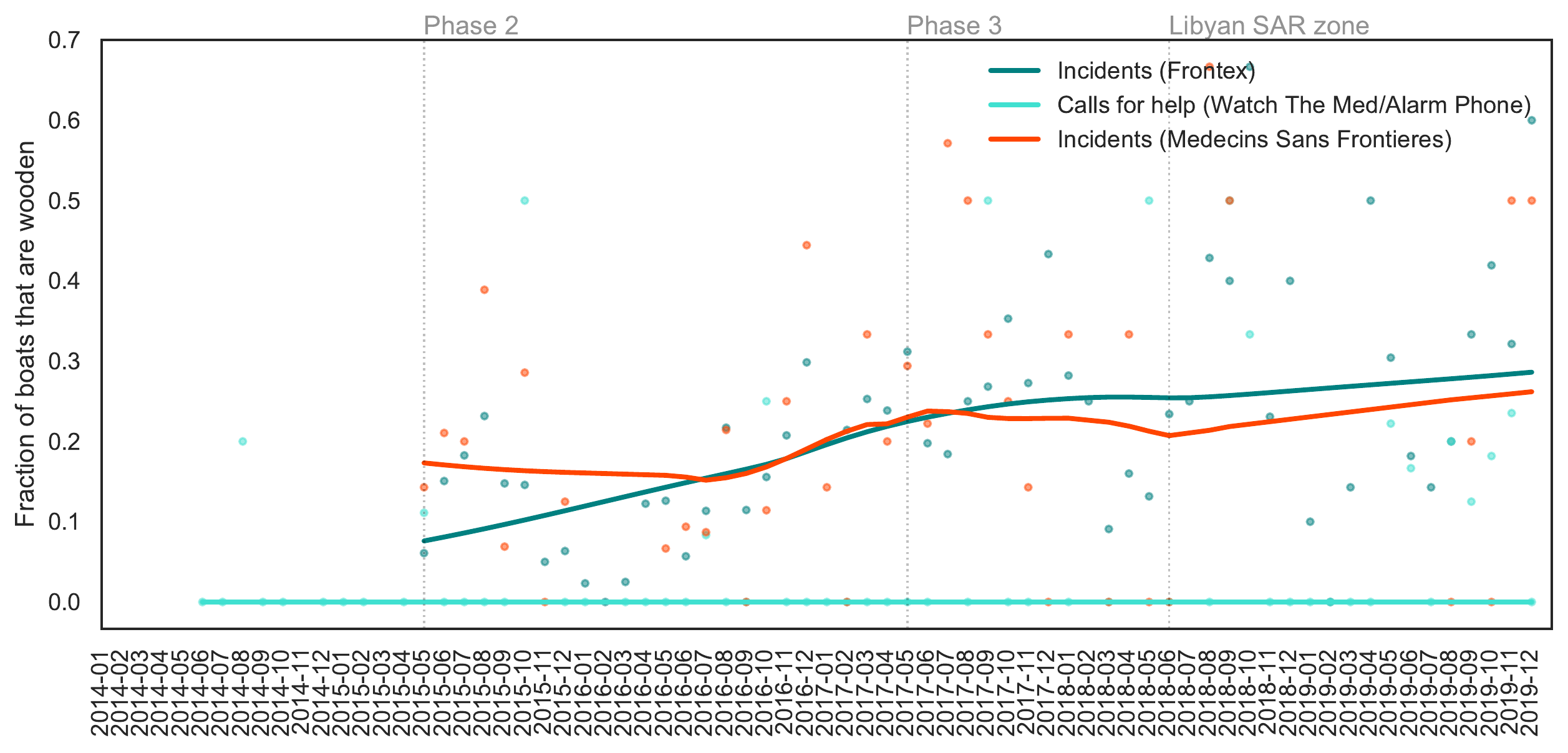}}
    	\caption{Monthly fraction of wooden boats}\label{fig:boat_type_wooden}
	\end{subfigure}
    \caption{The choice of boat type by month and dataset}\label{fig:boat_type_choice}
    \floatfoot{Each dot represents the monthly fraction of rubber or wooden boats for the respective dataset, whereas the lines represent locally weighted scatter plot smoothing (LOWESS) fits to the scatter plot trends. Note: The Frontex and Watch the Med datasets included boat types other than rubber or wooden. Therefore, the trends do not sum to one across both plots.}
\end{figure}

Next, we analyze the choice of boats over time, which is illustrated in Figure~\ref{fig:boat_type_choice}. In the Frontex and MSF data shown in Figure~\ref{fig:boat_type_rubber}, we can see that the use of rubber boats peaked in 2016, when NGO rescue activity was at its highest. While MSF continues to rescue a high share of rubber boats, in both datasets we can see an increasing tendency to select wooden boats, 
which continued through 2019. While the Watch The Med dataset appears to show an increase in the proportion of rubber rafts, this may be due to the fact that rubber rafts are increasingly likely to call Watch The Med for help because fewer of them are being independently discovered by NGOs or EU ships near the Libyan Coast, and because in earlier periods these boats had been more likely to call the Italian MRCC for help instead (see Appendix~\ref{subsec:data_comparison} for further discussion).

\FloatBarrier
\subsubsection{Variation in Outcomes (Boat Location) Over Time}

We have little visibility into the activities of the Libyan Coast Guard, since our incident datasets generally focus on NGO or EU-led rescues. In the absence of comprehensive data on interceptions, we focus on proxy outcomes based on the location of individual incidents.

Boats departing Libya pass from the Libyan SAR zone into EU (Italian/Maltese) SAR zones (these zones are illustrated in Figure \ref{fig:sar_zones_map}). In Phase 2, boats in either zone were highly likely to be rescued to Europe. However, starting in Phase 3 boats in the Libyan SAR zone were increasingly likely to be captured and returned to Libya, and crossing into the EU SAR zones was more and more important for securing rescue to Europe. Therefore, the average location of individual incidents in Phase 3 likely reflects how much effort smugglers are investing to avoid detection near the Libyan shore.

\begin{figure}
	\centerline{\includegraphics[width=5in]{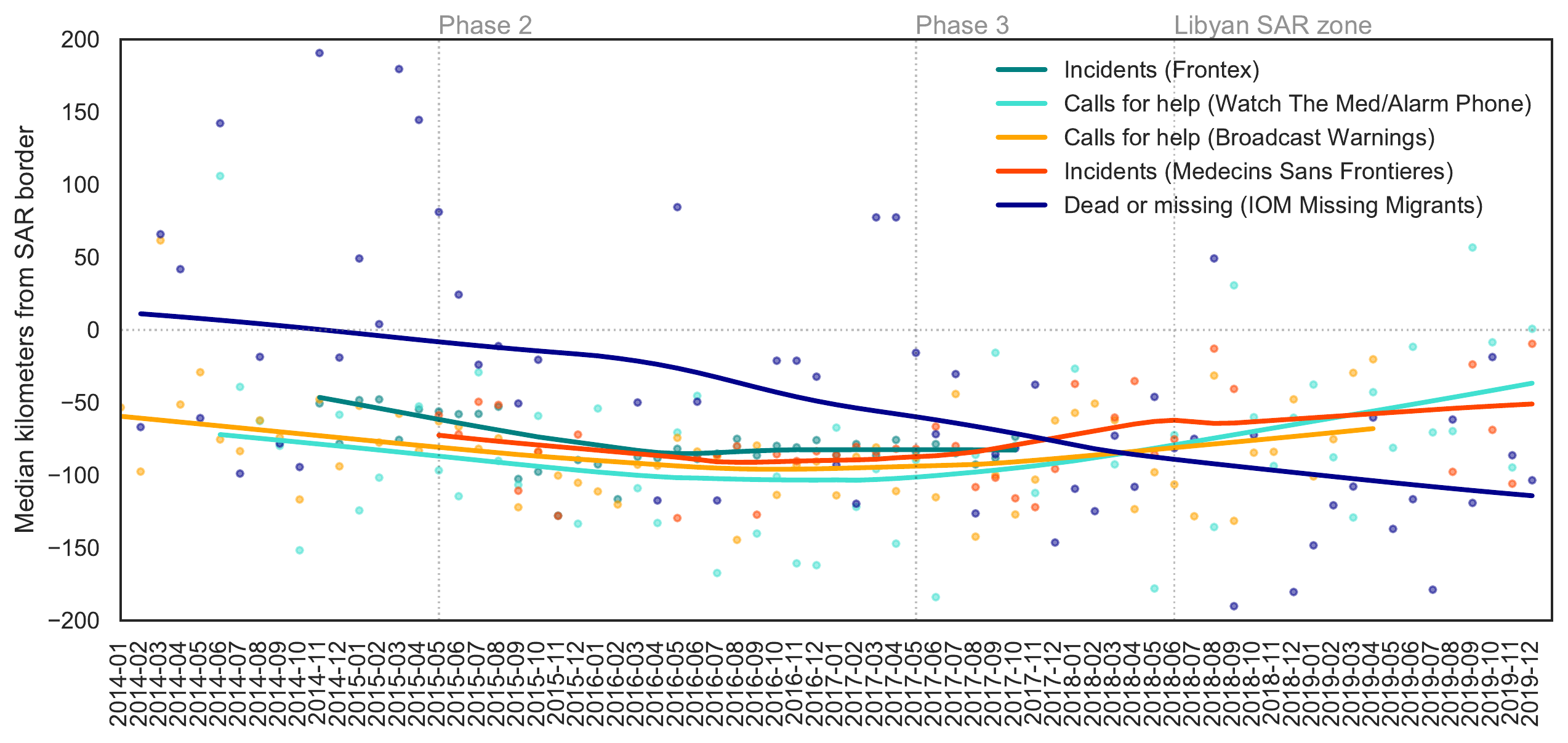}}
\caption{Distance from the Libyan SAR zone border - Incident location by month and dataset}\label{fig:dists}

\floatfoot{Each dot represents the monthly median distance or fraction of incidents for the respective data set, whereas the lines represent locally weighted scatter plot smoothing (LOWESS) fits to the scatter plot trends. In Panel (a), negative distances represent incidents on the Libyan side of the border, and positive distances represent incidents on the EU side of the border. In Panel (b), ``land'' is defined as the land mass of any country bordering the Mediterranean. In Panels (a) and (b), we note that the Frontex data set only included location data for a subset of time periods and for incidents that occurred \textit{outside} the Frontex operational area, that is, nearer to the coast of Libya.}
\end{figure}

Figure~\ref{fig:dists} plots the median distance to the SAR zone border over time, where negative distances represent incidents on the Libyan side of the border, and positive distances represent incidents on the EU side of the border. We can see that for almost all periods and datasets, the median incident occurs on the Libyan side of the border. Recorded incidents were closest to Libya in Phase 2 when NGO rescues were high, but have gradually moved towards the EU SAR zones as interceptions have grown. 

The fact that migrant boats are traveling farther out to sea before being rescued and/or identified by one of these data-collecting actors is likely a result of two different phenomena. On the one hand, a growing share of incidents near the shore are likely filtered out of our datasets by LCG interceptions. On the other hand, as the presence of rescue boats near the Libyan coast has declined, smugglers are aware that boats must go farther to increase their probability of rescue, and may adapt their strategies accordingly. As noted above, they appear to be launching smaller boats and shifting to wooden boats, but migrants on board may also be delaying their decision to call for help until they are farther away from the Libyan coast. Weak evidence for the latter hypothesis may be seen by comparing incident records from Watch The Med/Alarm Phone to the other incident datasets; in recent periods, calls to Watch The Med/Alarm Phone appear to show a slightly larger shift away from land and into the EU SAR zone. 

The one exception to these trends is the IOM missing migrants dataset, in which incidents appear to be moving closer to shore on average. It is important to note that this dataset \textit{does} include deaths recorded off the coast of Libya {(for example, IOM collects data on the number of dead and missing people from Libyan interceptions, and also reports incidents where bodies wash up on shore)} and therefore might contain a more representative sample of incidents in the Libyan SAR zone when interception rates are high. Deaths may be occurring increasingly near the coast for two reasons. First, it seems likely that LCG rescues may be more dangerous for migrants on average, due both to the lack of professionalism and expertise by the LCG, and to the fact that migrants sometimes conduct risky maneuvers to avoid capture by the LCG. Second, it is also likely that a larger proportion of boats near the coast go undetected due to the decrease in NGO patrol presence,  the slower response time of the LCG, and migrants' reluctance to call for help; this may lead to a growing number of sinking incidents in which bodies wash up on shore. 

\subsubsection{The Connection Between Strategic Inputs and Outcomes} 
Finally, we provide summary details on the connection between boat type, crowding, and incident outcomes. 
We assume that in Phases 1 and 2, reaching a European SAR zone or the Frontex operational area is relatively unimportant to migrants because almost all boats are rescued to Europe regardless of whether or not they are in the Libyan SAR zone. In Phase 3, however, we assume that migrants make the most effort to move away from the Libyan coast before being detected, since exiting the Libyan SAR zone and approaching Europe will make it more likely that they are rescued rather than returned to Libya. At the same time, we expect that the growing rate of interceptions in this period may make the payoffs to different boat sizes and types more pronounced, since a poor choice of boat size or type will leave migrants more vulnerable to interception. 

\begin{figure}[t]
	\begin{subfigure}{\textwidth}
    	\centerline{\includegraphics[width=5in]{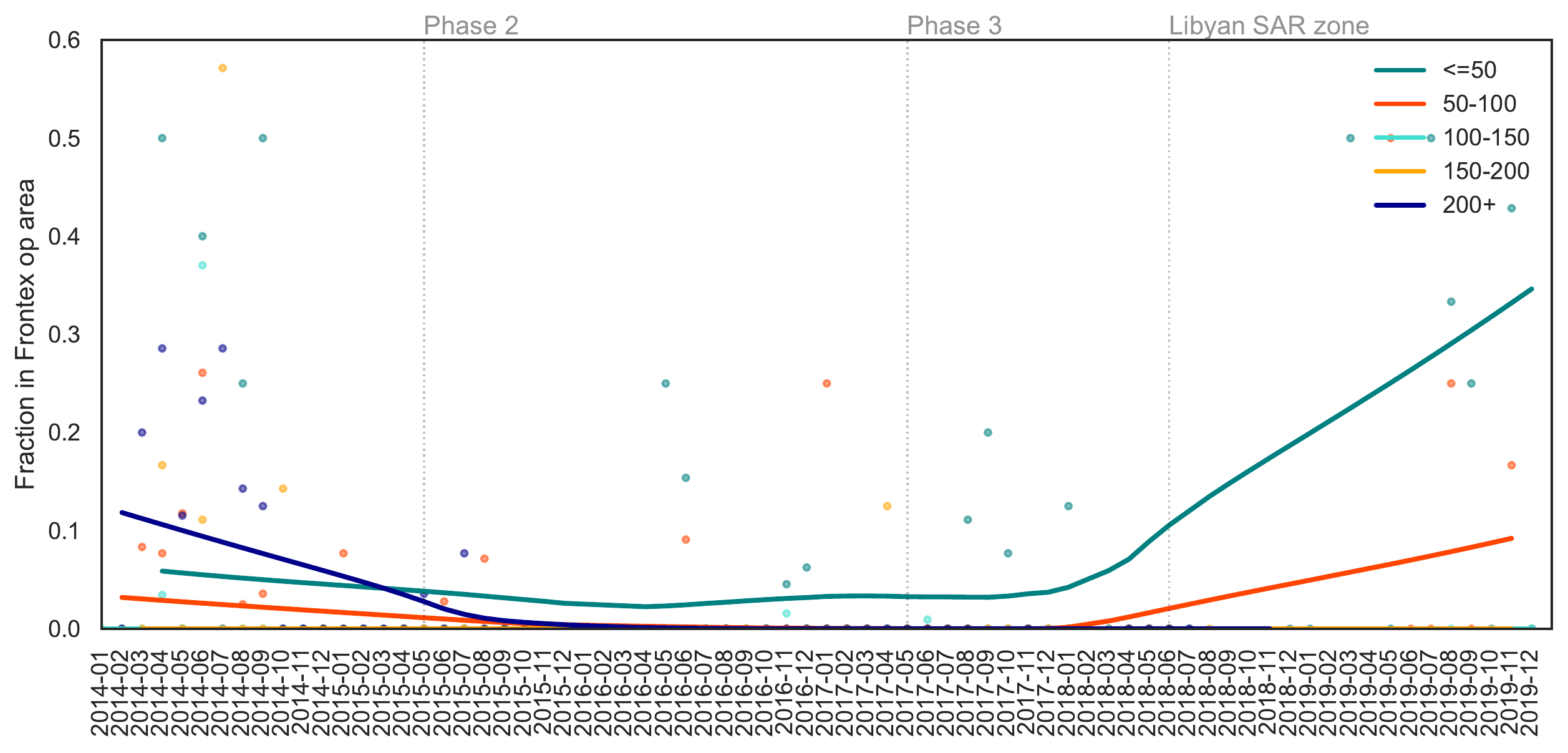}}
    	\caption{The probability that an incident is in the Frontex operational area, by number of people}\label{fig:p_in_x_n_ppl}
	\end{subfigure}
	\begin{subfigure}{\textwidth}
    	\centerline{\includegraphics[width=5in]{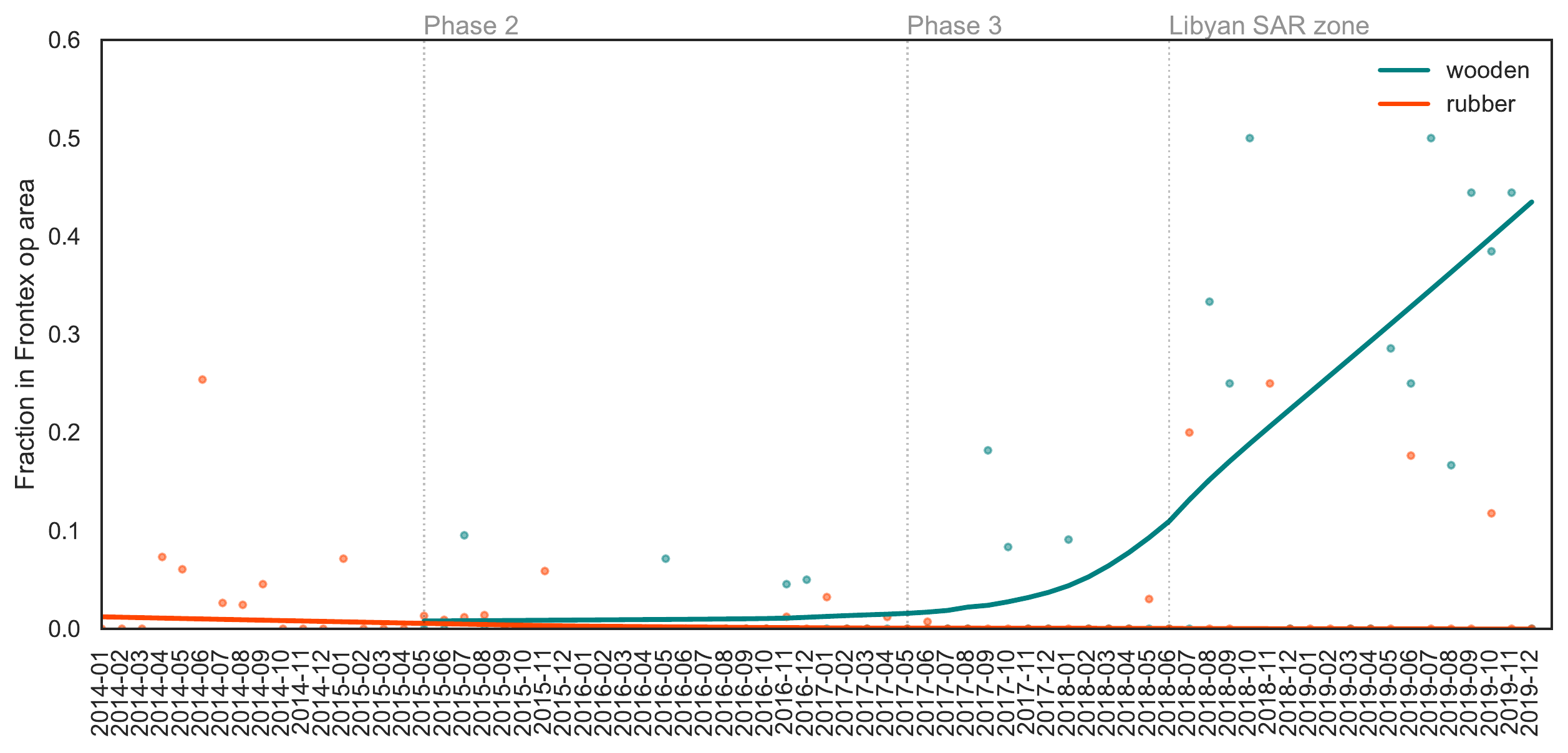}}
    	\caption{The probability that an incident is in the Frontex operational area, by boat type}\label{fig:p_in_x_boat_type}
	\end{subfigure}
\caption{Incident location by month and dataset, by number of people and boat type}\label{fig:probs}
\floatfoot{Each dot represents the monthly fraction of incidents in the Frontex operational area for the respective boat size or type, whereas the lines represent locally weighted scatter plot smoothing (LOWESS) fits to the scatter plot trends.}
\end{figure}

This is consistent with Figure~\ref{fig:probs},
which shows increasingly divergent outcomes by boat size and type in Phase 3. From Figure~\ref{fig:p_in_x_n_ppl}, we can see that smaller boats appear to have an advantage in reaching the Frontex operational area. This advantage is most pronounced for boats with less than 50 passengers, followed by boats with 50-100 passengers. From Figure~\ref{fig:p_in_x_boat_type}, we see that wooden boats are most successful in reaching the Frontex operational area; this is likely because they are more seaworthy in general, although we do see some rubber boats that are able to cross over successfully.\footnote{The Frontex dataset also includes a number of ``other'' boat types including: {fishing boats, motor boats, fiberglass boats, sailboats}, etc. On average, these ``other'' boats are {more successful in reaching the Frontex operational area}, but we do not analyze them because these boat types can be very heterogeneous, and because they represent only {a small portion (11\%) of incidents, relative to rubber boats (74\%) and wooden boats (14\%).} }

\subsection{Inferring Smuggler Strategy from Incident-Level Datasets}
In this section, we have analyzed smuggler strategy from two different angles. First, we have shown that the inputs chosen by smugglers (i.e., the size and type of boat) have varied over time. We also documented a corresponding shift in smuggling outcomes (i.e., the location of incidents involving migrants) and shown a correlation between these inputs and outcomes. This provides support for our main strategic model of boat size choice, which has been fit using incident-level data from Frontex.

We conclude with a note on the incident-level datasets analyzed above. Our analysis relies heavily on the Frontex dataset because it is the most comprehensive incident-level data source; we estimate that the number of people in the Frontex dataset closely corresponds to the number of arrivals to Europe reported by IOM (see Appendix~\ref{sec:justify_assumptions} for details). However, this dataset has a number of limitations, including that it is not publicly available; that it does not contain a full set of exact incident locations; and that, to the best of our knowledge, it does not cover interceptions by the Libyan Coast Guard. We also note that incident records from 2020 were not released for security reasons.

We analyze other incident-level datasets to address these shortcomings, comparing multiple datasets to demonstrate that incidents' average distances to shore are increasing and that smugglers are shifting towards the use of smaller boats and wooden boats. However, we note that each of these sources records different incident attributes, and each demonstrates different biases and shortcomings. Furthermore, for several of the datasets, the criteria by which incidents were identified or excluded is not entirely clear. Collecting data from multiple alternative sources also required substantial manual effort, and matching incidents across sources was challenging due to variations in the reported location, date, and number of people involved. Likely for this reason, we are unaware of other research that compares incident-level data from multiple different data sources in the region. Finally, we note that across all of these datasets, there is remarkably little transparency into recent activities off the coast of Libya, such as LCG interceptions and aerial surveillance flights run by European authorities to identify migrant boats that have not yet exited the Libyan SAR zone.  
\FloatBarrier
\newpage
\FloatBarrier
\section{Supplementary Details for the Incident-level Analysis}
\subsection{Justification of Assumptions}\label{sec:justify_assumptions}

Below, we justify the assumptions used in the primary incident-level analysis, which we outline in Section~\ref{subsec:assumptions}.

\textbf{Assumption 1:} \textit{Frontex's recorded incidents do not cover interceptions. However, Frontex data is a uniform and nearly comprehensive sample of rescue incidents in the region.}

\begin{figure}[htb]
        \centering
        \includegraphics[width=\linewidth]{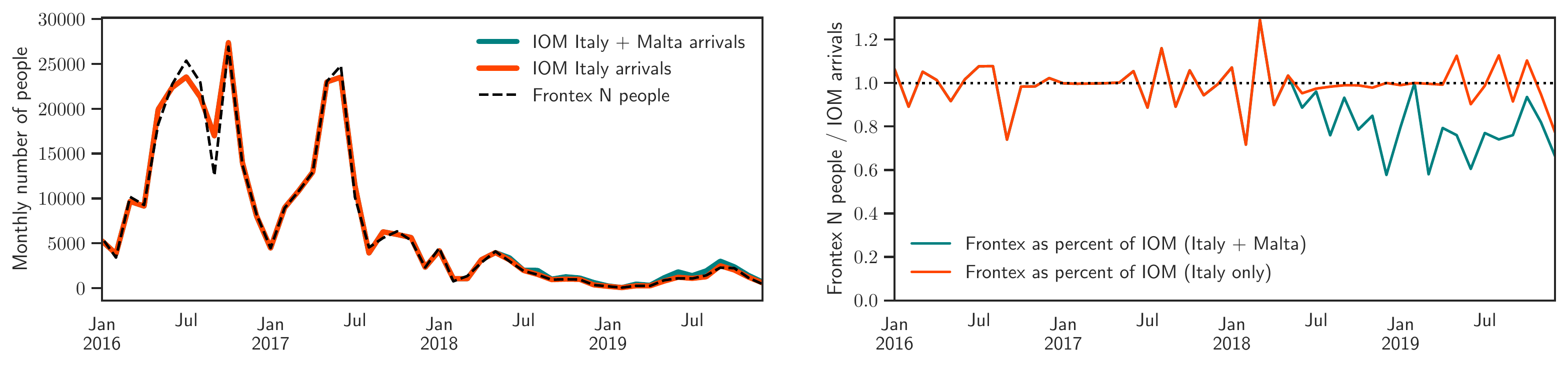}
        \caption{Comparison between people involved in Frontex incidents and total arrivals reported by IOM}
        \label{fig:representativeness_of_frontex_data}
    \end{figure}

To analyze the representativeness of Frontex's dataset, we compare the number of people involved in Frontex incidents to the total number of sea arrivals to Italy or Malta reported by the IOM. As shown in Figure~\ref{fig:representativeness_of_frontex_data}, while the datasets do not match exactly, the Frontex dataset appears to capture most people who reached Europe on this route during our study period. A discrepancy emerges in 2018, when the IOM data begins to incorporate data on arrivals to Malta as well as Italy; however, the coverage of the Frontex dataset remains around 70-80\%. Therefore, it seems reasonable to assume that the distribution of departure countries in the Frontex dataset approximates the distribution of departure countries in the IOM dataset.

\textbf{Assumption 2:} \textit{The smugglers' decision focuses on the probability of interception and does not independently weigh the probability of sinking, which is small.}

In the Frontex incident dataset, only {5 of the 4,365} incidents involving boats departing Libya suffered the loss of all people on board,  
and the overall fatality rate is just {0.2} percent. 
According to the IOM flows dataset, approximately {2\%} of all migrants attempting the crossing between 2016 - 2019 have gone dead or missing. 

In Appendix~\ref{sec:ttests}, we show that Frontex incidents (i.e., incidents in which a boat was rescued to Europe) were slightly less likely to involve dead or missing migrants in Phase 3, even as smaller boats were used and the incidents appear to have occurred farther from Libya.  Therefore, it appears that the strategic changes have not increased the risk of death for passengers, conditional on boats being rescued to Europe. 

In the IOM flows dataset, there appears to have been a rise in the fatality rate during Phase 3, but IOM does not report whether these fatalities are associated with the Libyan or the Tunisian route. On the Libyan route, we estimate that fatalities may be closely associated with the probability of interception, because (1) migrants may be exposed to risks in the course of LCG operations; (2) the LCG may have a slower overall response time to distress incidents when it is charged with a rescue; and (3)  the efforts of migrants to evade the LCG may lead them to sink without detection. Therefore, we expect that the probability of interception may act as a proxy for the risk of sinking.

\subsection{Additional Details on the Frontex Incident Dataset}\label{sec:frontex_descriptives}
\begin{figure}
    \includegraphics[height=2in]{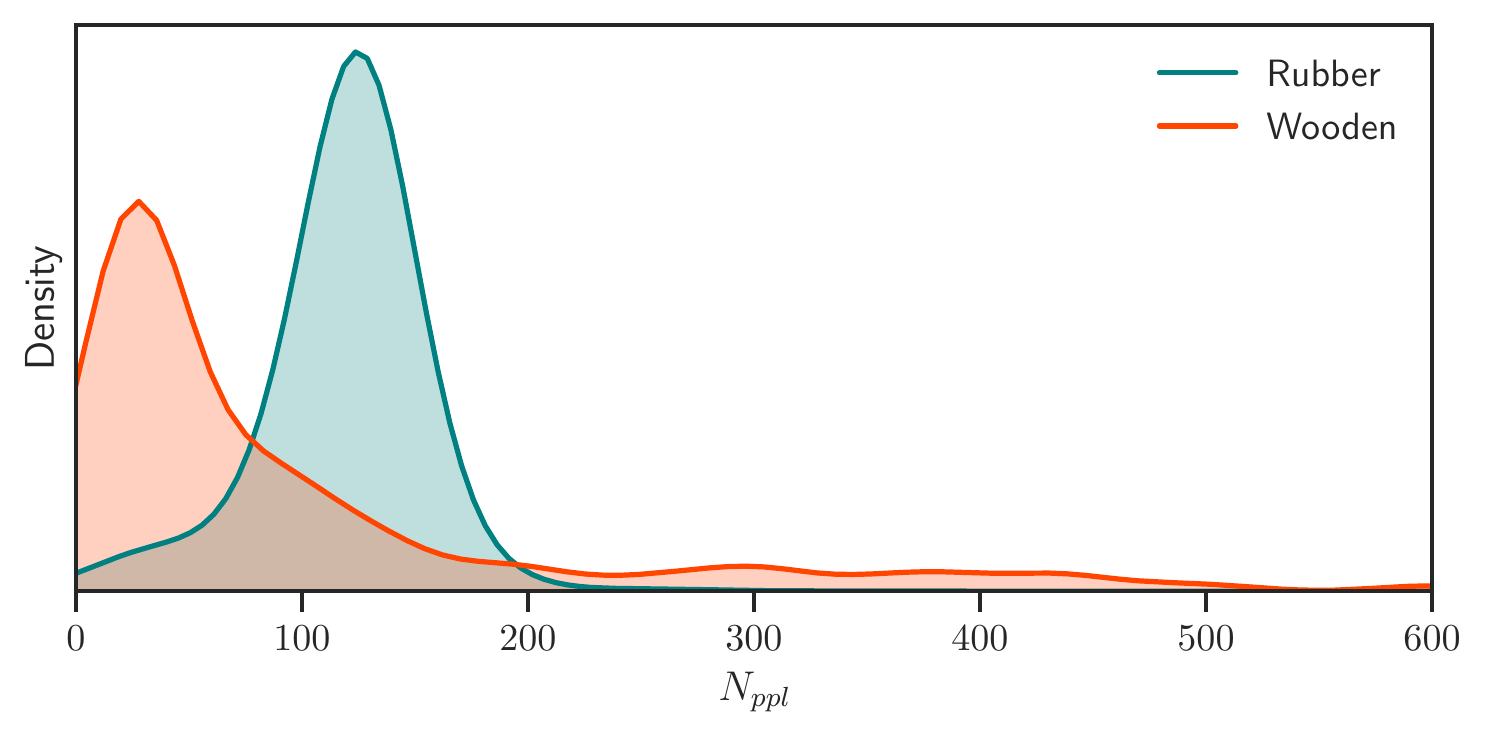}
    \includegraphics[height=2in]{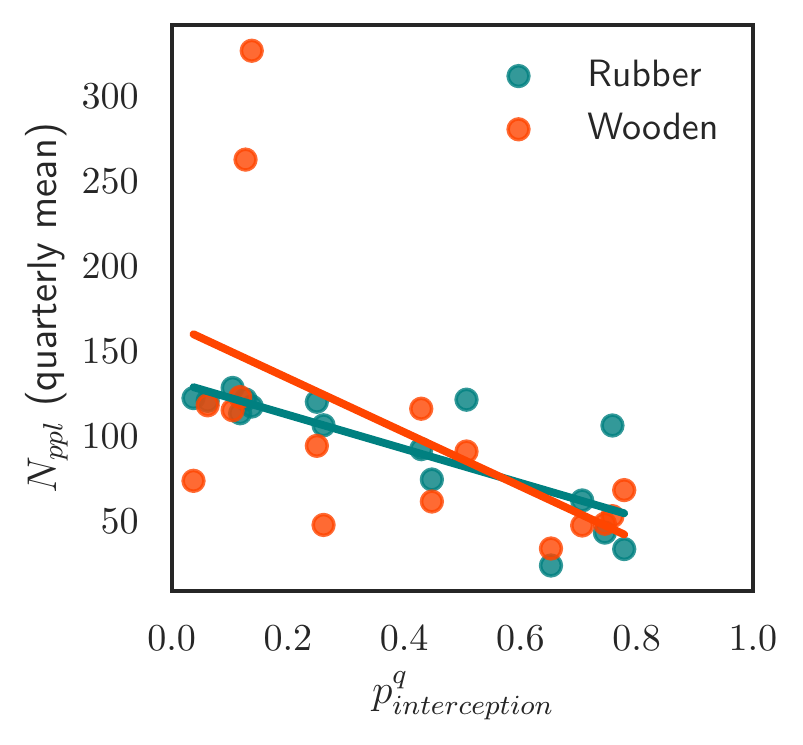}
    \caption{The distribution of boat sizes in the dataset (left) and the relationship between average boat size and the quarterly probability of interception (right)}
    \label{fig:frontex_descriptives}
    \floatfoot{Note that the long tail on the distribution of boat sizes extends beyond 600, but the axis range has been limited for clarity of presentation.}
\end{figure}

\FloatBarrier
Below, we briefly provide additional summary statistics on the Frontex incident dataset. In Figure~\ref{fig:frontex_descriptives}, we plot the empirical distribution of boat sizes in the dataset. We see that most wooden boats tend to be small, but that there is a long tail of extremely large boats. In contrast, rubber boats generally hold under 200 people, with a peak around 100 - 150 passengers. For this reason, we have focused our estimation on rubber boats. The right panel of the figure illustrates that for both types of boats, the average number of people on board is positively correlated with the quarterly probability of rescue, which is consistent with the hypothesis that smugglers are responding strategically to LCG interceptions by changing the size of the boats they launch.

Table~\ref{tbl:mnl_support} shows the quarterly number of Libyan incidents in the Frontex dataset by boat type and size. In general, there are more incidents involving rubber than wooden boats, which further motivates our decision to focus on rubber boats. We also note that there are far fewer incidents starting in 2017Q3; this is consistent with the fact that the overall volume of departures fell dramatically once the rate of LCG interceptions increased, and with the fact that an increasing number of boats were presumably intercepted by the LCG before they could be captured in the Frontex dataset.

\begin{table}[htb]
    {\footnotesize\begin{tabular}{|l|rrr|rrr|r|}
\hline
{} & \multicolumn{3}{l|}{Rubber} & \multicolumn{3}{l|}{Wooden} & Total \\
{} &   0$<$N$<$= 50 &  50$<$N$<$=100 & 100$<$N$<$=900 &   0$<$N$<$= 50 &  50$<$N$<$=100 & 100$<$N$<$=900 & \\
       &                &                &                &                &                &                &       \\
\hline
2016Q1 &              0 &             16 &            118 &              0 &              0 &              3 &   137 \\
2016Q2 &              9 &             15 &            243 &              6 &              1 &             21 &   295 \\
2016Q3 &             16 &             45 &            308 &             58 &              1 &             11 &   439 \\
2016Q4 &             20 &             30 &            250 &             51 &              4 &             20 &   375 \\
2017Q1 &             12 &             22 &            120 &             21 &             12 &             14 &   201 \\
2017Q2 &              6 &             48 &            278 &             32 &             37 &             41 &   442 \\
2017Q3 &              7 &             11 &             92 &             17 &              6 &              7 &   140 \\
2017Q4 &              5 &             17 &             42 &             33 &              3 &              4 &   104 \\
2018Q1 &              5 &             13 &             17 &              8 &              1 &              5 &    49 \\
2018Q2 &              4 &              8 &             33 &              7 &              3 &              3 &    58 \\
2018Q3 &              0 &              0 &              1 &              1 &              2 &              0 &     4 \\
2018Q4 &              3 &              0 &              0 &              2 &              1 &              0 &     6 \\
2019Q1 &              3 &              0 &              0 &              0 &              1 &              0 &     4 \\
2019Q2 &              4 &              4 &              0 &              2 &              2 &              0 &    12 \\
2019Q3 &              5 &              6 &              1 &              3 &              2 &              0 &    17 \\
2019Q4 &              3 &              8 &              3 &              7 &              2 &              1 &    24 \\
\hline
Total  &            102 &            243 &           1506 &            248 &             78 &            130 &  2307 \\
\hline
\end{tabular}
}
    \caption{The number of Libyan incidents in the Frontex dataset, by boat type and size}\label{tbl:mnl_support}
\end{table}
    
\FloatBarrier

\subsection{T-tests Supporting Strategic Shifts in the Frontex Dataset}\label{sec:ttests}

Next, we briefly compare the characteristics of Frontex incidents originating in Libya during Phase 2 and 3 using two-sample t-tests with unequal variances. This analysis is intended to support the descriptive plots included in Section~\ref{sec:incident_desc}. From Table~\ref{tab:ttests}, we can see that incidents in Phase 3 have a significantly lower average number of people per boat; are significantly less likely to involve rubber boats; and are significantly more likely to occur in the Frontex operational area. Incidents in Phase 3 are less likely to involve dead or missing migrants, and have fewer deaths on average. However, this latter result may be a function of boat size; when we test for differences in the average \textit{proportion} of dead or missing people per boat, we find no significant difference. 

\begin{table}[htb]
    \centering
    \begin{tabular}{|l|rrrr|}
\hline
Variable & Phase 2 Mean & Phase 3 Mean & Difference & P-value \\ 
\hline
Number of people per transport means & 134.857 &  108.978 & 25.878$^{***}$ & 0.000 \\ 
Boat type = rubber (vs. wooden) & .844 &  .731 & .113$^{***}$ & 0.000 \\ 
In Frontex operational area & .009 &  .052 & -.043$^{***}$ & 0.000 \\ 
Incident involved dead or missing & .053 &  .039 & .014$^{**}~$ & .0400 \\ 
Number of dead or missing & .331 &  .187 & .144$^{*}~~$ & .0899 \\ 
Fraction of dead or missing & .004 &  .004 & -.001$~~~~$ & .3196 \\ 
\hline
\end{tabular}

    \caption{T-tests of strategic shifts in the Frontex dataset}
    \label{tab:ttests}
	\floatfoot{Phase 2 has been defined as May 2015 - April 2017, whereas Phase 3 has been defined as  May 2017 - December 2019.}
\end{table}

\FloatBarrier
\subsection{Robustness to Alternative Weighting Schemes}\label{sec:robustness_mnl_weights}

To estimate the utility function using incident-level data, we used frequency weights in which each incident was weighted by the total number of incidents in its quarter: $w=\frac{1}{N^q}$. This ensured that each quarter was given equal weight in the estimation and effectively up-weighted incidents from later quarters, when there were fewer incidents observed.

To show that our results are not an artefact of the weighting scheme, in Columns (4)-(6) of Table~\ref{tbl:utility_all_models} we compare the frequency weights with two alternative weighting schemes. Column (4) shows unweighted estimates, whereas Column (5) shows estimates weighted according to the probability of rescue: $w = \frac{1}{p^{q}_{rescue}}$. The motivation for weighting according to the probability of rescue is that when the probability of rescue is low, each observed incident should be up-weighted because it represents other, unobserved incidents which were filtered out of the dataset by interceptions. 

When we compare estimates from these alternative weights to the frequency weights in Column (6), we see that the frequency weights lead to estimates that are more extreme in magnitude. That is, the baseline payoff to crowding ($\alpha_n$) is higher, whereas the penalty to interception ($\beta_n$) is more negative. This is unsurprising because, as noted above, the frequency weights place higher weights on incidents later in the dataset, when the probability of interception is higher and the strategic response is more evident.

In Columns (1)-(3) of Table~\ref{tbl:utility_all_models}, we also present the results of a simple model where utility consists only of the first ($\alpha_n$) term (i.e., the payoff to crowding) and ignores the penalty to interceptions ($\beta$); again, we consider all three weighting schemes. We can see that in this model, estimates of $\alpha_n$ are more conservative in magnitude than in the full model; this is presumably because the simpler model fails to independently capture the interception penalty associated with larger boats, and this most likely dampens the estimates of $\alpha_n$.

\begin{table}
\centerline{{
\def\sym#1{\ifmmode^{#1}\else\(^{#1}\)\fi}
\begin{tabular}{l*{6}{c}}
\toprule
                    &\multicolumn{1}{c}{(1)}&\multicolumn{1}{c}{(2)}&\multicolumn{1}{c}{(3)}&\multicolumn{1}{c}{(4)}&\multicolumn{1}{c}{(5)}&\multicolumn{1}{c}{(6)}\\
\midrule
$\alpha_{~50-100}$           &    0.868\sym{***}&    0.683\sym{***}&   -0.166         &    1.225\sym{***}&    1.340\sym{***}&    1.786\sym{***}\\
                    &  (0.118)         &  (0.134)         &  (0.270)         &  (0.177)         &  (0.198)         &  (0.413)         \\
\addlinespace
$\alpha_{~100+}$          &    2.692\sym{***}&    2.373\sym{***}&    0.886\sym{***}&    3.537\sym{***}&    3.704\sym{***}&    3.849\sym{***}\\
                    &  (0.102)         &  (0.117)         &  (0.264)         &  (0.158)         &  (0.176)         &  (0.604)         \\
\addlinespace
$\beta_{~50-100}$     &                  &                  &                  &   -1.740\sym{***}&   -2.178\sym{***}&   -3.587\sym{***}\\
                    &                  &                  &                  &  (0.609)         &  (0.676)         &  (0.955)         \\
\addlinespace
$\beta_{~100+}$    &                  &                  &                  &   -5.161\sym{***}&   -5.986\sym{***}&   -6.511\sym{***}\\
                    &                  &                  &                  &  (0.572)         &  (0.587)         &  (1.998)         \\
\midrule
weights & - & rescue & frequency & - & rescue & frequency \\
pseudo R2           &    0.459         &    0.397         &    0.106         &    0.483         &    0.458         &    0.268         \\
N Obs.              &    5,553         &    5,553         &    5,553         &    5,553         &    5,553         &    5,553         \\
\bottomrule
\multicolumn{7}{l}{\footnotesize Standard errors in parentheses}\\
\multicolumn{7}{l}{\footnotesize \sym{*} \(p<0.10\), \sym{**} \(p<0.05\), \sym{***} \(p<0.01\)}\\
\end{tabular}
}
} 
\caption{Results from the conditional logit model, under different weighting schemes}\label{tbl:utility_all_models}
\end{table}

\FloatBarrier

\FloatBarrier

\end{document}